\documentclass{aa}  

\usepackage{graphicx}
\usepackage{natbib}
\bibpunct{(}{)}{;}{a}{}{,}
\usepackage{txfonts}
\newcommand {\hi} {{\rm H}\,{\small\rm I}}

\newcommand {\hicap} {{\rm H}\,{\scriptsize\rm I}}

\newcommand {\cii} {{\rm C}\,{\small\rm II}}
\newcommand {\nii} {{\rm N}\,{\small\rm II}}

\newcommand {\kms} {\,{\rm km\,s}^{-1}}
\newcommand {\ergs} {\,{\rm erg\,s}^{-1}}
\newcommand {\erg} {\,{\rm erg}}
\newcommand {\pc} {\,{\rm pc}}

\newcommand {\kpc} {\,{\rm kpc}}

\newcommand {\cmmq}{\,{\rm cm^{-2}}}
\newcommand {\cmmc}{\,{\rm cm^{-3}}}
\newcommand {\kmskpc} {\,{\rm km\,s}^{-1}\,{\rm \kpc}^{-1}}

\newcommand {\de}{^{\circ}}
\newcommand{\rsun}{\,{\rm R}_\odot}

\newcommand {\msun}{\,{\rm M}_\odot}

\newcommand {\msunsqp}{\,{\rm M}_\odot \, {\rm pc}^{-2}}
\newcommand{\vsun}{\,{v}_\odot}

\newcommand{\yr}{\,{\rm yr}}

\newcommand{\K}{\,{\rm K}}

\newcommand {\msunspc}{\,{{\rm M}_\odot\,\rm pc}^{-2}}

\newcommand{\avg}[1]{\left< #1 \right>} 
\def\apjl{ApJ}%
\def\apjs{ApJS}%
\def\ao{Appl.~Opt.}%
\def\aap{A\&A}%
\begin{document} 
\title{Distribution and kinematics of atomic and molecular gas inside the Solar circle}
\titlerunning{Modelling the atomic and molecular gas inside the Solar circle}
\authorrunning{A. Marasco et al.}

   \author{A. Marasco\inst{1},
          F. Fraternali\inst{2,3},
          J.\,M. van der Hulst\inst{3}
          \and
          T. Oosterloo\inst{1,3}
          }

    \institute{ASTRON, Netherlands Institute for Radio Astronomy, 
		 Postbus 2, 7900 AA Dwingeloo, The Netherlands\\
		  \email{marasco@astron.nl}
	\and
             Department of Physics and Astronomy, University of Bologna, 
             via P. Gobetti 93/2, 40129 Bologna, Italy
   	 \and
    		Kapteyn Astronomical Institute, University of Groningen,
   		Postbus 800, 9700 AV Groningen, The Netherlands
             }

   \date{Received ; accepted}

 
\abstract
{The detailed distribution and kinematics of the atomic and the CO-bright molecular hydrogen in the disc of the Milky Way inside the Solar circle are derived under the assumptions of axisymmetry and pure circular motions.
We divide the Galactic disc into a series of rings, and assume that the gas in each ring is described by four parameters: its rotation velocity, velocity dispersion, midplane density and its scale height. We fit these parameters to the Galactic \hicap\ and $^{12}$CO (J=1-0) data by producing artificial \hicap\ and CO line-profiles and comparing them with the observations. Our approach allows us to fit all parameters to the data simultaneously without assuming a-priori a radial profile for one of the parameters.
We present the distribution and kinematics of the \hicap\ and H$_2$ in both the approaching (QIV) and the receding (QI) regions of the Galaxy.
Our best-fit models reproduces remarkably well the observed \hicap\ and CO longitude-velocity diagrams up to a few degrees of distance from the midplane. 
With the exception of the innermost $2.5\kpc$, QI and QIV show very similar kinematics.
The rotation curves traced by the \hicap\ and H$_2$ follow closely each other, flattening beyond $R\!=\!6.5\kpc$.
Both the \hicap\ and the H$_2$ surface densities show a) a deep depression at $0.5\!<\!R\!<\!2.5\kpc$, analogous to that shown by some nearby barred galaxies, b) local overdensities that can be interpreted in terms of spiral arms or ring-like features in the disk.
The \hicap\ (H$_2$) properties are fairly constant in the region outside the depression, with typical velocity dispersion of $8.9\pm1.1\,(4.4\pm1.2)\kms$, density of $0.43\pm0.11\,(0.42\pm0.22)\cmmc$ and HWHM scale height of $202\pm28\,(64\pm12)\pc$.
We also show that the \hicap\ opacity in the LAB data can be accounted for by using an `effective' spin temperature of $\sim150\K$: assuming an optically thin regime leads to underestimate the \hicap\ mass by about $30\%$.}
   \keywords{Galaxy: kinematics and dynamics -- Galaxy: structure -- Galaxy: disk -- solar neighborhood -- ISM: kinematics and dynamics}
   \maketitle
%

\section{Introduction}
For over 60 years, the properties of the neutral gas in the Galaxy have been studied extensively by several authors.
The pioneering work by \citet{Westerhout57} allowed imaging of the spatial distribution of the \hi\ in the Milky Way for the first time.
In Westerhout's work, the transformation from the observed space to the physical space was achieved by assuming an a-priori model for the Galactic velocity field, i.e. for the rotation curve. 
Later on, a similar analysis was repeated several times, with relatively minor modifications to the original procedure, by using the most up-to-date observations for both the atomic \citep{Kulkarni+82,Burton86,NakanishiSofue03,KalberlaDedes08} and the molecular hydrogen \citep{Bronfman+88,NakanishiSofue06}.
These works have revealed that the gas distribution in the Milky Way is quite complex: the \hi\ disc is flared and warped \citep{Levine+06} and a multi-arm spiral structure is visible in both the atomic and the molecular hydrogen \citep[e.g.][]{Dame+86,Levine+06scie}.
Insights on the dark matter distribution in the disc can be obtained when the mapping is coupled to a self-consistent mass model \citep{Kalberla+07}.

Despite its many successes, this approach has a number of shortcomings.
Perhaps the most severe of them was pointed out by \citet{Burton74}, who noticed that the derived gas density distribution is dramatically sensitive to small changes in the assumed velocity field, as velocity perturbations by only a few $\kms$ can mimic variations in density up to a factor of a hundred.
An additional complication is introduced by the so-called `near-far problem', which occurs in the Galactic region inside the solar circle (hereafter, the `inner' Galaxy) where the mapping is not unique: for each sight-line, two distinct distances - one in front and one beyond the tangent point - can be associated to the same line-of-sight velocity.
This produces ambiguity in the reconstruction of the density field, and even though clever solutions have been proposed \citep[e.g.][]{NakanishiSofue03}, the detailed gas distribution in the inner Galaxy remains quite uncertain.
Finally, not all these studies show a direct comparison between the synthetic line-profiles - produced by the derived gas density distribution moving in the modelled velocity field - and the data, which is key to assessing the goodness of the model.

Since any prejudice on the rotation curve can significantly affect the derived gas distribution, it is mandatory to determine the former independently from the latter.
The rotation curve in the inner Galaxy is classically derived by using the so-called \emph{tangent point} method \citep{Kwee+54}.
This method assumes that the gas has pure circular motions, and it is based on extracting the terminal-velocity from the line profiles and assigning it to gas located at the tangent-point radius $R_{\rm tp}\!=\!\rsun\sin(l)$, where the near-far ambiguity vanishes.  
However, the identification of the terminal-velocity in the profiles is not straightforward, because gradients in gas density and velocity dispersion along the line-of-sight have a large impact on the tail of the profiles.
Hence, the terminal velocity is often derived by assuming that the velocity dispersion and the density of the gas are constant for a wide range of line-of-sight velocities \citep{ShaneBiegerSmith66, Celnik+79,RohlfsKreitschmann87, Malhotra95}.
This leads to a certain degree of circularity in the method, as an assumption on the gas density profile must be made in order to derive the terminal velocities, and so the rotation curve, which thereafter can be used to image the spatial distribution of the gas itself.

In this paper, we present a novel approach to derive the kinematics and the distribution of the atomic and molecular hydrogen in the inner Galaxy via the modelling of the observed emission-line profiles.
Our modelling is based on the following assumption: the Galactic disc is axisymmetric and the gas moves on co-planar and concentric circular orbits, without coherent radial or vertical motions.
While this assumptions may appear crude, it allows us to fit \emph{simultaneousy} both the gas kinematics (rotation velocity and velocity dispersion) and the gas distribution (midplane density and scale height) of our model to the data.
This approach circumvents most of the aforementioned issues, and provides a simple model of the inner Galactic disc that is in remarkable agreement with the observations.
Note that the modelling of the gas kinematics in nearby galaxies is classically done with the so-called `tilted ring' technique \citep[e.g.][]{Rogstad+74,Begeman87}, which is based on the same assumption adopted here\footnote{Except for the departures from co-planarity due to the presence of warps, which are not important in this context.}.

This paper is structured as follows.
In Section \ref{method} we describe in details our models and present the method adopted to fit them to the data.
In Sections \ref{resultshi} and \ref{resultsh2} we show our results on the kinematics and distribution of the \hi\ and H$_2$ as derived by our modelling.
We discuss our results in Section \ref{discussion} and present our conclusions in Section \ref{conclusions}.

In this work, we use two coordinate systems.
The first is the Galactic coordinate system $(l,b,v)$, where $l$ and $b$ are the Galactic longitude and latitude and $v$ is the line-of-sight velocity in the local standard of rest.
The second is a cylindrical coordinate system $(R,\phi,z)$, where $R$ and $z$ are the distances perpendicular to and along the rotation axis of the Galaxy and $\phi$ is the azimuthal coordinate, assumed positive in the direction of the Galaxy rotation.
We use the Galactic constants $\rsun\!=\!8.3\kpc$ and $\vsun\!=\!240\kms$ \citep{McMillan11}.
In Appendix \ref{Vsun220} we show how our results change if we assume a different value for $\vsun$.

\section{Method}\label{method}
\begin{figure*}[tb]
\centering
\includegraphics[width=0.85\textwidth]{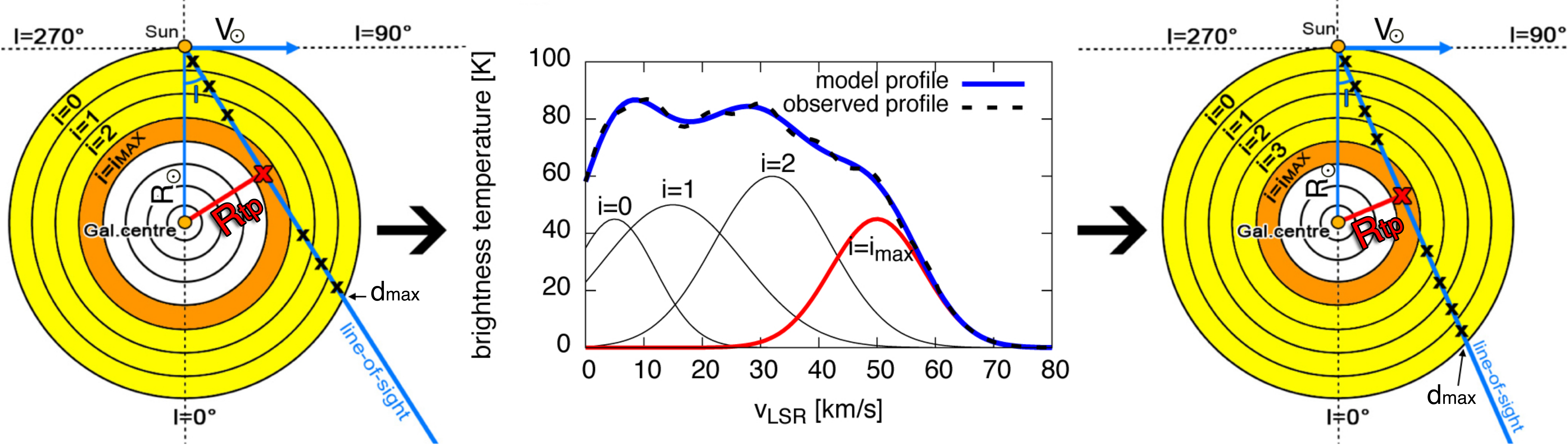}
\caption{Scheme of the iterative fitting method used in this work. The \emph{first panel} sketches the inner Milky Way divided into rings. The gas emission-line profile at longitude $l$ is given by the sum of the contributions of all rings with $0\le i\le i_{\rm max}$ (\emph{second panel}), the parameters of the last ring can be found by fitting the last Gaussian of the model profile (thick red curve) to the data (dashed curve). We iterate this procedure by decreasing $l$ at each step (\emph{third panel}).}
\label{FIGmethod}
\end{figure*}

To model the atomic and the molecular hydrogen components of the Milky Way, we make use of the Leiden-Argentine-Bonn (LAB) all-sky 21-cm survey \citep{Kalberla+05} and of the $^{12}$CO ($J\!=\!1\!-\!0$ rotational transition line) survey of \citet{Dame+01}. 
We use an hanning-smoothed version of the LAB data, with an angular (velocity) resolution of $\sim0.6\de$ ($\sim2\kms$) and voxel size of $0.5\de\times0.5\de\times2\kms$.
The CO data of \citet{Dame+01} have instead angular (velocity) resolution of $\sim0.15\de$ ($\sim1.3\kms$) and voxel size of $0.125\de\times0.125\de\times1.3\kms$.
The different angular and spectral resolutions may complicate the interpretation of our results, thus we prefer to bring both datasets to a common resolution.
In order to increase the signal-to-noise ratio and reduce the number of points to fit, we smooth both datacubes to an angular resolution\footnote{intended as the FWHM of a Gaussian kernel} of $1\de\times1\de$.
The CO datacube was also smoothed to a velocity resolution of $2\kms$, and re-binned to match the voxel size of the LAB data.
For the rest of the paper we will be using these `degraded' versions of the datacubes.

We focus on the regions where the emission-lines are produced by the inner ($R\!<\!\rsun$) Galaxy: $0\de\!<\!l\!<\!90\de$ and $v\!>\!0$, or \emph{first quadrant} (QI), and at $270\de\!<\!l\!<\!360\de$ and $v\!<\!0$, or \emph{fourth quadrant} (QIV). 
Because of the Galaxy rotation, QI appears to be receding from an observer placed at the location of the Sun, whereas QIV appears to be approaching. 
We model QI and QIV separately.

We assume that the Galactic disc can be decomposed into a series of concentric and co-planar rings, centred at the Galactic centre.
Following \citet{Olling95}, the vertical distribution of the gas is approximated as a Gaussian.
The gas in each ring is described by four parameters: the rotation velocity $v_{\phi}$, the velocity dispersion $\sigma$, the midplane volume density $n_0$ and the  scale height $h_s$, which is the half width at half maximum (HWHM) of the Gaussian distribution (so that $n_0 e^{-z^2/(1.44 h_s^2)}$ gives the gas density $n$ as a function of $z$ for a given ring).
We assume that the gas kinematics does not change with $z$ (but see Section \ref{extrapl}).
These four parameters are then fit to the data: our fitting strategy consists of building a synthetic observation of our model (Section \ref{lineprofile}) and comparing it with the data (Section \ref{fitting}).

We model the inner disc by using $180$ rings for each Galactic quadrant, each ring is at distance $\rsun|\sin(l)|$ from the Galactic centre: this gives us two points per resolution element per quadrant.
We do not model the warp in the Galactic disc as its amplitude is negligible in the regions considered \citep{Levine+06}.

In the following, we describe how we build synthetic datacubes for the \hi\ emission of our model, and how we compare them to the observations.
An equivalent procedure is used for the molecular gas, with a few differences discussed in Section \ref{resultsh2}.

\subsection{Modelling the gas on the midplane} \label{midplane}
We first focus on the \hi\ located on the midplane ($b\!=\!0\de$) and fit the parameters $v_{\phi}$, $\sigma$ and $n_0$ of each ring to the LAB data.

Our iterative method, similar to that adopted by \citet{Kregel+04} to study the \hi\ kinematics in nearby edge-on galaxies, is sketched in Fig.\,\ref{FIGmethod}.
We consider an observer located on the midplane at $R\!=\!\rsun$ and we are interested in modelling the \hi\ brightness temperature profile $T_{\rm B}(v)$ at a given Galactic longitude $l$ and at latitude $b\!=\!0\de$.
This profile is built from the contribution of all rings with $0\!\le\!i\!\le\!i_{\rm max}$ along the line-of-sight, where $R_0\!\equiv\!\rsun$ and $R_{i_{\rm max}}\!\equiv\!\rsun\sin(l)\!=\!R_{\rm tp}$.
Note that rings at radii smaller than $R_{\rm tp}$ do not contribute to this profile.
For each new $l$ we fix $v_{\phi,i}$, $\sigma_i$ and $n_{0,i}$ of all rings with $i<i_{\rm max}$ to the values determined by the previous iterations, we evaluate a synthetic brightness temperature profile $T_{\rm B}(v)$ (see Section \ref{lineprofile}) and fit it to the observed profile by varying only the parameters of the ring at $i=i_{\rm max}$. 
Details of the fitting procedure are given in Section \ref{fitting}.
We apply this method to determine the values of the parameters for all rings recursively, starting from $\rsun$ ($l\!=\!90\de$ for QI, or $l\!=\!270\de$ for QIV) down to the Galactic centre ($l\!=\!0\de$ or $360\de$).

The parameters of the first ring in the iteration can not be determined and must be assumed: we fix them to $v_{\phi,0}\!=\!\vsun$, $\sigma_0\!=\!8\kms$ and $n_{0,0}\!=\!0.5$ cm$^{-3}$.
We tested that a factor of two variation in $n_{0,0}$ or $\sigma_0$ has virtually no impact on our results. 
This happens because the ring number density is the largest around $\rsun$, thus in the solar neighborhood the model parameters stabilise quickly regardless the values chosen for the outermost ring.

Our approach differs substantially from the classical tangent point method, where a) each line profile is treated independently, and b) the emission at the terminal-velocity is always associated to gas at the tangent point.
In fact, under the assumptions of axisymmetry and circular rotation, (a) and (b) are redundant.
Our iterative fitting procedure intrinsically accounts for the fact that line profiles at different longitudes are not independent, but instead originate from a common underlying disc.
Also, for any sight-line we do \emph{not} force the gas at the tangent point radius (i.e. at ring $i=i_{\rm max}$) to contribute to the highest-velocity region of the line profile, because the fit to the innermost ring is unconstrained.
We argue that, within the framework of axisymmetry, ours is the best possible approach for modelling the data without making any a-priori assumption on the trends of the various parameters.

\subsection{Modelling the gas above the midplane}\label{abovemidplane}
Once the parameters for the midplane are determined, we proceed to compute the scale height of the gas in each ring.
We focus on gas at $b\neq0$ and adopt the same iterative method of Section \ref{midplane}.
The difference here is that, instead of modelling a single line-profile for each longitude $l$, we model the latitude-velocity (\emph{b-v}) slice $T_{\rm B}(b,v)$ and we fit it to the LAB data by adjusting the scale height of the tangent-point ring.

The main caveat of this procedure lies in the assumption that the gas is in cylindrical rotation, i.e. the rotation velocity of the gas in each ring does not change with $z$.
However, \citet{MarascoFraternali11} showed that a layer of slow-rotating extra-planar \hi\ is present in the Milky Way.
Therefore, we decide to confine our fit to latitudes $|b|\!\le\!5\de$, which allows us to minimise contaminations from this slow-rotating component while maintainiong enough data points to study the vertical distribution of the gas at most radii (see Section \ref{extrapl} for further details).
Profiles at $|b|\!>\!5\de$ are relevant only for modelling regions close to $\rsun$: for instance, for a disc half-thickness of $100\pc$, a sight-line a $b\!=\!6\de$ would `pierce' through the disc already at $R\!=\!7.3\kpc$ and the line profile would contain little information on the gas located at smaller radii.
As contamination from extra-planar \hi\ should be significant close to $\rsun$, we prefer not to model the high-latitude profiles in the data.
We refer the reader to \citet{MarascoFraternali11} for a model of the Galactic \hi\ emission at high latitudes.

\subsection{Simulating a line profile} \label{lineprofile}
In our model, the brightness temperature profile $T_{\rm B}(v)$ in a given direction of the sky $(l,b)$ is determined in two steps.
The first consists in deriving the \hi\ column density profile $N_{\rm HI}(v)$.
Gas at a generic distance $d$ with respect to an observer placed at the Sun location (see Fig.\,\ref{FIGmethod}) has the following cylindrical coordinates:
\begin{displaymath} 
	\left\{ \begin{array}{ll}
	R(d) = & \sqrt{\rsun^2 + d^2 \cos(b)^2 - 2d\rsun\cos(l)\cos(b)}\\ 
	z(d) = & d\sin(b)\,,
	\end{array} \right. 
\end{displaymath}
and its line-of-sight velocity is given by 
\begin{equation} \label{vlos}
v(d) = \left(v_\phi(R(d))\frac{\rsun}{R(d)} - \vsun\right)\sin(l)\cos(b)\,.
\end{equation}
Note that eq.(\ref{vlos}) is valid for circular motions only.
This gas will contribute to the total column density in $(l,b)$ by an amount $\Delta N_{\rm HI} = n(R,z)\,\Delta d$, being $\Delta d$ a generic step along the line-of-sight.
We assume that this column density is spread in the velocity domain by following a Gaussian distribution with mean equal to $v(d)$ and standard deviation equal to $\sigma(d)$, which is the velocity dispersion of the gas at radius $R(d)$.
In practice, we consider a series of steps along the line of sight $0<d_i<d_{\rm max}$, where $d_{\rm max}$ is the distance at which the line of sight intercepts the Solar circle (see Fig.\,\ref{FIGmethod}), and evaluate the \hi\ column density profile as
\begin{equation} \label{NHI}
N_{\rm HI}(v) = \frac{\Delta v}{\sqrt{2\pi}}\sum_{d_i < d_{\rm max}} \frac{n(d_i)}{\sigma(d_i)} \exp\left(-\frac{(v-v(d_i))^2}{2\sigma^2(d_i)}\right) \Delta d_i
\end{equation} 
where $v$, $\sigma$ and $n$ are derived by linear interpolation of the various rings and $\Delta v$ is the channel separation in the datacube ($2\kms$).
We assume that $\Delta d_i$ is either $10\pc$ or the distance along the line-of-sight between two consecutive rings, whichever is the smallest. 

As a second step, we must convert $N_{\rm HI}(v)$ to a brightness temperature $T_{\rm B}(v)$.
This is a delicate step, which we begin to tackle by first reminding some basic equations for the \hi\ radiative transfer.
In the absence of background sources, the solution to the equation of radiative transfer for an extended (i.e., filling entirely the telescope beam), homogeneous and isothermal layer of \hi\ at a given line-of-sight velocity $v$ is
\begin{equation}\label{radiative}
T_{\rm B}(v) = T_{\rm S} \left(1-e^{-\tau(v)}\right), 
\end{equation}
where $\tau(v)$ is the \hi\ optical depth and $T_{\rm S}$ is the \hi\ excitation (or spin) temperature.
Following \citet[][pag.\,473]{bm98}, the \hi\ column density between two generic line-of-sight velocities $v_1$ and $v_2$ can be expressed in terms of the gas spin temperature and the optical depth as
\begin{equation}\label{wellknownformula}
N_{\rm HI} = c\int_{v_1}^{v_2} T_{\rm S} \tau(v)\,\rm{d}v  
\end{equation}
where $c\!=\!1.823\times10^{18}$ cm$^{-2}$ K$^{-1}$ km$^{-1}$\,s.
If we differentiate eq.(\ref{wellknownformula}) we can easily re-arrange for the optical depth:
\begin{equation}\label{tau}
\tau(v) = \frac{N_{\rm HI}(v,v+\Delta v)}{c\,T_{\rm S}\,\Delta v},
\end{equation} 
where the quantity $N_{\rm HI}(v,v+\Delta v)/\Delta v$ has units of $\cmmq\,{\rm km}^{-1}\,\rm{s}$ and is in practice given by eq.\,(\ref{NHI}).
In the optical thin limit $\tau\!\ll\!1$ (i.e., low column density and/or high spin temperature), eq.\,(\ref{radiative}) gives $T_{\rm B}\!\simeq\! T_{\rm S}\tau\!=\!N_{\rm HI}/c$, so the measured brightness temperature is proportional to the intrinsic column density per unit velocity.
In the optical thick limit $\tau\!\gg\!1$, instead, $T_{\rm B}\!\simeq\!T_s$.
Thus eq.\,(\ref{radiative}) implies that the brightness temperature cannot exceed the gas spin temperature.

The interstellar medium (ISM), however, is not isothermal and homogenous, but instead consists of an ensemble of clouds, shells and filaments with different sizes, temperatures and densities, roughly in pressure equilibrium \citep[e.g.][]{Wolfire+03}.
Radiative transfer in such a complex medium can not readily be described by eq.\,(\ref{radiative}) and (\ref{tau}).
In general, there is no unique relation between column density and brightness temperature: given any $N_{\rm HI}(v)$, the observed brightness temperature profile will depend on the physical properties of the clouds that make up the ISM, on their distribution along the line of sight and on the observing setup \citep[e.g.][see also Appendix \ref{A_multiphase_clumpy} of this work]{DickeyLockman90}.
However, thanks to the many theoretical \citep{McKeeOstriker77,Wolfire+03} and observational \citep{Gibson+00,HT03b,Dickey+03,StrasserTaylor04} studies on the Galactic ISM carried in the last decades, we have matured a good understanding of the physical properties of the neutral gas.
Using results from these studies, in Appendix \ref{appendix_NHI_Tb} we build a model for the multiphase Galactic ISM and compute the exact $N_{\rm HI}(v)-T_{\rm B}$ relation for an observing setup analogous to that of the LAB survey. 
We show that such a relation can be still approximately described by eq.\,(\ref{radiative}) and (\ref{tau}), using an `effective' spin temperature of $\sim152\K$, the peak brightness temperature measured in the LAB survey.
This value is 2-3 times larger than the typical $T_{\rm S}$ determined in studies of \hi\ absorptions against background continuum sources.
For instance, \citet{Dickey+03} and \citet{HT03b} found median $T_{\rm S}$ of $65\K$ and $50\K$ respectively.
These studies, however, attempt to characterise solely the coldest and densest phase of the \hi, which is the only phase responsible for the absorptions but contributes only in part to the overall 21-cm emission.  
Our aim is to have the simplest possible prescription for the conversion between $N_{\rm HI}(v)$ and $T_{\rm B}(v)$ that best describes the 21-cm LAB data, thus our effective spin temperature of $152\K$ is a better motivated choice and we adopt it as our fiducial $T_{\rm S}$ for the purpose of computing eq.\,(\ref{radiative}) and (\ref{tau}).
We caution the reader that this value must be interpreted neither as the typical $T_{\rm S}$ of the Galactic \hi\ or as representative for one particular phase of the neutral gas.
It is, in fact, only a practical choice.
In Section \ref{fiducial_vs_thin}, we discuss how our results change for different $T_{\rm S}$.

In summary, the \hi\ line profile is derived by first evaluating eq.(\ref{NHI}), then eq.(\ref{tau}) and finally eq.(\ref{radiative}).

\subsection{Fitting technique}\label{fitting}
In order to fit the ring parameters $\mathbf x  = (v_\phi, \sigma, n_0, h_s)$ to the data, we must define a way to compare the model \hi\ line profile, $T_{\rm B}^{\rm mod}(\mathbf x, v)$, with the observed profile $T_{\rm B}^{\rm obs}(v)$.
This comparison is performed via a Monte Carlo Markov Chain (MCMC) algorithm.
MCMC samples the parameter space in a way proportional to the probability of the model given the observations $P(T_{\rm B}^{\rm mod}(\mathbf x)|T_{\rm B}^{\rm obs})$, which is proportional to the product of the {\it likelihood} and the {\it prior} $P(T_{\rm B}^{\rm obs}|T_{\rm B}^{\rm mod}(\mathbf x)) \times P(T_{\rm B}^{\rm mod}(\mathbf x))$ (Bayes theorem).
We define the likelihood as:
\begin{eqnarray}\label{likelyhood}
P(T_{\rm B}^{\rm dat}|T_{\rm B}^{\rm mod} ({\mathbf x})) &\propto& \prod_{k=1}^{n.\,pixels} \exp \left(-\frac{|T_{\rm B}^{\rm mod}(\mathbf x,v_k)-T_{\rm B}^{\rm obs}(v_k)|}{\epsilon} \right) \nonumber\\
 &=& \exp \left(-\sum_{k=1}^{n.\,pixels} \frac{|T_{\rm B}^{\rm mod}(\mathbf x,v_k)-T_{\rm B}^{\rm obs}(v_k)|}{\epsilon} \right) \nonumber\\
  &=& \exp\left(-\mathcal{R}(\mathbf x)/\epsilon\right)
\end{eqnarray}
\noindent
where $\mathcal{R}(\mathbf x)$ is the \emph{residual} between the model and the data, $\epsilon$ is the error bar in the data (assumed to be constant throughout the whole dataset) and the sum is extended to all pixels in the data velocity domain, which extends from $-400\kms$ to $+400\kms$. 
Note that the choice of $\mathbf x$ that minimizes $\mathcal{R}$ is also the one that maximizes the likelihood.
The value of $\epsilon$ is important as it affects the width of the posterior probability distribution, which gives the uncertainty on the fit parameters.
In principle $\epsilon$ should be the rms-noise of the LAB survey but, given that our model is a simple axisymmetric approximation of the Galactic disc, the requirements that it fits the data to the noise level is not realistic and would lead to strongly underestimate the uncertainty associated to our results.
Therefore we define a ad-hoc error $\epsilon^*$ as the ratio between the minimum value of $\mathcal{R}({\mathbf x})$ and the number of pixels used to calculate it, and for each sight-line we use either $\epsilon^*$ or the LAB rms-noise, whichever is the largest.

As fitting boundary conditions, we require that $0<v_\phi<300\kms$, $n_0>0$, $h_s>0$. 
In addition, we require that $\sigma\ge\sqrt{k_{\rm B}T_{\rm S}/m_{\rm H}}$, i.e. the gas total velocity dispersion must be at least equal to the thermal velocity dispersion, being this latter $1.5\kms$ for our fiducial model ($T_{\rm S}=152\K$). 
We set our prior to be 1 in the allowed parameter ranges, and 0 otherwise.

In practice, when fitting the midplane emission (Section \ref{midplane}), we use 2000 iterations of the chain to evaluate $\epsilon^*$ and further 8000 iterations to sample the full posterior probability distribution for the parameters $v_\phi$, $\sigma$ and $n_0$.
We use the mean and the standard deviation of these distributions as representative for the central value and the error bar of our parameters.
The emission above the midplane (Section \ref{abovemidplane}) is fit with a similar procedure, but we use 100 iterations for $\epsilon^*$ and further 400 to sample the posterior probability, given that the midplane parameters are already set and we only fit for $h_s$.

In Appendix \ref{testing} we test our fitting technique on mock \hi\ observations of an idealised disc, and we demonstrate that it can recover the system input parameters with very good accuracy.
	
\section{Results: atomic hydrogen}\label{resultshi}
\begin{figure*}[tbh]
\centering
\includegraphics[width=0.8\textwidth]{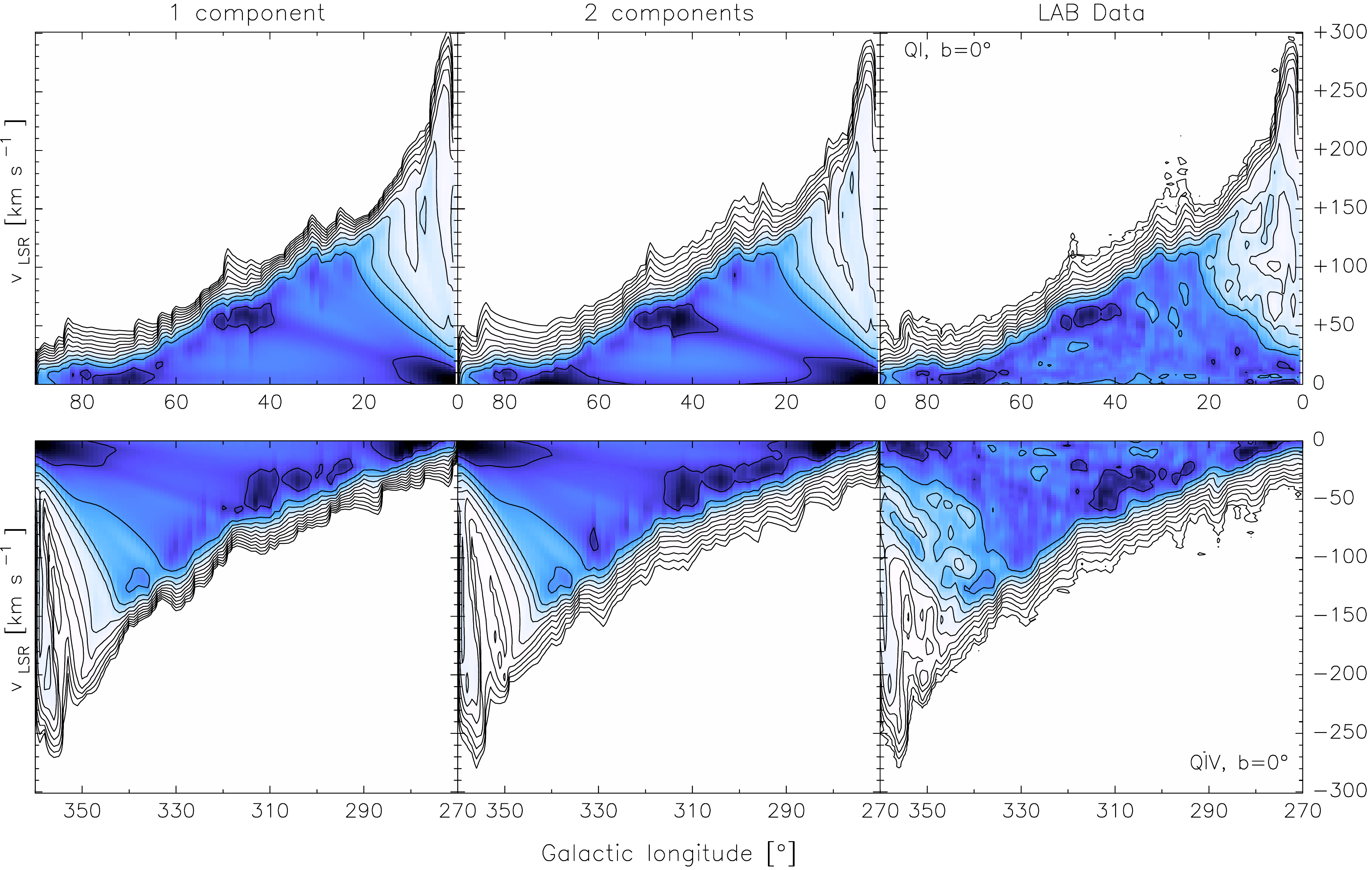}
\caption{$l-v$ diagrams at $b=0\de$ for the \hicap\ emission in the regions QI (\emph{top row}) and QIV (\emph{bottom row}) of the Milky Way.
The leftmost panels show our one-component model, the central panels show our two-component model, the rightmost panels show the LAB data.
Brightness temperature contours range from $0.1\K$ (or $4$ times the rms noise) to $102.4\K$ in multiples of $2$. }
\label{HI_midplane}
\end{figure*}
\begin{figure*}[tbh]
\centering
\includegraphics[width=0.8\textwidth]{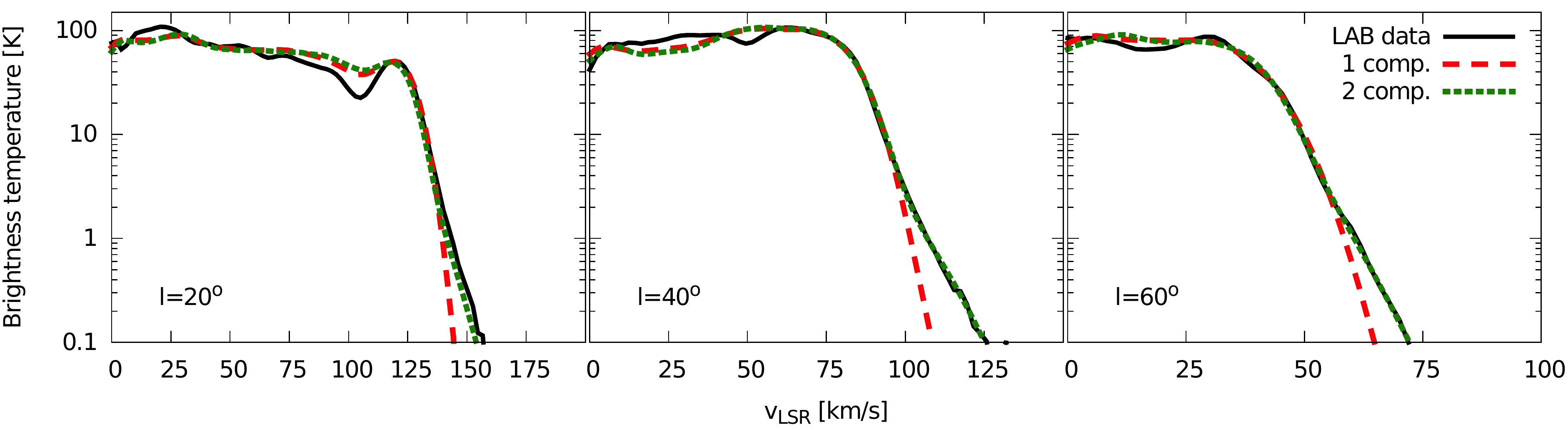}
\caption{\hicap\ brightness temperature profiles taken at three representative longitudes (as indicated in the bottom left corner of each panel) for our one-component model (long-dashed line), our two-component model (short dashed line) and for the LAB data (solid line). All profiles are at $b\!=\!0\de$. Both models adopt $T_{\rm S}=152\K$. The two component model fits the tails of the observed profiles significantly better.}
\label{lineprofiles}
\end{figure*}
\begin{figure}[tbh]
\centering
\includegraphics[width=0.5\textwidth]{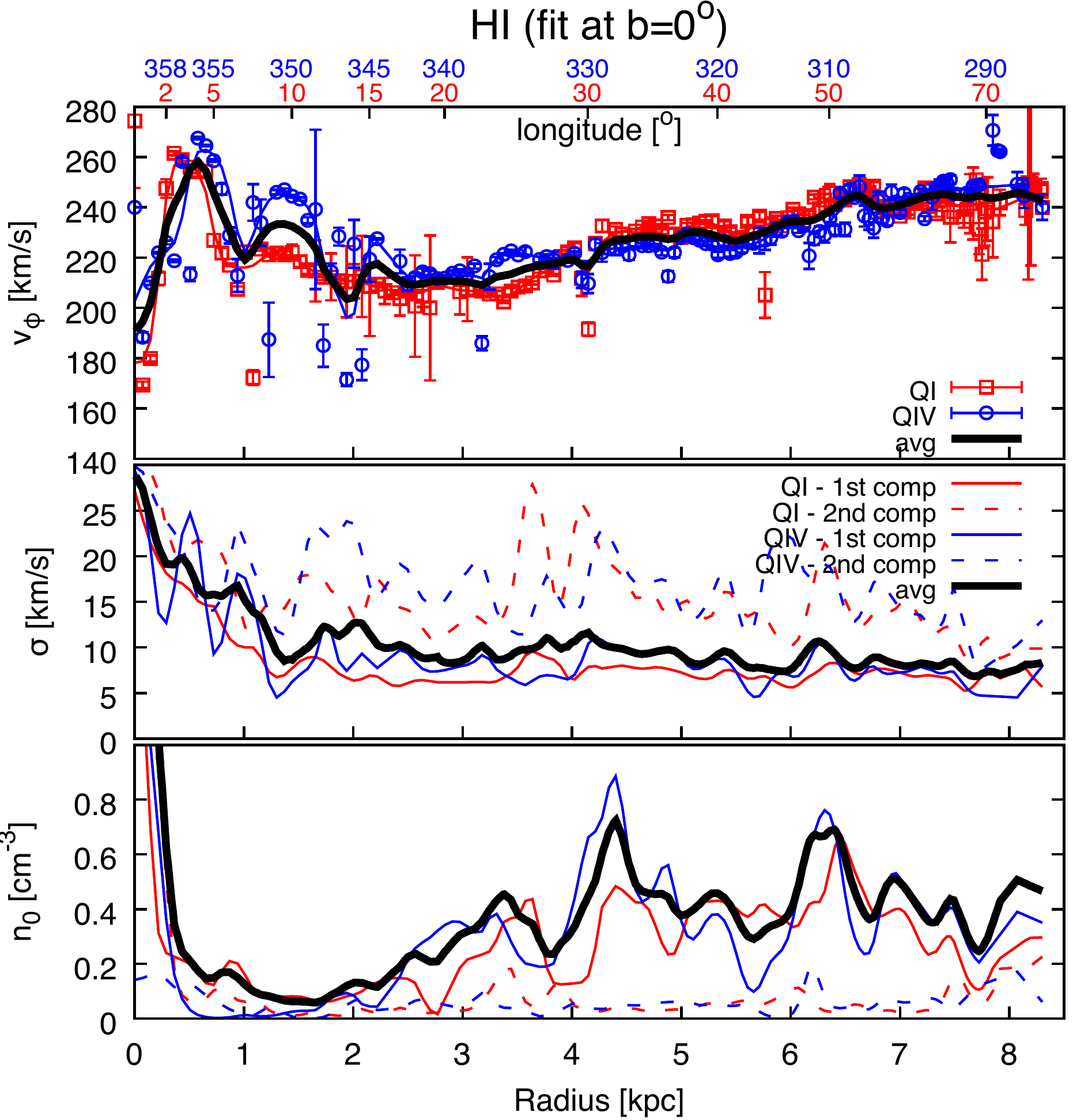}
\caption{Rotation velocity (\emph{top panel}), velocity dispersion (\emph{central panel}) and midplane volume density (\emph{bottom panel}) profiles for our fiducial \hicap\ disc model of the Milky Way. Profiles for QI and QIV are shown in \emph{red} and \emph{blue} respectively. In the central and bottom panels, \emph{solid-lines} represent the first, low velocity dispersion component, while \emph{dashed-lines} represent the second component at high velocity dispersion.
The top panel shows both the single best-fit values (points with error-bars) and the profile derived by smoothing the latter at $0.2\kpc$ of resolution (see text).
For the sake of clarity, only the smoothed profiles are shown in the other panels.
The thick solid black lines represent the averaged smoothed profiles (computed via eq.\,(\ref{avgsigma}) for $\sigma$ and eq.\,(\ref{avgn0}) for $n_0$).}
\label{result_2c_240}
\end{figure}
\begin{figure*}[tbh]
\centering
\includegraphics[width=\textwidth]{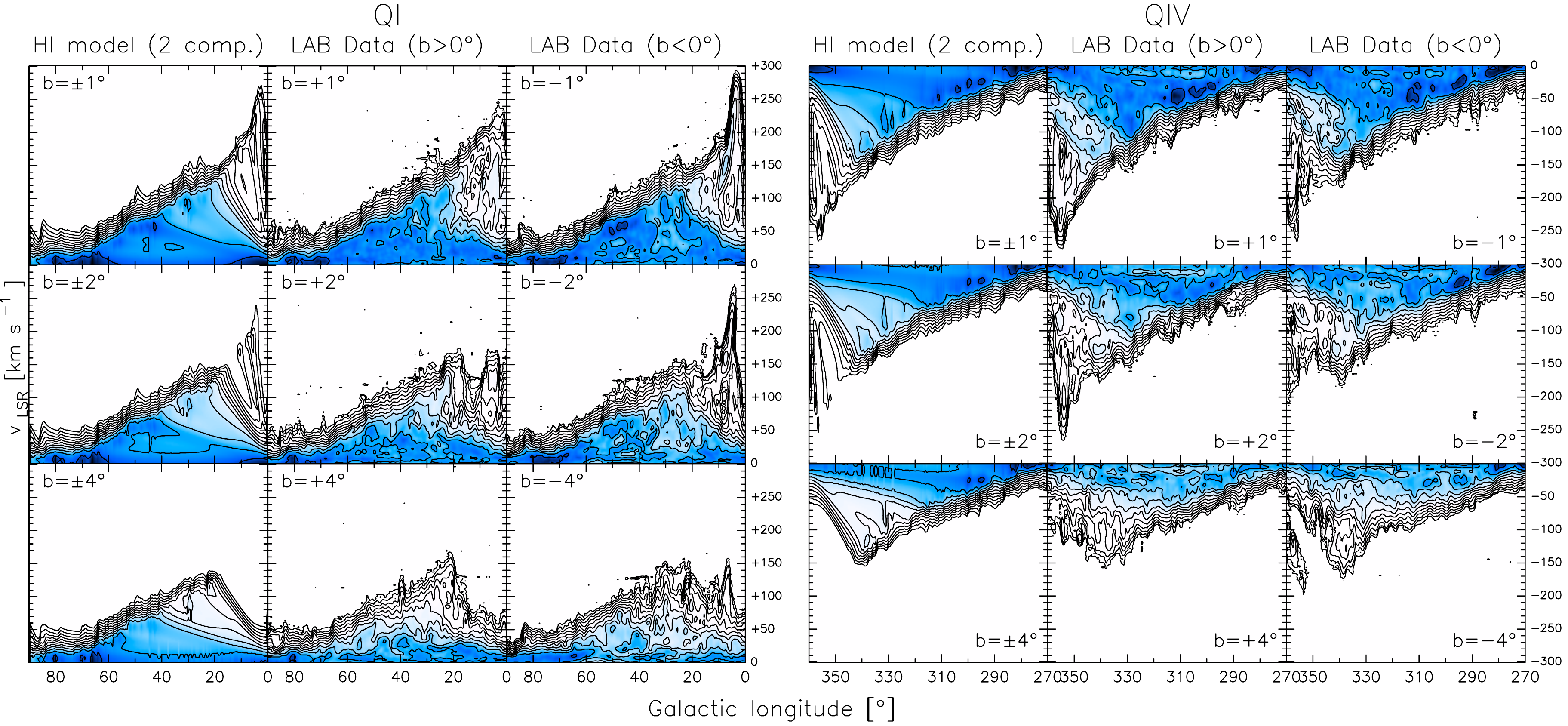}
\caption{\hicap\ $l\!-\!v$ diagrams above the midplane. \emph{Left-hand panels:} $l\!-\!v$ diagrams taken in the receding quadrant QI at different latitudes (as indicated on the top-left corner of each panel) for our fiducial model (first column) and for the LAB data (second and third column). Brightness temperature contours range from $0.1\K$ to $102.4\K$ in multiples of $2$.
\emph{Right-hand panels:} the same, but for the approaching quadrant QIV.}
\label{HIabove}
\end{figure*}	
\begin{figure}[tbh]
\centering
\includegraphics[width=0.45\textwidth]{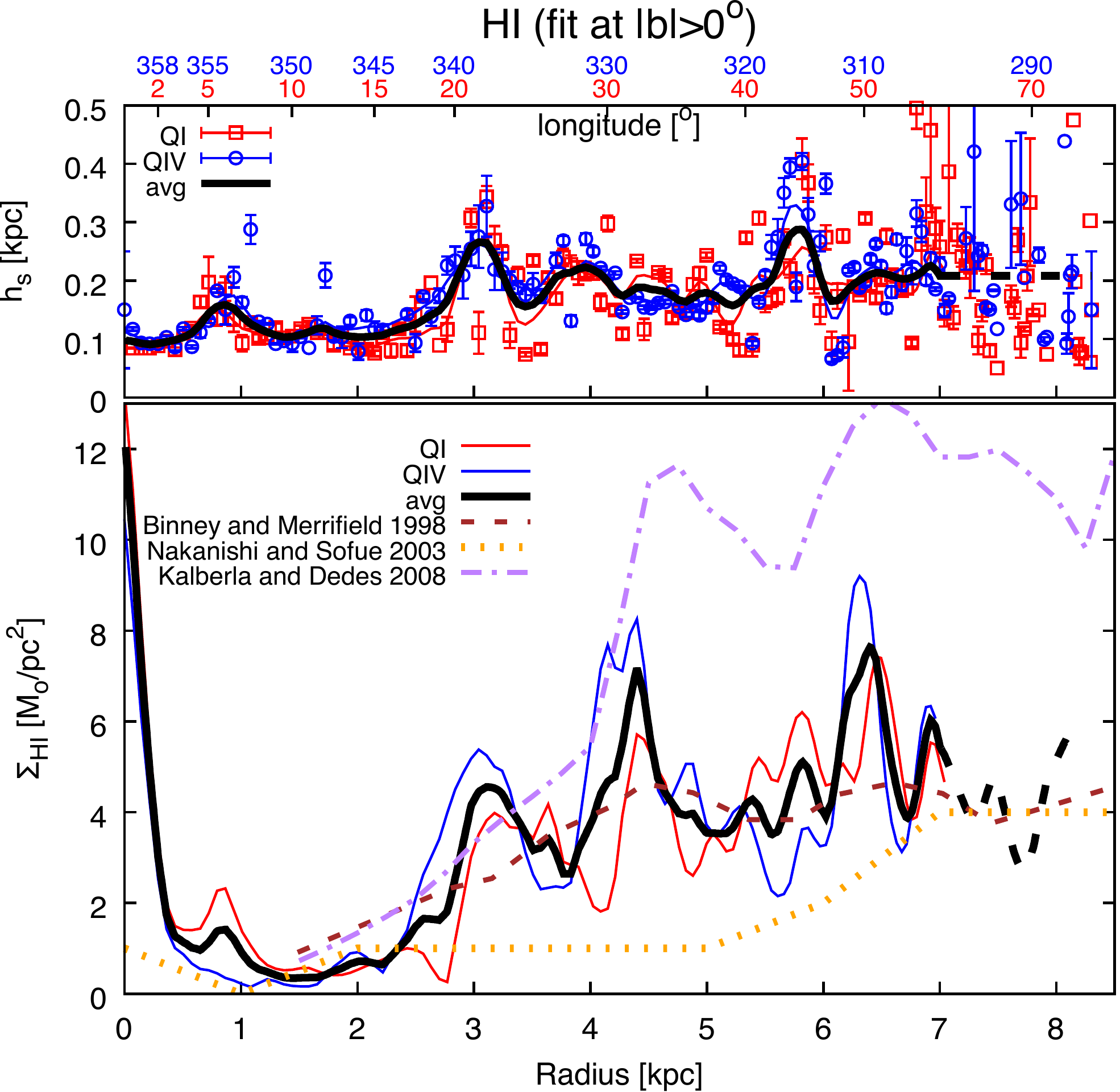}
\caption{\hicap\ scale height (\emph{top panel}) and surface density (\emph{bottom panel}) profiles for our fiducial \hi\ model of the Milky Way. Profiles for QI and QIV are shown in \emph{red} and \emph{blue} respectively. The top panel shows the best-fit values to the data (points with error-bars) and the profiles derived by smoothing the latter to $0.2\kpc$ of resolution (thin solid lines), while only the smoothed profiles are shown in the bottom panel.
The thick solid black lines show the mean profiles.
For comparison, we show the density profiles determined by \citet{bm98} (\emph{dashed line}), \citet{NakanishiSofue03} (\emph{dotted line}) and \citet{KalberlaDedes08} (\emph{dot-dashed line}), all re-scaled to $\rsun\!=\!8.3\kpc$.}
\label{scaleheight_2c_240}
\end{figure}
	\subsection{Midplane: one-component model}\label{HI1comp}	
Fig.\,\ref{HI_midplane} shows the longitude-velocity ($l\!-\!v$) diagrams at latitude $b\!=\!0\de$ predicted by our model (leftmost panels) for the receding and the approaching sides of the disc, and compares them with the observations (rightmost panels).
The agreement with the LAB \hi\ data is surprisingly good in most locations.
In particular we are able to reproduce the brighter \hi\ features visible in the data, which often occur at velocities different from the terminal velocities (like the bright \hi\ spot visible around $l=350\de, v=-10\kms$).
However, the tail of the \hi\ emission is more extended in the data than in our model, which suggests that the model underestimates the velocity dispersion of the gas close to the tangent points.
This is not caused by our choice of the spin temperature ($152\K$), as the optically thin model (not shown here) looks identical to that shown in the left panels of Fig.\,\ref{HI_midplane}. 
A possibility is that, as the \hi\ brightness temperatures in the data span over three order of magnitudes, our fitting technique (Section \ref{fitting}) is biased towards the brightest part of the profiles and is less sensitive to the faint \hi\ emission.
To test this, we re-performed our fit by substituting $T_{\rm B}$ with $\log(T_{\rm B})$ in eq.\,(\ref{likelyhood}).
This compresses significantly the brightness temperature dynamical range.
We found that, in this way, the \hi\ tails are reproduced significantly better, but at the expenses of the brightest regions.

We conclude that our base model can not reproduce the brightest and faintest regions of the \hi\ Galactic disc simultaneously, and consider the inclusion of a second gas component in our disk model.
	
	\subsection{Midplane: two-component model} \label{HI2comp}
We now assume that the \hi\ in the Galactic disc can be described by two components that share the same rotational velocity but have different densities and velocity dispersions.
This increases by two the number of free parameters for each ring.
Both components are fit simultaneously to the data by using the procedure described in Section \ref{fitting}, but we substitute $T_{\rm B}$ with $\log(T_{\rm B})$ in eq.\,(\ref{likelyhood}) to account for the large brightness temperature dynamical range.

The $l\!-\!v$ diagrams for the best-fit model with two-components are shown in the middle panels of Fig.\,\ref{HI_midplane}.
We have assumed that both components have $T_{\rm S}$ of $152\K$, but this choice does not affect the appearance of the $\!l\!-\!v\!$ diagrams. 
The differences with respect the single component model are minimal, except that now the tail of the \hi\ emission is more extended and the data are reproduced remarkably well at all locations and in both quadrants.
This can be better appreciated in Fig.\,\ref{lineprofiles}, which shows three representative \hi\ line profiles at $l\!=\!20\de,40\de$ and $60\de$ for the data and for both our models.
Clearly, both models are in good agreement with the data, but the two-component model reproduces the tails of the profiles significantly better.
Hereafter, we will refer to this two-component, best-fit disc model with $T_{\rm S}\!=\!152\K$ as our \emph{fiducial} model.

Fig.\,\ref{result_2c_240} shows the rotation curve, the velocity dispersion and midplane density profiles for our fiducial model.
Profiles for the approaching and the receding sides of the galaxy are shown separately, along with an average value representing a full axisymmetric model.
The panel on top in Fig.\,\ref{result_2c_240} shows the best-fit values for rotation velocity (points + error-bars) as derived via our MCMC routine. 
A smooth rotation curve (solid thin lines) is derived by smoothing these points with a Gaussian kernel with $0.2\kpc$ of FWHM.
In the smoothing procedure, each point has been given a weight inversely proportional to its error-bar to ensure that the most uncertain values do not affect our final profiles.
From this panel, it is clear that the \hi\ rotation curves in QI and QIV are very similar to one another. 
Such similarity in $v_{\phi}$ points toward a high degree of axisymmetry in the inner Galactic disc.
Differences in velocity larger than $\sim10\kms$ occur only at $R\!<\!2.5\kpc$: at such small radii, we expect the Galactic bar to have a non-negligible impact on the kinematics of both gas \citep[e.g.][]{Fux99} and stars \citep[e.g.][]{Portail+17}.
Further discussion on the effect of the bar can be found in Section \ref{effectofbar}.

The central and bottom panels of Fig.\,\ref{result_2c_240} show the $\sigma$ and $n_0$ profiles separately for the two components, and for both QI and QIV.
Here, we show only the smoothed profiles.
There is a clear distinction between the two components: the first one (component A) has low velocity dispersion ($\sim8\kms$) and high density ($\sim0.3-0.5$ cm$^{-3}$), while the other (component B) has high $\sigma$ ($15-20\kms$) and low $n_0$ ($\sim0.04$ cm$^{-3}$).
As mentioned, the high-$\sigma$ component is the one needed to reproduce the faint tail of the line profiles.
The contribution of this component to the \hi\ mass within the solar circle is $15\%$ in QI and $20\%$ in QIV.
The average velocity dispersion and density profiles ($\avg{\sigma}$ and $\avg{n_0}$) are determined as
\begin{eqnarray}
\avg{\sigma} & = & \frac{1}{2}\left[\left(\frac{n_{0,A} \sigma_{A}^2 + n_{0,B} \sigma_{B}^2}{n_{0,A}+n_{0,B}}\right)^\frac{1}{2}_{\rm QI} +  \left(\frac{n_{0,A} \sigma_{A}^2 + n_{0,B} \sigma_{B}^2}{n_{0,A}+n_{0,B}}\right)^\frac{1}{2}_{\rm QIV}\right]\label{avgsigma}\\
\avg{n_0} & = & \frac{1}{2}\left[\left(n_{0,A}+n_{0,B}\right)_{\rm QI} + \left(n_{0,A}+n_{0,B}\right)_{\rm QIV}\right]\,\label{avgn0}.
\end{eqnarray}
The average $\sigma$ is very large in Galactic centre, where it reaches values of $25\!-\!30\kms$, but it decreases quickly with $R$, dropping down to $\sim10\kms$ around $R=1.5\kpc$. 
For larger $R$, $\avg{\sigma}$ slowly decreases down to values of $\sim8\kms$ at $R\!=\!\rsun$.
Similar trends for $\sigma(R)$ have been derived also in external galaxies \citep{Tamburro+09}.
The averaged midplane density peaks in the Galactic centre but then shows a large depression at $0.5\!<\!R\!<\!2.5\kpc$.
At larger $R$, $\avg{n_0}$ oscillates around a typical value of $0.4$ cm$^{-3}$.
The density profiles of QI and QIV show a series of peaks and depressions, and are discussed in more depth in Section \ref{spiralarms}.

We discuss in more depth the nature of the second component in Section \ref{2compdiscussion}.
Here, we stress that the averaged radial profiles of this new model are remarkably similar to those determined for our single component model (not shown here).
The rotation velocities, in particular, are practically the same at all radii.
The averaged velocity dispersions alone show some minor differences, which are however never larger than $\sim2\kms$ and typically below $1\kms$.
This not only suggests that the models presented are indeed the best possible fit to the data, but also indicates the overall robustness of our fitting procedure.

	\subsection{Scale height and column density}	\label{HIscaleheight}
Our fiducial model is the one that best reproduces the Galactic \hi\ distribution and kinematics at $b=0\de$. 
We can now study how the gas distributes above the midplane using the method described in Section \ref{abovemidplane}.
Since we believe that the two \hi\ components do not represent physically distinct phases of the Galaxy's ISM - like discussed in Section \ref{HI2comp} - we will not treat them separately and fit them both with a single scale height.

Fig.\,\ref{HIabove} shows the $l\!-\!v$ diagrams taken above and below the midplane at $b\!=\!\pm1,\pm2,\pm4\de$ for our fiducial case and for the LAB data.
In general, the model follows nicely the trend shown by the data, where the high-velocity, low-level \hi\ emission visible close to the galactic center ($l\!<\!20\de$ in QI and $l\!>\!340\de$ in QIV) vanishes at $|b|\!\ge\!2\de$ because \hi\ above the midplane is lacking in the innermost regions of the disc.
In the model, $l-v$ diagrams at opposite latitudes are identical by construction and are shown in the left-most panels of Fig.\,\ref{HIabove}.
This is not the case for the data, where  $l-v$ diagrams at opposite latitudes show some differences (in particular at $b=\pm2\de$) that are likely caused by the presence of the Galactic bar and by the vertical displacement of the Sun with respect to the Galaxy midplane \citep[e.g.][]{HumphreysLarsen95}.
Our model does not take these features into account and predicts $l-v$ distributions that are in between those shown by the data.
Note also that the model does not reproduce the data at $b\ne0\de$ as accurately as on the midplane.
This because we are assuming that the \hi\ rotation velocity and velocity dispersion do not change with $z$, and attempt to fit a simple Gaussian for the vertical gas distribution of each ring.
We discuss this further in Section \ref{extrapl}.

Fig.\,\ref{scaleheight_2c_240} shows the resulting scale height and \hi\ surface density profiles for our fiducial model. 
The \hi\ scale height increases inside-out, moving from $\sim100\pc$ for $R\!<\!3\kpc$ to $\sim200\pc$ at larger radii.
Also in this case the regions QI and QIV show very similar trends, pointing towards a high degree of symmetry in the inner disc.
Unfortunately, a precise value for the scale height at $R\!>\!7\kpc$ is difficult to derive, as the \hi\ profiles at $l\!>\!60\de$ (QI) and $l\!<\!300\de$ (QIV) change very little with $b$.
Therefore, for the purpose of determining the surface density at radii larger than $7\kpc$, we fixed the scale height to a value of $217\pc$, derived by averaging the approaching and receding $h_s$ at $R\!=\!7\kpc$.
The bottom panel of Fig.\,\ref{scaleheight_2c_240} compares the \hi\ surface density $\Sigma_{\rm HI}$ predicted by our model (thick solid line) with a selection of those presented in the literature (dashed lines).
In general, the $\Sigma_{\rm HI}$ of our model is in very good agreement with that determined by \citet{bm98}, while it lies above that determined by \citet{NakanishiSofue03} and well below that predicted by \citep{KalberlaDedes08}.
In addition, we predict the \hi\ depression at $0.5\!<\!R\!<\!2.5\kpc$ to be deeper than all previous determinations, and a steep rise in the inner few hundred parsecs.

\section{Results: molecular hydrogen}\label{resultsh2}
As it is often done in the literature, we assume that most molecular hydrogen is traced by the CO ($J=1-0$) emission line, whose source of excitation is primarily collisions with H$_2$ molecules, and use the CO data of \citet{Dame+01} to model the Galactic H$_2$ inside the Solar circle.
We caution that there exists a fraction of molecular hydrogen that is not coupled with CO, and therefore remains `invisible' in our data \citep[the so-called `CO-dark' H$_2$,][]{Langer+14,Tang+16}.
What we model here is instead the `CO-bright' phase of the molecular hydrogen.

The modelling of the Galactic CO follows a prescription similar to that adopted for the \hi, with the following differences.
First, we use only a single gas component.
Second, we do \emph{not} consider the effect of the optical thickness.
Even though molecular clouds are optically thick in the CO $J\!=\!1\!-\!0$ rotational line, it is commonly assumed that their ensemble is not: clouds do not shade each other because they are well separated in both space and velocity \citep[e.g.][]{GordonBurton76,bm98}.
Thus a typical CO line-profile is produced by the sum of the profiles of each individual cloud.
Assuming that the population of molecular clouds does not vary within the inner Galaxy, the integral of the CO line profile between two line-of-sight velocities $v_1$ and $v_2$ can be considered proportional to the H$_2$ column density:
\begin{equation}\label{NH2}
N_{\rm H_2} = X_{\rm CO}\int_{v_1}^{v_2}T_{\rm B}(v)\,{\rm d}v 	
\end{equation}
where $T_{\rm B}$ is the brightness temperature of the CO line and $X_{\rm CO}$ is the CO-to-H$_2$ mass conversion factor (or simply $X_{\rm CO}$ factor).
Following \citet{Bolatto+13}, we adopt $X_{\rm CO} = 2.0\times10^{20}$ cm$^{-2}$ K$^{-1}$ km$^{-1}$\,s, with an uncertainty of $\pm30\%$.
Eq.\,(\ref{NH2}) can be inverted to derive $T_{\rm B}(v)$, and the quantity $N_{\rm H_2}(v)$ can be evaluated with a formula analogue to eq.\,(\ref{NHI}). 
We assume $v_\phi\!=\!\vsun$, $\sigma\!=\!4\kms$ and $n_0\!=\!0.5$ cm$^{-3}$ for the first ring of our model \citep[see][]{GordonBurton76}. 
Finally, in our fit we do not impose a lower limit on the velocity dispersion because of the low kinetic temperature and large molecular weight of the CO.

\begin{figure}[tbh]
\centering
\includegraphics[width=0.49\textwidth]{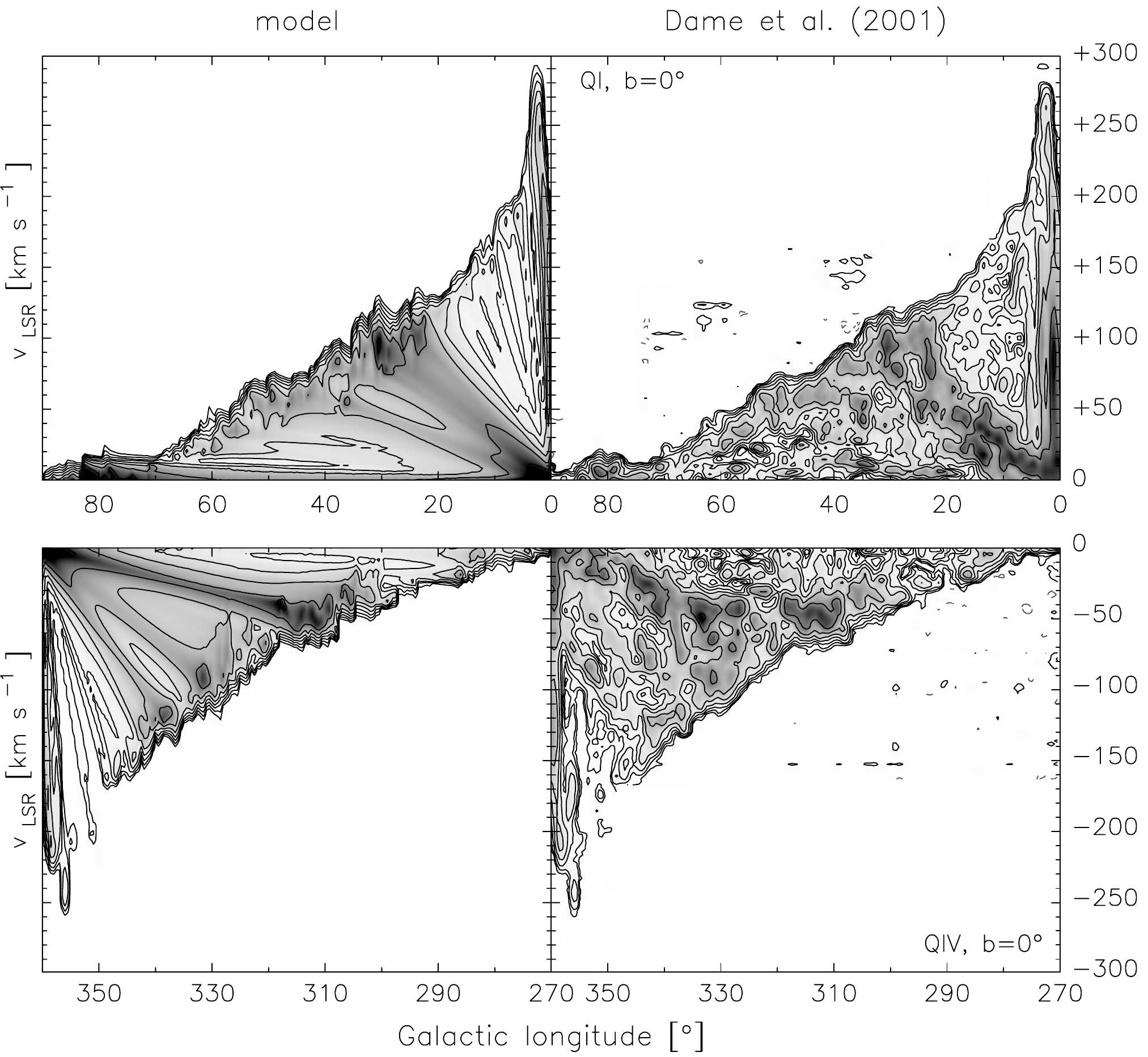}
\caption{$l-v$ diagrams at $b=0\de$ for the $J\!=\!\!1\!-\!0$ CO emission line in the regions QI (\emph{top row}) and QIV (\emph{bottom row}) of the Milky Way.
The panels on the left show our model, the panels on the right show the CO data of \citet{Dame+01}.
Both dataset are smoothed to the angular resolution of $1\de\times1\de$ and the velocity resolution of $2\kms$. 
Brightness temperature contours range from $0.03\K$ (or $3$ times the rms noise) to $15.36\K$ in multiples of $2$. 
}
\label{CO_midplane}
\end{figure}

\begin{figure}[tbh]
\centering
\includegraphics[width=0.5\textwidth]{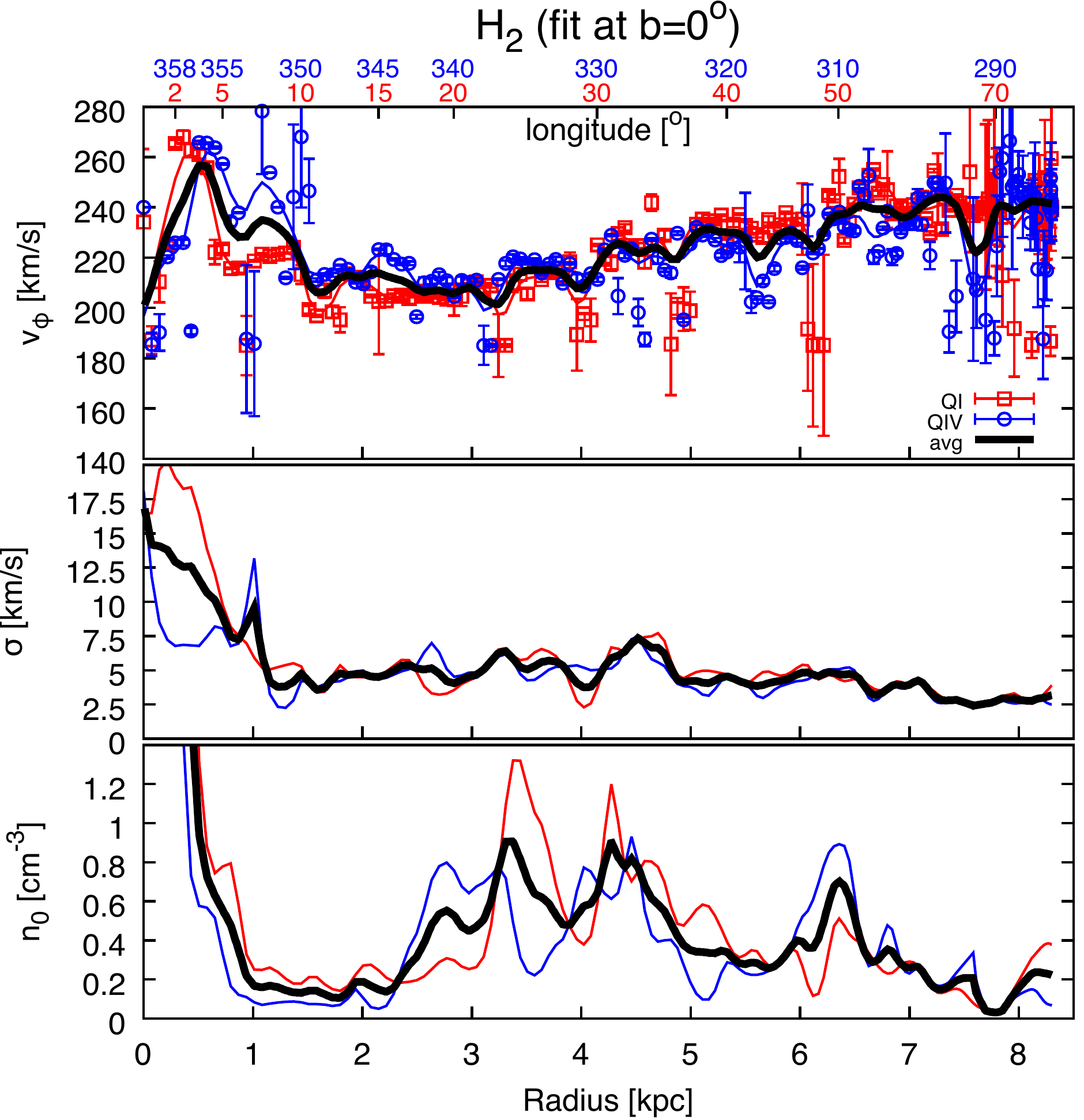}
\caption{As in Fig.\,\ref{result_2c_240}, but for our model of Galactic H$_2$ as derived from our fit to the CO data of \citet{Dame+01}.}
\label{result_CO}
\end{figure}

\begin{figure*}[tbh]
\centering
\includegraphics[width=\textwidth]{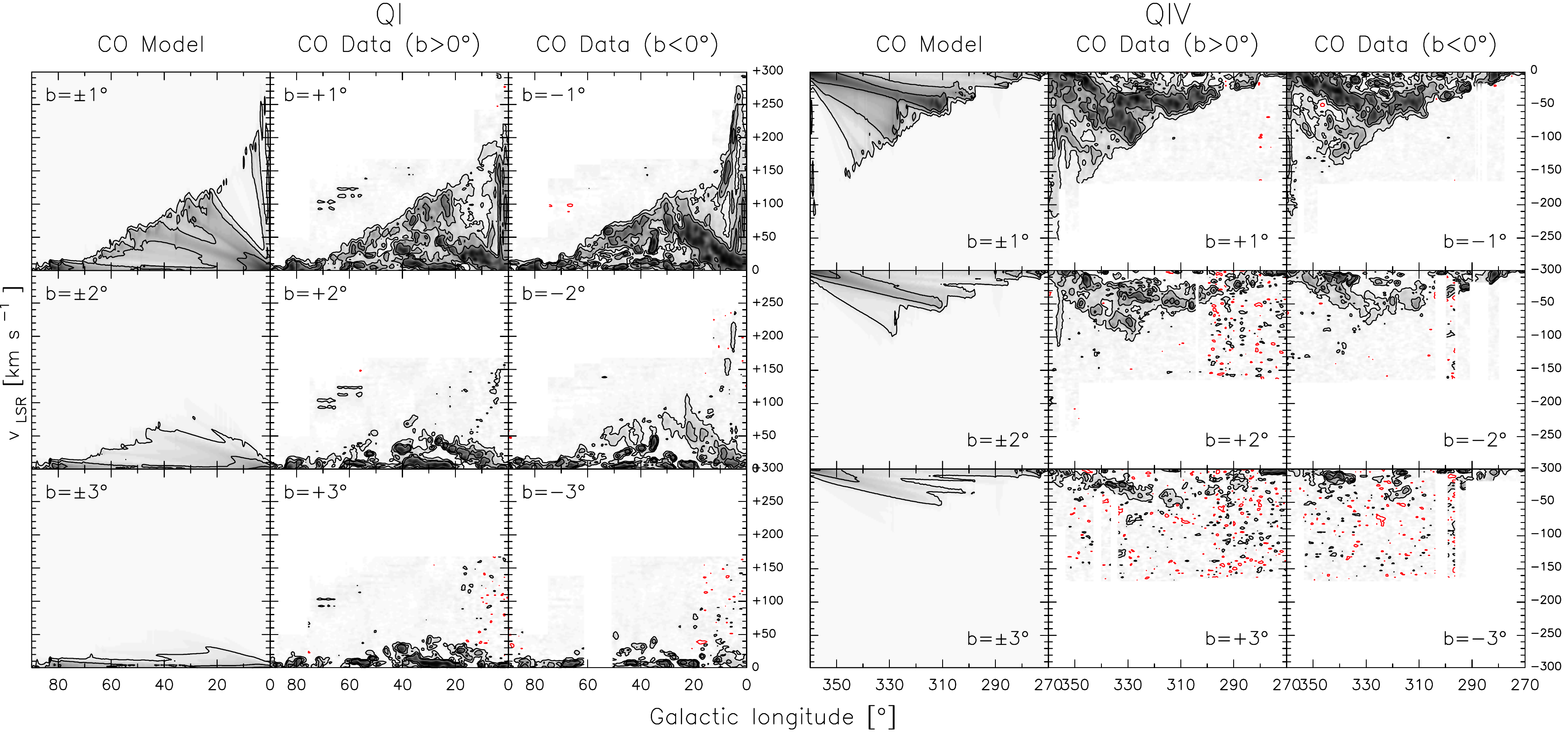}
\caption{As in Fig.\,\ref{HIabove}, but for our model of the CO emission. Black contours are at $T_{\rm B}$ of 0.05, 0.2 and 0.5 K. The red contour is at $T_{\rm B}\!=\!-0.05\K$.}
\label{COabove}
\end{figure*}	

\begin{figure}[tbh]
\centering
\includegraphics[width=0.5\textwidth]{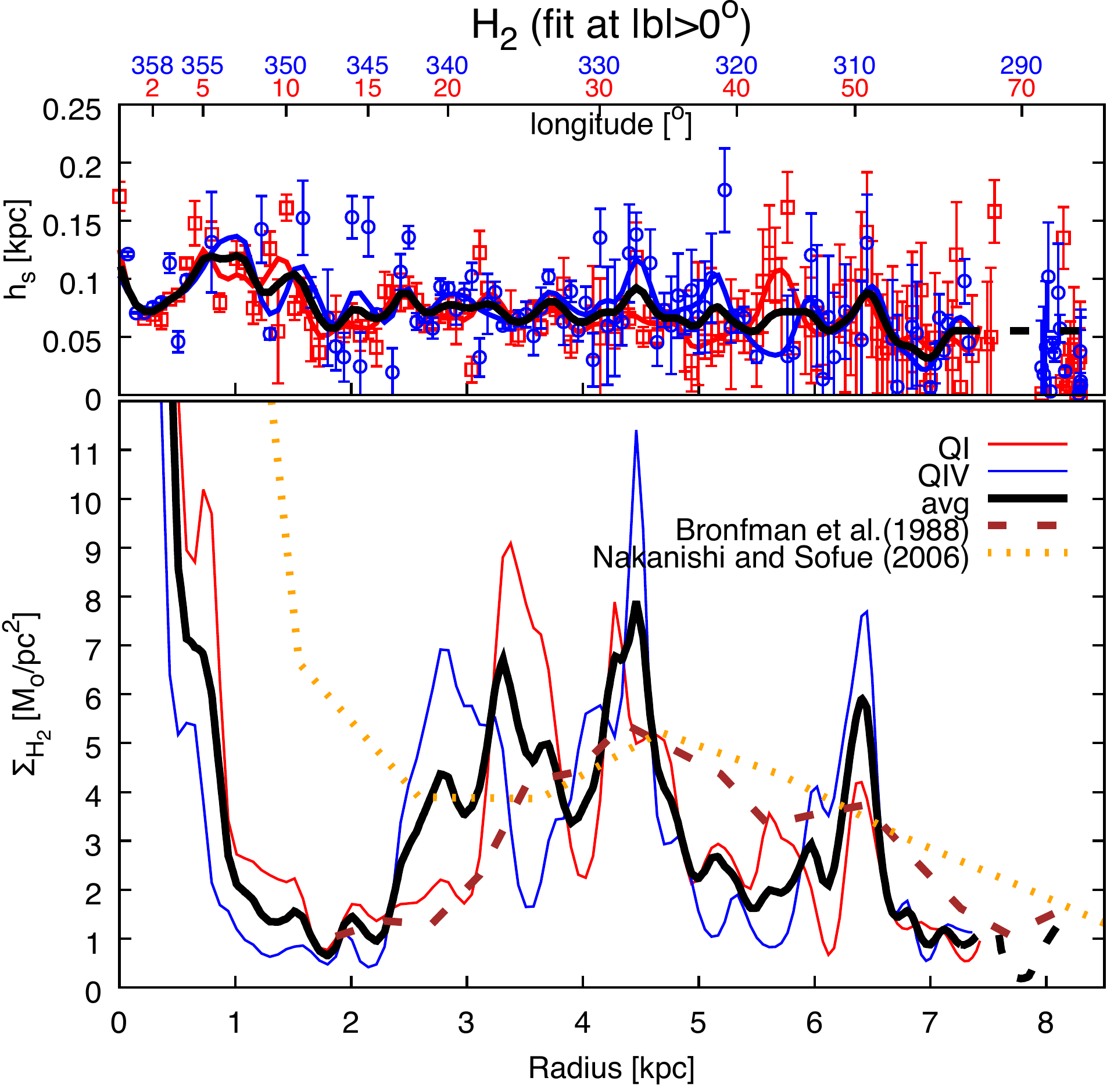}
\caption{As in Fig.\,\ref{result_2c_240}, but for our model of Galactic H$_2$ as derived from our fit to the CO data of \citet{Dame+01}. In the bottom panel, the dashed and the dotted lines show the H$_2$ profiles derived by \citet{Bronfman+88} and \citet{NakanishiSofue06} respectively, both re-calibrated to $\rsun\!=\!8.3\kpc$ and $X_{\rm CO} = 2.0\times10^{20}$ cm$^{-2}$ K$^{-1}$ km$^{-1}$\,s.}
\label{scaleheight_CO}
\end{figure}

Fig.\,\ref{CO_midplane} shows the $l-v$ diagrams of the midplane CO emission for our model (on the left) and for the data (on the right). 
Comparing the \hi\ and the CO data reveals some important differences.
As molecular gas is organised mostly in clumps which are distributed along spiral arms, the CO emission appears to be more patchy (in the $l-v$ space) than the \hi\ emission, despite the fact that both datacubes have been smoothed to the same resolution.
Clearly, this patchiness can not be properly reproduced by our model, which assumes a smooth and axisymmetric distribution of molecular gas. 
Apart from this difference, the model reproduces quite well the pattern shown by the data.
This is especially evident in QI, where several features that are not located in the proximity of the tangent velocity - and therefore are not directly fit - are well predicted by the model.
The most evident of these features originates around $l\!=\!0\de, v\!=\!0\kms$ and crosses the diagram diagonally up to $l\!\sim\!30\de, v\!=\!120\kms$, representing part of the Galactic `molecular ring' \citep[e.g.][]{Clemens+88}.
As we can reproduce reasonably well both the peaks and the tails of the observed CO profiles, there is no need to include a second component in the line-fitting.

The rotation curve, the velocity dispersion and midplane density profiles for the Galactic H$_2$ are shown in Fig.\,\ref{result_CO}.
As already noticed for the \hi, also the molecular gas of the Milky Way shows a remarkable degree of symmetry between the approaching and the receding sides of the disc. 
For both $v_{\phi}$ and $\sigma$, the main differences between the two sides arise at $R\!<\!1.5\kpc$, where non-circular motions induced by the bar play an important role.  
The H$_2$ velocity dispersion peaks at the Galaxy center with a value of $15\!-\!20\kms$. 
Similarly to the atomic hydrogen, $\sigma$ drops quickly within the inner $1\kpc$ down to a value $\sim4\kms$, and it slowly decreases at larger radii.
Despite the molecular gas being concentrated in spiral arms, the midplane H$_2$ densities in QI and QIV follow a similar global trend.  
The exception is that, in a few locations ($R=2.7,3.5,5.2\kpc$), peaks in QI correspond to depressions in QIV (or viceversa).
The atomic gas shows similar features (Section \ref{HI2comp}), though their locations do not perfectly overlap with those of the H$_2$. 

We now model the CO distribution in the direction perpendicular to the midplane.
Our procedure is analogous to that used in Section \ref{HIscaleheight} for the \hi, with two differences.
First, we limit the fit to the latitudes $|b|\!<\!3\de$, because a) the signal-to-noise of the data decreases significantly at higher latitudes; and b) in the regions of our interest, the $l\!-\!v$ diagrams at $|b|\!>\!3\de$ are similar to those at $|b|\!=\!3\de$ and do not seem to contain additional information on the vertical distribution of the CO.
Second, we impose $h_s\!<\!200\pc$ as a prior.
Our fitting technique does not converge well if we leave the CO scale height completely free to vary, and it is reasonable to assume that the disc of molecular gas is thinner than the \hi\ disc in our Galaxy. 

Fig.\,\ref{COabove} compares some representative $l\!-\!v$ diagrams extracted from our best-fit model at different latitudes with those of the data.
Note that the noise in the data increases significantly with $|b|$.
As expected, the main difference between the model and the data is that the latter is more clumpy and shows more structures, suggesting that the molecular gas is not diffusely distributed in the disc as the model assumes.  
However, the model reproduces quite well the observed drop in CO emission as a function of latitude, indicating that the scale heights that we are determining are reliable.

The molecular hydrogen scale height and surface density profiles are shown in Fig.\,\ref{scaleheight_CO}.
The scale heights in the two quadrants show a very similar trend with $R$.
The average $h_s$ declines slowly with $R$, varying from $\sim\!100\pc$ in the Galactic centre to $\sim\!50\pc$ in the solar neighborhood.
As for the atomic hydrogen, also in this case we prefer to fix the scale height at $R\!>\!7\kpc$ to the value of $52\pc$, determined by averaging the approaching and the receding $h_s$ obtained at $R\!=\!7\kpc$.
In a recent review, \citet{HeyerDame15} have presented a compilation of H$_2$ scale-height profiles derived in the last 30 years via different methods applied to different datasets.
In all cases, the H$_2$ HWHM scale-height stays about flat in the region $3\!<\!R\!<\!8\kpc$, with mean and standard deviation of $50\pm11\pc$, compatible with the values determined here (see also Table \ref{table_mean}).
The H$_2$ surface density profile indicates that most gas is concentrated either close to the Galactic centre or in the region $2.5\!<\!R\!<\!5.0\kpc$, while a large depression occurs at $1\!<\!R\!<\!2.3\kpc$ which was already noticed by \citet{Sanders+84}.
The average $\Sigma_{\rm H2}$ is characterised by three distinct peaks: the first around $R\!=\!3.3\kpc$, the second at $R\!=\!4.5\kpc$ and the third at $R\!=\!6.4\kpc$.
The first peak is however a by-product of our averaging, as the CO distribution in QI and QIV in the region $2\!<\!R\!<\!4\kpc$ is opposite in nature.
The other two peaks are instead `genuine', being present in both the approaching and the receding regions - although the third is especially prominent in QIV.
Our H$_2$ surface density profile is compatible with that determined by \citet{Bronfman+88} (dashed line in Fig.\,\ref{scaleheight_CO}), but shows more structures: peaks are higher and depressions are deeper.
The agreement with the profile derived by \citet{NakanishiSofue06} is however less good, especially at $R\!<\!2.5\kpc$. 
Note that the Bronfman et al. and the Sofue et al.'s profiles do not agree with each other at $R\!<\!3\kpc$. 
 
\section{Discussion}\label{discussion}
\subsection{Comparison between atomic and molecular gas}\label{compare_HI_H2}
\begin{figure*}[tbh]
\centering
\includegraphics[width=1.0\textwidth]{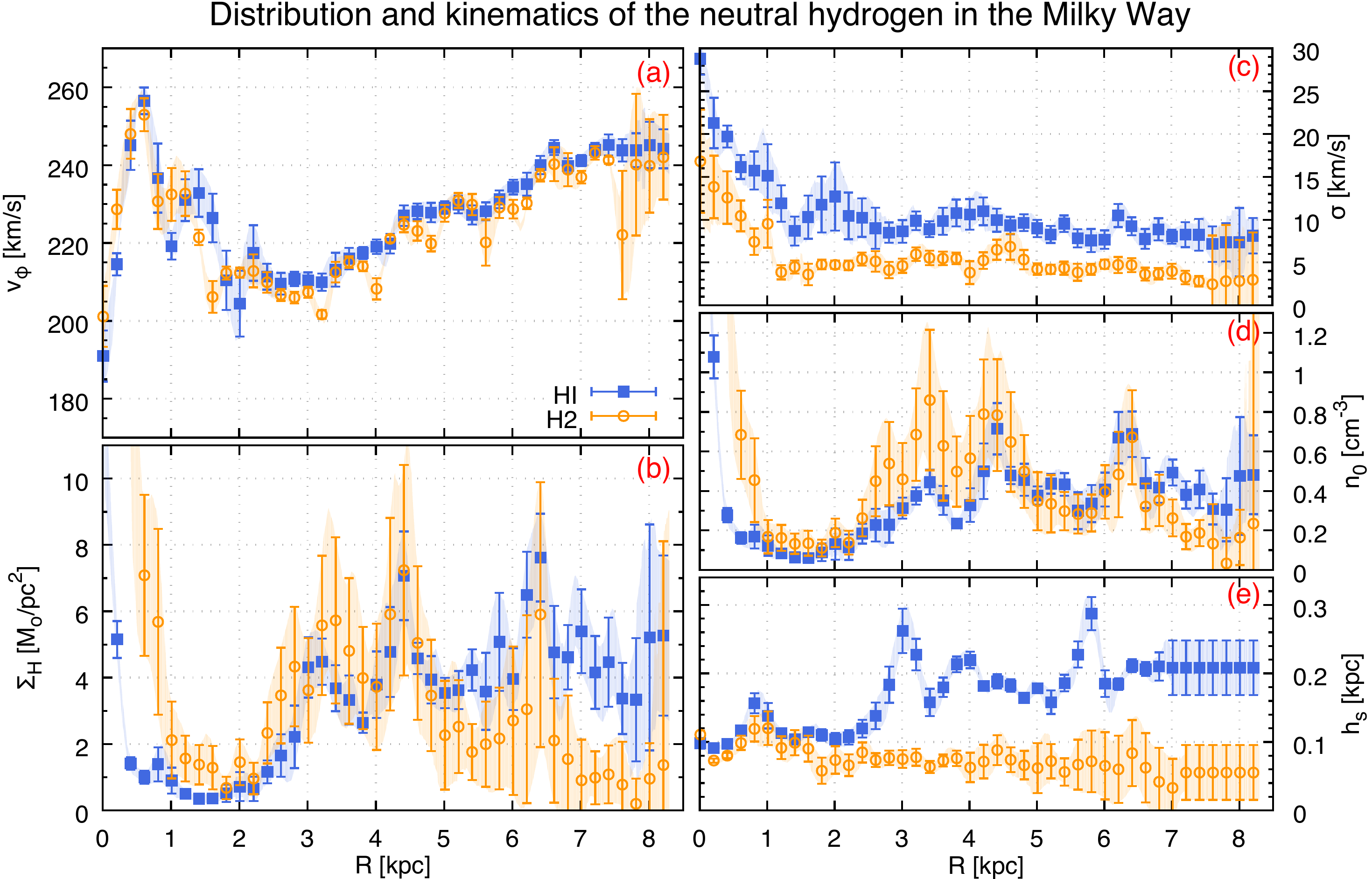}
\caption{Distribution and kinematics of the atomic (blue filled squares) and CO-bright molecular (brown empty circles) hydrogen in the inner disc of the Milky Way, as derived from our modelling. The panels show the profiles for the rotation velocity ($a$), surface density ($b$), velocity dispersion ($c$), midplane density ($d$) and HWHM scale height ($e$). Each profile is smoothed to the resolution of $0.2\kpc$. One point per resolution element is plotted. Error-bars show the $1\sigma$ uncertainty on the measurements, and account for both the approaching-receding asymmetry and the MCMC error-bars derived via our fitting technique (see the text for details). An additional $30\%$ error on the H$_2$ midplane and surface densities is included to account for the uncertainties in the $X_{\rm CO}$-factor \citep{Bolatto+13}. An ad-hoc $40\pc$ error is assumed for the scale height at $R\!>\!7\kpc$.}
\label{HI_vs_H2}
\end{figure*}

The results obtained in this work are summarized in Fig.\,\ref{HI_vs_H2}, which shows the radial profiles for the rotation velocity (panel $a$), surface density (panel $b$), velocity dispersion (panel $c$), midplane density (panel $d$) and half width at half maximum scale height (panel $e$) for both the atomic and the CO-bright molecular hydrogen of the Milky Way.
As in the previous Sections, each profile is derived by averaging the approaching and the receding parts, which are smoothed to the spatial resolution of $0.2\kpc$, while eq.\,(\ref{avgsigma}) and (\ref{avgn0}) are used to compute the average \hi\ velocity dispersion and midplane density.
We show one point per resolution element.

The uncertainty $\sigma_{\rm tot}$ on each quantity, which is represented by the error-bars in Fig.\,\ref{HI_vs_H2}, is derived by summing in quadrature two terms.
The first term is $\sigma_{\rm asym}$, which accounts for the asymmetries between the two regions QI and QIV.
Following \citet{Swaters99}, we estimate $\sigma_{\rm asym}$ as the difference between the (smoothed) approaching and receding profiles divided by 4.
For the \hi\ velocity dispersion and midplane density, $\sigma_{\rm asym}$ is computed by subtracting - instead of adding up - the two terms in the square brackets of eq.\,(\ref{avgsigma}) and (\ref{avgn0}), and dividing the result by 4.
The second term is $\sigma_{\rm fit}$, which takes into account the MCMC error-bars given by the standard deviations of the posterior probability distributions as described in Section \ref{fitting}.
For a given parameter at a given $R$, $\sigma_{\rm fit}$ is computed as the median MCMC error-bar within $0.2\kpc$ from that radius.
Error-bars from QI and QIV are both included in the computation of $\sigma_{\rm fit}$.
An additional $30\%$ uncertainty on the H$_2$ midplane and column density is summed in quadrature to account for the uncertainty in the $X_{\rm CO}$ factor \citep{Bolatto+13}.
Finally, a generous ad-hoc 40 pc error is assumed for both the \hi\ and H$_2$ scale heights at $R\!>\!7\kpc$, where our modelling is poor and $h_s$ is fixed to a constant value.

The similarities between the atomic and molecular gas in terms of kinematics and distribution are striking.
The \hi\ and H$_2$ rotation curves remarkably overlap with each other.
They both rise quickly, peak to $\sim260\kms$ at $0.6\kpc$ from the centre, decline down to $210\kms$ at $R\!=\!2.5$, and then gently rise again. 
The rotation velocity flattens around $R\!=\!6.5\kpc$ to a value that depends on the choice of $\vsun$, which here is $240\kms$ (see Appendix \ref{Vsun220} for the rotation curves re-scaled to $\vsun\!=\!220\kms$).
The main difference between the two components is that that the rotation traced by the H$_2$ shows several `bumps' that pulls the velocity downwards with respect to the values determined for the \hi.
Our interpretation is that, given that the molecular gas a) is less diffuse and has a more `patchy' distribution with respect to the \hi, and b) is likely to be more concentrated around the Galaxy spiral arms, there is no guarantee that some CO emission will always occur at the tangent point for all line-of-sights.  
When this does not happen, our fitting algorithm interprets the lack of emission in terms of a drop in $v_\phi$. 
These bumps would therefore be artificial and caused by the fact that H$_2$ is not homogeneously distributed within the disc.
However, the magnitude of these velocity drops is very small, typically less than $10\kms$, suggesting that the molecular gas in the Galactic disc can be treated as a diffuse and continuous component, at least as a first approximation.

The \hi\ and H$_2$ surface density profiles are characterised by large error-bars, due to the propagation of the uncertainties on the scale heights and midplane densities.
In spite of that, they show several similarities: they both peak around the Galactic centre, have a deep depression in the region $0.5\!<\!R\!<\!2.5\kpc$ and then rise again at larger distances.
We find all these correlations to be very significant, especially because they have been derived by a blind fit to two totally independent datasets.
The surface density in the Galactic centre is dominated by the molecular gas, but its precise value is very uncertain: we estimate $12\pm2\msunsqp$ for the \hi\ and $115\pm40\msunsqp$ for the H$_2$.
The \hi\ density profile at $R\!>\!3\kpc$ is fairly flat, between $3$ and $7\msunsqp$, whereas the H$_2$ density declines with radius and abruptly drops down to $\sim1\msunsqp$ at $R\!>\!7\kpc$.
Apart from the general trend, both surface density profiles show two clear peaks at $R\!=\!4.5\kpc$ and $R\!=\!6.4\kpc$.
In Section \ref{spiralarms} we discuss two possible origin for these features: one is that they are due to actual ring-like overdensities in the gas distribution, the other is that they arise because, at those particular locations, the line-of-sight intersects the regions of spiral arms, where the gas density is above the mean. 
Interestingly, there is little trace of these features in the velocity dispersion profiles or in the H$_2$ scale height profile, while the \hi\ scale height peaks at slightly lower radii with respect to the density enhancements.
An additional peak can be seen around $R\!\simeq\!3\kpc$ but, as mentioned in Section \ref{resultsh2}, at $2\!<\!R\!<\!4\kpc$ the H$_2$ distribution in QI and QIV are anti-correlated thus the correspondence of the average H$_2$ profile with the \hi\ profile in this regions could be, to some extent, coincidental.

\begin{table*}[tbh]
\caption{Properties of the Galactic atomic and molecular hydrogen averaged in the region $3\!<\!R<\!8.3\kpc$. $\sigma$: velocity dispersion; $n_0$: midplane density; $h_s$: HWHM scale height; $\Sigma$: face-on surface density. Each entry shows the mean value computed in the region considered, followed by its standard deviation and by the mean uncertainty associated to the measurements. Scale heights are derived assuming Gaussian profiles for the vertical distribution of the gas. The last row shows the total \hicap\ and H$_2$ masses within the Solar circle, determined as discussed in the text. Masses and densities do not include the contribution from Helium.
}              
\label{table_mean}      
\centering                                      
\begin{tabular}{l c c c}          
\hline\hline                        
Par. & units & \hicap & H$_2$ \\    
\hline                                   
$\sigma$	&	$\kms$		&	$8.9\pm1.1\pm1.5$		&	$4.4\pm1.2\pm1.5$\\
$n_0$	&	$\cmmc$		&	$0.43\pm0.11\pm0.10$	&	$0.42\pm0.22\pm0.21$\\
$h_s$	&	$\pc$		&	$202\pm28\pm21$		&	$64\pm12\pm32$\\
$\Sigma$	&	$\msunspc$	&	$4.5\pm1.1\pm1.6$		&	$3.0\pm1.9\pm2.0$\\
\hline             
$M_{\rm H}(\!<\!\rsun)$	&	$10^8\msun$		&	$9.0\pm0.5$		&	$6.7\pm0.7$\\
\hline
\end{tabular}
\end{table*}

\begin{table*}[tbh]
\caption{Properties of the Galactic atomic and molecular hydrogen extrapolated to $R\!=\!\rsun\!=\!8.3\kpc$. Values for the scale heights (in parenthesis) are fixed to those determined at $R\!=\!7\kpc$, for which an ad-hoc uncertainty of $40\pc$ is assumed.}              
\label{table_Rsun}      
\centering                                      
\begin{tabular}{l c c c}          
\hline\hline                        
Par. & units & \hicap & H$_2$ \\    
\hline                                   
$\sigma$	&	$\kms$		&	$7.8\pm0.8$		&	$3.7\pm1.2$\\
$n_0$	&	$\cmmc$		&	$0.41\pm0.06$		&	$0.23\pm0.11$\\
$h_s$	&	$\pc$		&	$(217\pm40)$		&	$(52\pm40)$\\
$\Sigma$	&	$\msunspc$	&	$4.5\pm0.7$		&	$1.3\pm0.7$\\
\hline                                             
\end{tabular}
\end{table*}

Table \ref{table_mean} shows the typical properties of the Galactic atomic and molecular hydrogen averaged in the region $3\!<\!R<\!8.3\kpc$, where fluctuations in the various quantities are limited, along with their total masses within the Solar circle. 
For each model parameter we show the mean value computed in the region considered, the standard deviation around the mean, and the mean uncertainty - i.e., the mean error bar shown in Fig.\,\ref{HI_vs_H2} - in that region.
Values for the rotation velocity depends on the choice of $\vsun$ (see Appendix \ref{Vsun220}), and therefore are not shown.
Note that, given that the molecular gas density is a decreasing function of $R$, its mean value is not truly representative for the density in the region considered.
The total masses within the Solar circle are computed as the mean and the standard deviation of $10^4$ different realisations, each derived by randomly extracting the \hi\ and H$_2$ surface densities profiles from the values (points with error bar) shown in panel (b) of Fig.\,\ref{HI_vs_H2}\footnote{For this purpose, each point with error bar is interpreted as the mean and the standard deviation of a Gaussian distribution}.
Our fiducial value for the H$_2$ mass within the Solar circle is in broad agreement with that of \citet{HeyerDame15}.
We stress that these masses, even when corrected for the Helium content, do not represent a complete census of the gas content in the inner disc of the Milky Way. 
Diffuse ionized gas contributes in mass around $1/3$ of the neutral atomic component \citep{Reynolds93}. 
Also, studies of [\cii] emission-lines have revealed the presence of a CO-dark H$_2$ component that may encompass a significant fraction ($20-50\%$) of the molecular gas in the Galaxy \citep{Langer+14}. 
Clearly, these components are missed by our analysis.

In order to derive the gas properties in the Solar neighborhood it is best to extrapolate from measurements at lower radii, as the uncertainties associated to the measurements at $R\!=\!\rsun$ are large.
The extrapolation is computed as follows.
First, for each parameter radial profile we produce $10^4$ realisations by randomly extracting the parameter values in the region $R\!>\!7.0\kpc$ from the points with error bars shown in Fig.\,\ref{HI_vs_H2}.
Each realisation is then fit with a straight line, and the value of the fit extrapolated at $R\!=\!8.3\kpc$ is recorded.
In Table \ref{table_Rsun} we show, for the various parameters, the mean and the standard deviations of these extrapolated values.
Clearly, the properties of the \hi\ in the Solar neighborhood are similar to those determined at lower radii, while the molecular gas is less abundant at $R\!=\!\rsun$ with respect to the rest of the inner disk.

\subsection{An independent estimate for the local \hi\ column density}
Recently, \citet{McKee+15} re-analised archival data in order to provide a census of surface densities for different baryonic components in the Solar neighborhood.
Using previous results by \citet{Heiles76} and \citet{DickeyLockman90}, and adopting a correction for the optical depth based on \citet{StrasserTaylor04}, they derive an \hi\ surface density in the Solar neighborhood $\Sigma_{\rm HI,\odot}$ of $7.8\pm1.2\msunsqp$ (from their Table 2, after removing the $40\%$ correction for the He and heavier elements), which is in tension with our determination of $4.5\pm0.7\msunsqp$.
They also quote a fiducial value for the local H$_2$ surface density of $0.7\pm0.2\msun$, which is compatible with our estimate given our large error-bar.

One might indeed doubt the values presented in our Table \ref{table_Rsun}, given that our modelling technique is less robust in the proximity of the Solar circle.
For this reason, we now provide a model-independent estimate of $\Sigma_{\rm HI,\odot}$ using an approach similar to that adopted by \citet{Heiles76}, which is based on a simple geometrical relation between the column density perpendicular to the disc and that measured towards a generic line of sight.

First, we mask in the LAB datacube the \hi\ emission produced by high and intermediate-velocity clouds, which can potentially contaminate our computation.
This masking follows the procedure described by \citet{MarascoFraternali11} in their Section 2.1, and is based on a $35\kms$ cut in `deviation' velocity \citep[see][]{Wakker91b}.
As discussed in \citet{MarascoFraternali11}, the masking is very conservative and indeed we find that it has only a marginal impact on our results, as by neglecting it the surface density increases by only $\sim0.2\msunsqp$.  
Next, we produce an all-sky \hi\ column density map by integrating in velocity all line profiles in the (masked) LAB data.
From this map, at any given latitude $b$ we compute a median column density $\widetilde{\Sigma}_{\rm HI}(b)$ by considering many sight-lines: consistently with our previous analysis, we focus separately on $0\de\!<\!l\!<\!90\de$ (region QI) and on $270\de\!<\!l\!<\!360\de$ (region QIV), for which we compute two distinct $\widetilde{\Sigma}_{\rm HI}(b)$.
Under the assumption of a constant gas scale-height within $\sim1\kpc$ from the Sun, $\Sigma_{\rm HI,\odot}$ can be computed as
\begin{equation}\label{NHI_local}
\Sigma_{\rm HI,\odot} = 2\times\sin(|b|)\times\left(\mathcal{D}\,\widetilde{\Sigma}_{\rm HI}(b) + \Sigma_{\rm HI,LB}(b)\right)
\end{equation}
where $\mathcal{D}$ is the correction for the \hi\ optical depth ($\mathcal{D}\!\ge\!1$) and $\Sigma_{\rm HI,LB}(b)$ is a sight-line dependent correction due to the Local Bubble, a severely gas-depleted region that surrounds the Solar neighborhood \citep[e.g.][]{CoxReynolds87}.
Here we adopt $\mathcal{D}\!=\!1.3$, in agreement with our findings (see Section \ref{fiducial_vs_thin}) and with those of \citet{StrasserTaylor04}.
Given the complex shape of the Local Bubble, the rightmost term in eq.\,(\ref{NHI_local}) is difficult to determine and we ignore it for a moment.

\begin{figure}[tbh]
\centering
\includegraphics[width=0.5\textwidth]{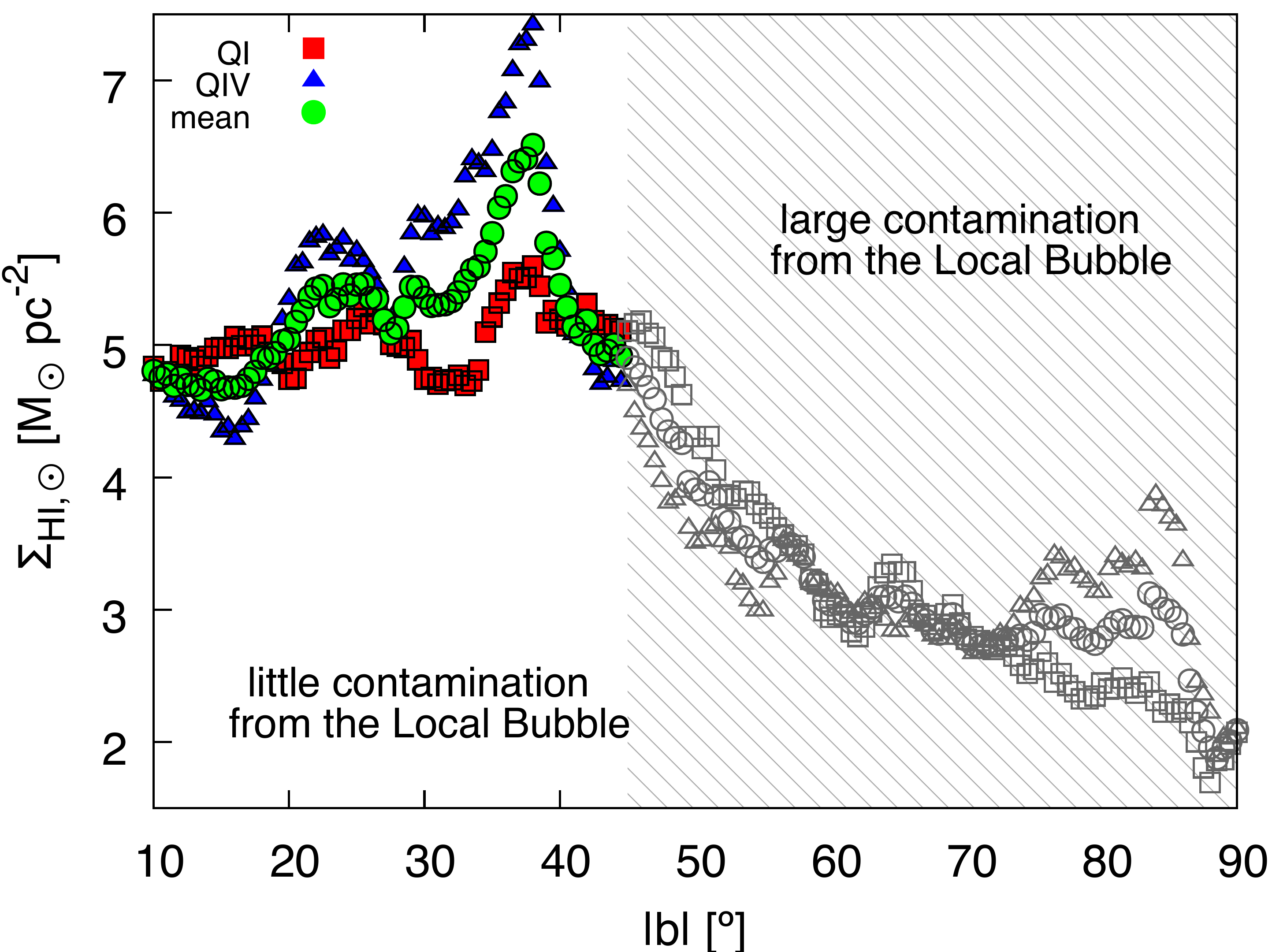}
\caption{\hicap\ column density in the Solar neighborhood determined via eq.\,(\ref{NHI_local}) at different Galactic latitudes $b$, based on the LAB data. 
\hicap\ emission from high and intermediate-velocity clouds have been masked.
The regions QI (squares) and QIV (triangles) are shown separately. The mean column density is shown by circles.
Values at $|b|\!>\!45\de$ (empty symbols) are strongly contaminated by the Local Bubble. Using only data at $10\de\!<\!|b|\!<\!45\de$ (filled symbols) we infer $\Sigma_{\rm HI,\odot}\!=\!5.3\pm0.5\msunsqp$ (accounting for the \hicap\ opacity, but not correcting for He).}
\label{NHI_local_fig}
\end{figure}

In Fig.\,\ref{NHI_local_fig} we show how $\Sigma_{\rm HI,\odot}$ varies as a function of $|b|$, for $|b|>10\de$, and for regions QI and QIV separately.
Ideally, $\Sigma_{\rm HI,\odot}$ should stay constant at all latitudes.
Instead, there is a clear discontinuity around $|b|\!\sim\!45\de$, above which the mean column density (shown by green circles in Fig.\,\ref{NHI_local_fig}) drops by a factor of $\sim2$: this can be readily explained as due to the `hourglass' shape of the Local Bubble \citep{Lallement+03} that - extending preferentially in the direction perpendicular to the midplane - drastically contaminates the column density measurements at high latitudes.
At lower latitudes, however, the contamination should be largely reduced as the Bubble `wall' is much closer to the Sun.
For instance, assuming a distance to Bubble wall of $60\pc$ \citep{Lallement+03}, we find an overall correction to $\Sigma_{\rm HI,\odot}$ of about $0.2\msunspc$ at $b\!=\!10\de$ using the mean \hi\ disc parameters of Table \ref{table_mean}.
In the region $10\de\!<\!|b|\!<\!45\de$ the mean $\Sigma_{\rm HI,\odot}$ seems approximately flat: here we derive an \hi\ surface density of $5.3\pm0.5\pm0.2\msunsqp$, where the first term is the mean in the region considered, followed by the standard deviation and by the mean uncertainty computed as the difference between QI and QIV divided by 4.
We find a slightly lower and more precise value ($5.0\pm0.3\pm0.1\msunsqp$) if we also exclude data at $28\de\!<\!|b|\!<\!\de40$, where QI and QIV show a significantly different trend.

These estimates are compatible with - but slightly larger than - the surface density reported in Table \ref{table_Rsun}.
While this suggests that the value of $\Sigma_{\rm HI,\odot}$ shown in Table \ref{table_Rsun} may be slightly underestimated, in Section \ref{extrapl} we argue that the whole \hi\ surface density profile may be overestimated because of contamination from extra-planar gas. 
In the light of these considerations, we still take $\Sigma_{\rm HI,\odot}\!=\!4.5\pm0.7$ as our fiducial value.

\subsection{On the nature of the second \hi\ component}\label{2compdiscussion}
It is hard to tell whether or not the additional low-density, high-$\sigma$ component used in Section \ref{HI2comp} to fit the Galactic \hi\ data is spurious in nature, i.e. it simply serves the purpose of absorbing the insufficiencies of our model, or has true physical meaning.
On the one hand, the two components are \emph{not} related to the classical `cold' and `warm' neutral ISM \citep[e.g.][]{Wolfire+03}, as their velocity dispersion is too high to be caused by pure thermal broadening.
On the other hands, double-component \hi\ profiles are observed also in external galaxies: \citet{Ianjamasimanana+12} studied stacked profiles in The \hi\ Nearby Galaxy Survey \citep[THINGS;][]{Walter+08} and found that they can be decomposed into a narrow and a broad component, with velocity dispersions analogous to those found here.

One possibility is that the second, high-$\sigma$ component is required to account for the various non-axisymmetric features along the line of sight, like streaming motions in the spiral arms or stochastic substructures.
These features modify locally the gas density and velocity dispersion and might induce deviations from the pure differential rotation.
All these effects contaminate the \hi\ line profiles in a way that is difficult to predict, but that can be globally taken into account by including a second, `artificial' \hi\ component with a different density and velocity dispersion.

Another possibility is that diffuse star formation in the disc injects turbulence in a fraction of the \hi\ medium. 
In this case, the high-$\sigma$ component would represent a genuine, turbulent phase of the Galactic ISM.
The kinetic energy associated to our second component is $6.3\times10^{53}\erg$, assuming a constant velocity dispersion of $\sim15\kms$ and accounting for the He content.
Assuming a turbulence dissipation timescale of $10^7\yr$ \citep{Tamburro+09}, an energy injection rate of $2\times10^{39}\ergs$ would be required to maintain the gas in this state.
This rate can be easily achieved by supernova feedback: for a Galactic supernova rate of $2$ events per century \citep{Diehl+06}, and assuming the canonical $1\times10^{51}\erg$ per event, a feedback efficiency of only $0.3\%$ would suffice to match this requirement.
Thus this second scenario is energetically plausible.

A third possibility is that this high-$\sigma$ component is the midplane counterpart of the high-latitude Galactic extra-planar \hi.
Extra-planar gas is indeed more turbulent than the gas in the disc and, overall, contributes only to a small fraction ($5\!-\!10\%$) of the total \hi\ mass \citep{Fraternali+01,Oosterloo+07,MarascoFraternali11}.
If the extra-planar \hi\ is produced by the Galactic fountain mechanism \citep{MFB12}, the neutral phase of the returning fountain flow could be observable on the midplane as a low-density, high-$\sigma$ component\footnote{In the model of \citet{MFB12} the fountain clouds are ejected from the disc fully ionized, thus \hicap\ emission is visible preferentially during the descending phase of their trajectories.}.
In this scenario, as in that previously discussed, it is feedback from star formation which ultimately produces the second component.
A possible argument against this scenario is that, if the fountain clouds that return back to the disc have substantially lower rotational speed with respect to the midplane gas, their emission should be systematically shifted to velocities lower than the terminal velocity, and therefore their contribution to the \hi\ profiles should be limited to the brightest regions of the $l\!-\!v$ diagram and not appear in its high-velocity, low-$T_{\rm B}$ tail.

\subsection{The effect of the bar}\label{effectofbar}
Perhaps the main weakness of our model is that it is based on the assumption that the gas follows pure circular orbits, which would be correct only if the gravitational potential was perfectly axisymmetric.
It is well established, however, that the Milky Way is a barred galaxy \citep[e.g.][]{Dwek+95,Binney+97}.
The first attempt to characterise the gas distribution and kinematics in the central regions of the Galaxy was made by \citet{LisztBurton80}, who proposed a scenario where the gas in the innermost 2 kpc moves along elliptical paths in a ringlike structure which is tilted with respect to disc midplane \citep[see also][]{Ferriere+07}.
Later on, numerical simulations of gas flows in a barred potential have revealed that, in the presence of a bar, the gas kinematics and distribution become asymmetric and non-stationary \citep{Fux99,RFC08,Sormani+15}.
As a consequence, the exact shape of the \hi\ and CO longitude-velocity diagrams in the innermost regions of the Galaxy - say $340\de\!<\!l\!<\!20\de$ - will depend on the details of the bar shape and pattern speed, on its orientation with respect to our view-point, and on the observing epoch.

In general, the rotation curve at $R\!\lesssim\!2.5\kpc$ shown in Fig.\,\ref{HI_vs_H2} should not be taken as representative for the circular velocity in that region.
The circular velocity should rise more gently than what can be derived from a simple tangent-point analysis of the l-v diagrams: this has been first recognized by \citet{Binney+91}, and remarked recently by dynamical models of the inner Galactic disc based on the density and kinematics of stars \citep{Portail+17} (see also Fig.\,\ref{MWvsnearbyVrot}).
Unfortunately, the only `correct' approach to study the gas dynamics in the innermost few kpc of the Galaxy is to follow the gas flow via hydrodynamical simulations and adjust the potential parameters in order to reproduce the details of the l-v diagrams, like in \citet{SormaniMagorrian15}.
This is, however, beyond the scope of the current work.

\subsection{The effect of the spiral arms}\label{spiralarms}
\begin{figure}[tbh]
\centering
\includegraphics[width=0.5\textwidth]{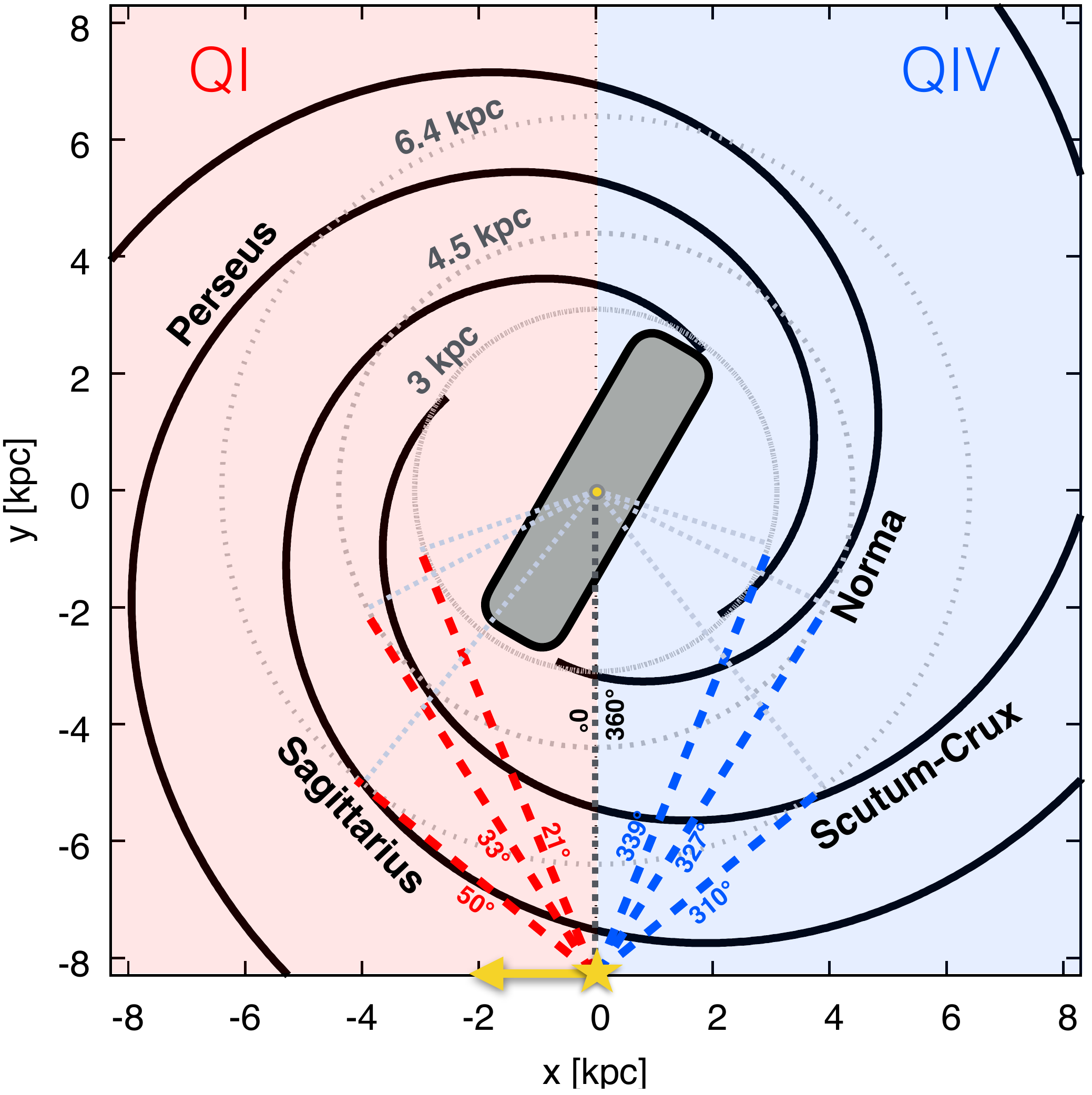}
\caption{Face-on view of the Galactic inner disc with the four-arm spiral pattern of \citet{Steiman-Cameron+10} overlaid on top. The Sun is at coordinates $(x,y)\!=\!(0,-8,3)\kpc$. The disk rotates clockwise, the approaching (QIV) and the receding (QI) sides are shaded in blue and in red respectively. An indicative 3-kpc bar is shown, oriented at $30\de$ with respect to the $l\!=\!0\de$ sight-line \citep{Wegg+15}.
The three circles (grey dotted lines) show the radii of the gas overdensities discussed in the text, while the red and blue dashed lines show the sight-lines that are tangent to these circles. The gas overdensities at $R\!=\!4.5,6.4\kpc$ are consistent with the sight-lines intercepting the spiral arms.}
\label{arms_peaks}
\end{figure} 

As already mentioned, the atomic and the molecular gas distributions show two prominent peaks at $R\!=\!4.5,6.4\kpc$. 
Remarkably, these peaks can be seen in both the receding and the approaching regions of the disc (Fig.\,\ref{scaleheight_2c_240} and \ref{scaleheight_CO}), and they correspond to distinctive `bumps' visible in the $l-v$ diagrams around $l\!=\!30\de,50\de$ (QI) and $l\!=\!310\de,330\de$ (QIV).
We noted that a third gas overdensity can be seen around $R\!=\!3\kpc$: it is visibile in both QI and QIV in \hi, but it is separated in two distinct peaks in H$_2$.
The fact that at least two of these overdensities are visible in both quadrants is significant.
The simplest explanation for these features would be the presence of two distinct ring-like gas overdensities within the Galactic disc at radii of $4.5$ and $6.4\kpc$.
Ring-like features are observed in the neutral gas of nearby galaxies, like NGC 3184 or NGC 2841 \citep{Walter+08}.

Another possibility is that these features arise when the line-of-sight intersects the regions of spiral arms, where the gas density is above the mean.
Unlike the previous scenario, in this case the density enhancement would be local, thus the derived column density would not be representative for the mean column density at that radius.
\citet{Burton71} already noticed the `bumps' in the \hi\ $l-v$ diagram of QI, and interpreted them in the framework of the density-wave theory \citep{Lin+69} as streaming motions induced by the Galactic spiral arms.
According to \citet{Burton71}, streaming motions alone are sufficient to explain the irregularities in the $l-v$ diagram of QI, even in the presence of a uniform gas density distribution (see his model V).
Clearly, our model does not include self-consistently non-circular motions induced by the spiral arms. 
However, we can check whether or not the sight-lines that are responsible for the peaks in the density distributions intersect the Galactic spiral arms at the locations where the overdensities occur.

In Fig.\,\ref{arms_peaks} we show a face-on view of the inner disc of the Milky Way with the four-arm spiral pattern of \citet{Steiman-Cameron+10} overlaid on top.
The spiral arm model is based on all-sky intensity maps of [\cii] and [\nii] lines, which trace the cool phase of the Galactic ISM.
The dashed red and blue lines in Fig.\,\ref{arms_peaks} represent the sight-lines (in QI and QIV respectively) for which the tangent point radius ($\rsun \sin(l)$) is $3,4.5$ or $6.4\kpc$. 
Clearly, the gas overdensity at $6.4\kpc$ can be explained as the $l\!=\!50\de$ sigh-line in QI intercepts the Sagittarius arm, while the $l\!=\!310\de$ sight-line in QIV intercepts the Scutum-Crux arm.
The overdensity at $4.5\kpc$ appears to be produced as the sight-lines in QI and QIV intercept the Scutum-Crux and the Norma arm respectively, although in this case the correspondence between spiral arms and tangent points is less precise.
Finally, the feature around $3\kpc$ is likely related to the long-known 3-kpc arm \citep{vanWoerden+57,DameThaddeus08} (not shown in Fig.\,\ref{arms_peaks}), or more in general to the region where the Perseus and the Scutum-Crux arms connect to the edge of the Galactic Bar.


\subsection{The effect of the spin temperature}\label{fiducial_vs_thin}
\begin{figure}[tb]
\centering
\includegraphics[width=0.49\textwidth]{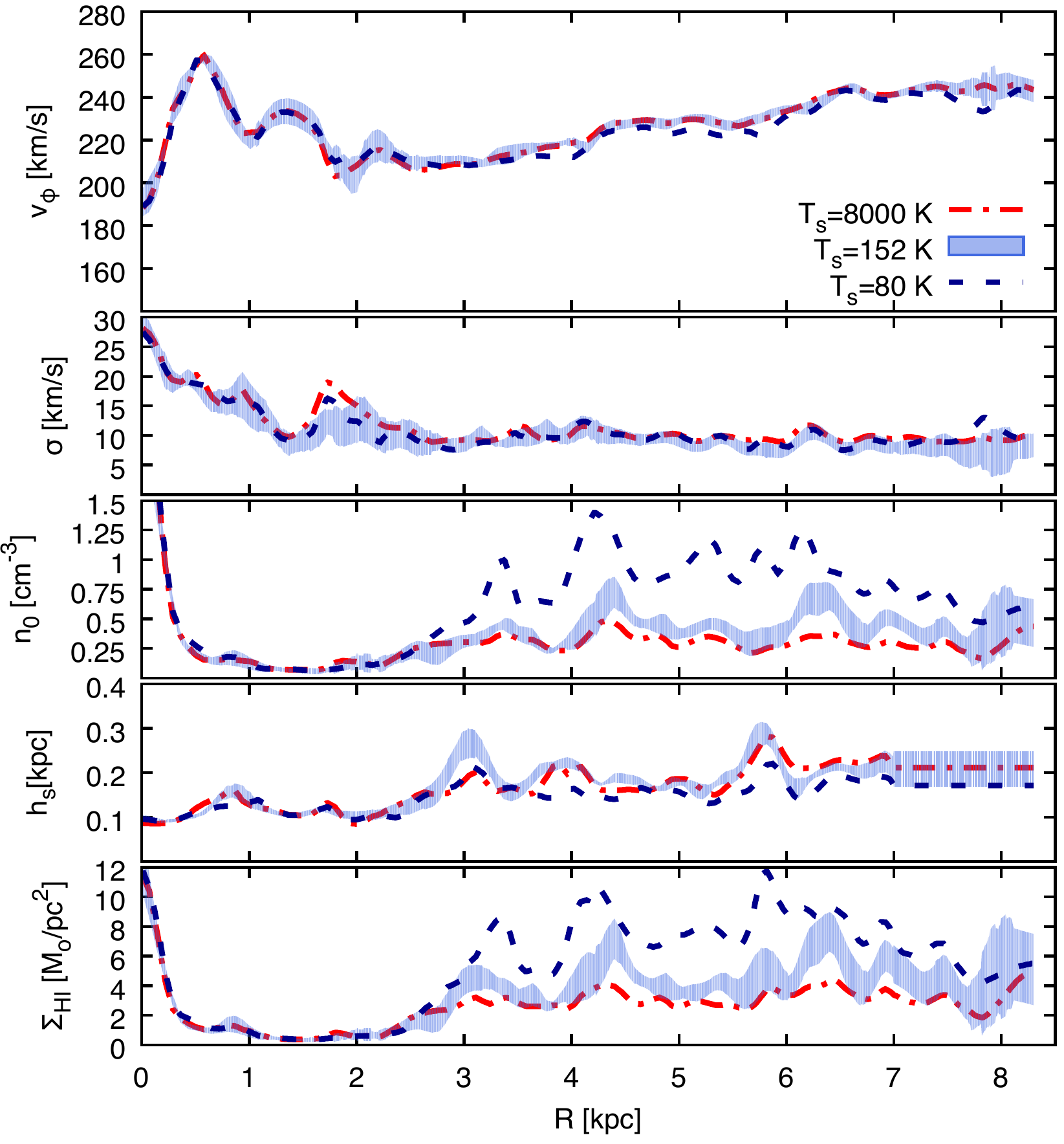}
\caption{
Comparison between our fiducial model (shaded regions) and two other models with different optical depths: an optically thin case ($T_{\rm S}\!=\!8000\K$, dot-dashed lines) and a more optically thick one ($T_{\rm S}\!=\!80\K$, dashed lines).
From \emph{top} to \emph{bottom}: rotation velocity, velocity dispersion, midplane density, scale height and surface density as a function of $R$.
Only the averaged profiles are shown, smoothed at the spatial resolution of $0.2\kpc$.
Different regimes predict significantly different column densities.
}
\label{thick_vs_thin}
\end{figure}

We now discuss the effect of different \hi\ spin temperatures on our results.
In Fig.\,\ref{thick_vs_thin} we compare the properties of the \hi\ disc derived by our fiducial model ($T_{\rm S}\!=\!152\K$, shaded region) with those predicted by our best-fit optically thin ($T_{\rm S}\!=\!8000\K$, dot-dashed line) and our best-fit optically thick ($T_{\rm S}\!=\!80\K$, dashed line) cases.
Clearly, differences concern mainly the gas density.
In the optically thin regime, $n_0$ is smaller because the \hi\ emission is not partially self-absorbed like in the fiducial case, so less gas is required to produce the same flux. 
The opposite happens in the optically thick regime.
Note that the density profiles predicted by these three regimes are not consistent with each other.
The \hi\ velocity dispersions predicted by the three regimes are virtually identical, despite the fact that \hi\ at higher temperature has a larger thermal broadening, corresponding to a higher velocity dispersion floor (see Section \ref{fitting}).

Since the scale heights of the three models are almost the same at all radii, the difference in the midplane density implies a difference in surface density and total mass.
The model with $T_{\rm S}\!=\!80\K$ reaches column densities up to $\sim\!10\msunsqp$, similar to those found by \citet[][see also Fig.\,\ref{scaleheight_2c_240}]{KalberlaDedes08}, but we exclude that such a low $T_{\rm S}$ is representative for the typical spin temperature of the Galactic \hi, as we discuss in Appendix \ref{appendix_NHI_Tb}.
The \hi\ masses within the solar circle predicted by these models are $6.4\times10^8\msun$ for the optically thin regime and $1.4\times10^9\msun$ for the optically thick one, which corresponds respectively to $30\%$ less and $52\%$ more than the mass predicted by our fiducial model ($9.1\times10^8\msun$).
Interestingly, a $30\%$ discrepancy in \hi\ mass between an opaque and an optically thin regime has also been found by \citet{StrasserTaylor04} in our Galaxy and by \citet{Braun+09} in M31.
Note that different spin temperatures have little impact on the \hi\ surface density in the solar neighborhood, which sits robustly around $5\msunsqp$, albeit we reiterate that this value is quite uncertain given that the scale height at $R\!>\!7\kpc$ can not be accurately determined.

\subsection{The effect of the extra-planar \hi}\label{extrapl}
\begin{figure}[tb]
\centering
\includegraphics[width=0.49\textwidth]{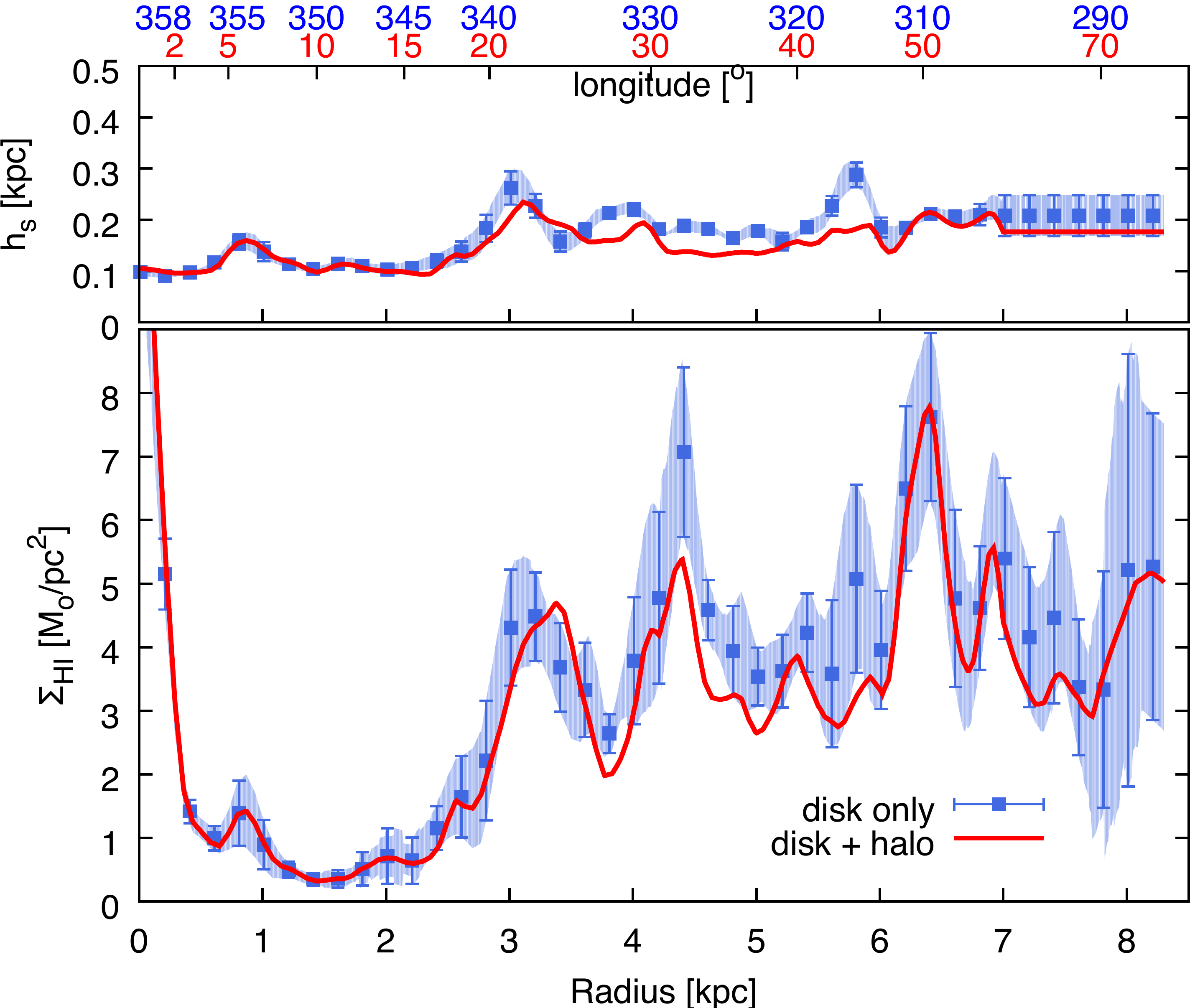}
\caption{
Galactic \hicap\ scale heights and surface density profiles predicted by our fiducial model (points with error-bars) and by a model that accounts for the extra-planar \hicap \citep[from][solid lines]{Marasco+13}. Only the averaged profiles are shown, smoothed at the spatial resolution of $0.2\kpc$. The model that include the extra-planar \hicap\ predicts slightly lower surface densities, but it is roughly consistent with the fiducial case. 
}
\label{scaleheight_HI_extraplanar}
\end{figure}

The \hi\ scale height and column density have been derived by assuming that the gas kinematics above and below the midplane are the same as those at $z\!=\!0$. 
However, \citet{MarascoFraternali11} found that the Milky Way is surrounded by a faint ($\sim\!5\!-\!10\%$ of the total \hi\ mass) layer of slow-rotating, inflowing extra-planar \hi, analogous to that found in external galaxies like NGC 2403 \citep{Fraternali+02}, NGC 891 \citep{Oosterloo+07} and NGC 3198 \citep{Gentile+13}.
This medium extends up to a few kpc above the disc and is thought to be produced by the Galactic fountain mechanism \citep{MFB12}.
As we test in Appendix \ref{testing}, neglecting this additional component leads to overestimate the \hi\ scale height and, as a consequence, the \hi\ column density.
A way to tackle this problem is to fit for the scale height by including in the model an additional component that represents the Galactic extra-planar \hi.
The parameters of this component are taken from \citet{Marasco+13}.
This approach has been tested on mock \hi\ observations of a system made of a disc plus an extra-planar component, and leads to the correct recovery of the input disc scale height (see Appendix \ref{testing} for the details).

In Fig.\,\ref{scaleheight_HI_extraplanar} we compare the scale height and surface density profile of our fiducial model, which does not include the extra-planar \hi, with those derived by including this additional component.
Differences are small, and arise only at $R\!>\!3-4\kpc$.
The surface density predicted by this new model is slightly lower than, but still consistent with, that predicted by our fiducial case.
The disc \hi\ mass within the inner Galaxy decreases by about $12\%$ (from $9.0$ to $7.9\times10^8\msun$) when including the extra-planar gas, which is less than the difference that we would derive by neglecting the \hi\ opacity in our calculations (Section \ref{fiducial_vs_thin}).
We conclude that the presence of the extra-planar \hi\ has only a modest impact on our results. 
By analogy with Table \ref{table_mean} and \ref{table_Rsun}, in Table \ref{table_mean_extrapl} and \ref{table_Rsun_extrapl} we report, respectively, the typical gas properties averaged in the region $3\!<\!R\!<\!8.3\kpc$ and those extrapolated to $R\!=\!\rsun$ for this new model that includes the extra-planar gas.

\begin{table}[tbh]
\caption{As in Table \ref{table_mean}, but correcting for the presence of the Galactic extra-planar \hicap\ as discussed in the text. Values for $n_0$ and $\sigma$ are the same as those reported in Table \ref{table_mean}.
}              
\label{table_mean_extrapl}      
\centering                                      
\begin{tabular}{l c c}          
\hline\hline                        
Par. & units & \hicap\\    
\hline                                   
$h_s$	&	$\pc$		&	$172\pm24\pm21$\\
$\Sigma$	&	$\msunspc$	&	$3.9\pm1.1\pm1.2$\\
\hline             
$M_{\rm H}(\!<\!\rsun)$	&	$10^8\msun$		&	$7.9\pm0.5$\\
\hline
\end{tabular}
\end{table}

\begin{table}[tbh]
\caption{As in Table \ref{table_Rsun}, but correcting for the presence of the Galactic extra-planar \hicap\ as discussed in the text. Values for $n_0$ and $\sigma$ are the same as those reported in Table \ref{table_Rsun}.}              
\label{table_Rsun_extrapl}      
\centering                                      
\begin{tabular}{l c c}          
\hline\hline                        
Par. & units & \hicap\\    
\hline                                   
$h_s$	&	$\pc$		&	$(176\pm40)$\\
$\Sigma$	&	$\msunspc$	&	$4.0\pm0.7$\\
\hline             
\end{tabular}
\end{table}

\subsection{Comparison with external galaxies}\label{MW_vs_nearby}
\begin{figure*}[tb]
\centering
\includegraphics[width=1.0\textwidth]{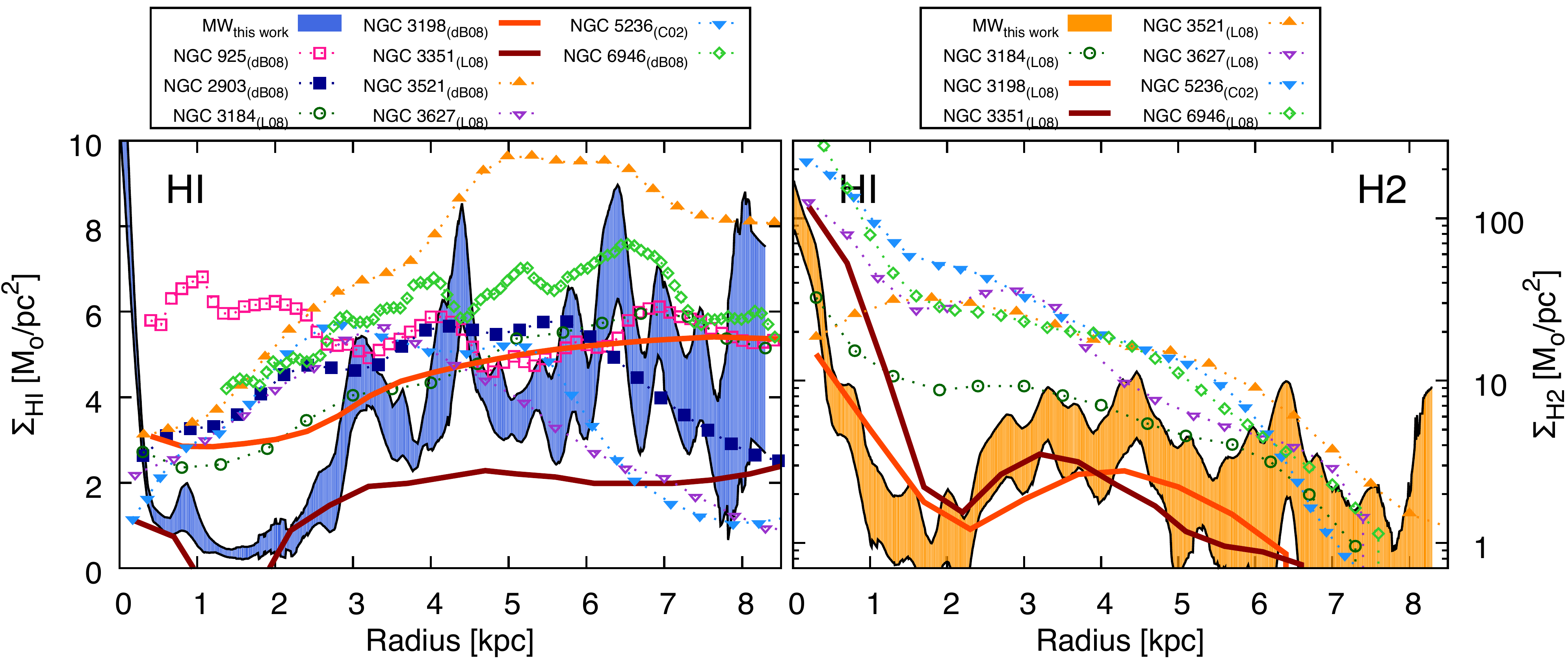}
\caption{Comparison between the \hicap\ (left panel) and H$_2$ (right panel) surface density profiles of the Milky Way (shaded regions), as determined in this work, and of a sample of nearby barred spirals (symbols and lines) from \citet[][L08]{Leroy+08} and \citet[][dB08]{deBlok+08}. Data for NGC 5236 are from \citet[][C02]{Crosthwaite+02}. NGC 3351 and NGC 3198 (solid lines) have `SB' morphological classification (from CR3), the other galaxies are classified as SAB. Note the different scales used for $\Sigma_{\rm HI}$ and $\Sigma_{\rm H2}$.}
\label{MWvsnearby}
\end{figure*}
How do the gas properties of our Galaxy compare with those determined in nearby spirals?
It is not trivial to find in the literature a proper comparison sample, which should contain spirals with a Milky-Way-like morphology and with spatially resolved \hi\ and H$_2$ data.
Here, we have focused on the sample of \citet{Leroy+08}, who have derived the atomic and molecular hydrogen surface density profiles in a 23 nearby galaxies using \hi\ data from THINGS \citep{Walter+08}, CO (J=1-0) data from the BIMA Survey of Nearby Galaxies \citep[BIMA SONG;][]{Helfer+03}, and CO (J=2-1) data from the HERA CO-Line Extragalactic Survey \citep[HERACLES;][]{Leroy+09}.
Within the sample of Leroy et al., we have selected those systems that are classified as SAB or SB according to RC3 \citep{RC3}: NGC 3184, NGC 3198, NGC 3351, NGC 3521, NGC 3627 and NGC 6946.
In addition, we have included in our sample the \hi\ surface density profiles of NGC 925 and NGC 2903, both SAB systems, from \citet{deBlok+08}.
Finally, we have complemented our sample with the \hi\ and H$_2$ density profiles of the SAB galaxy NGC 5236 (M83)\footnote{for which we assumed a distance of 4.6 Mpc \citep{Saha+06}.} from \citet{Crosthwaite+02}.
All systems have Numerical Hubble T values ranging from 3 to 7, equivalent to \emph{b}-to-\emph{d} stages in the classical de Vaucouleurs classification.

In Fig.\,\ref{MWvsnearby} we compare the \hi\ and H$_2$ surface density profiles of the systems in our sample with those of the Milky Way.
Overall, the \hi\ radial distribution of our Galaxy does not differ from that of the other spirals, especially at $R\!>\!3\kpc$.
With the exception of NGC 925, all galaxies show a depression in their inner ($R\!\lesssim\!3\kpc$) \hi\ content, analogous to - but less prominent than - that of the Milky Way.
Differences are more evident when it comes to the molecular gas content, as the Milky Way appears to be one of the most H$_2$-poor galaxies in the sample, along with NGC 3198 and NGC 3351 (shown by solid lines in Fig.\,\ref{MWvsnearby}).
Interestingly, these two galaxies are the only SB systems in our sample, and they both have an H$_2$ distribution analogous to that of the Milky Way, including a depression around $R\sim2\kpc$ similar to that of our Galaxy.
Both NGC 3198 and NGC 3351 are however less massive than the Milky Way, having $v_{\rm flat}$ (i.e., the rotational speed in the region where the rotation curve is flat) of $150\kms$ \citep{deBlok+08} and $196\kms$ \citep{Leroy+08} respectively\footnote{although the value quoted for NGC 3351 is quite uncertain given the low inclination of this galaxy.}.
We note that virtually all galaxies show some features in their H$_2$ surface density profile around $R\!=\!2\!-\!3\kpc$.

Even though our galaxy sample is small, our analysis suggests that the way the cold gas is distributed in our Galaxy is not untypical for a barred spiral, and in particular for an SB system.
There is general consensus that gas flow in a barred potential can produce a depression in the gas surface density in the regions of the bar \citep[e.g.][]{Athanassoula92}.
However, the details of the gas distribution within and around the bar depend on the physical condition of the bar, in particular on its pattern speed and exact shape \citep[e.g.][]{Sormani+15}.
We argue that variations in the bar shape and speed can justify the variety in surface density profiles shown by the systems in our sample.

\begin{figure}[tbh]
\centering
\includegraphics[width=0.48\textwidth]{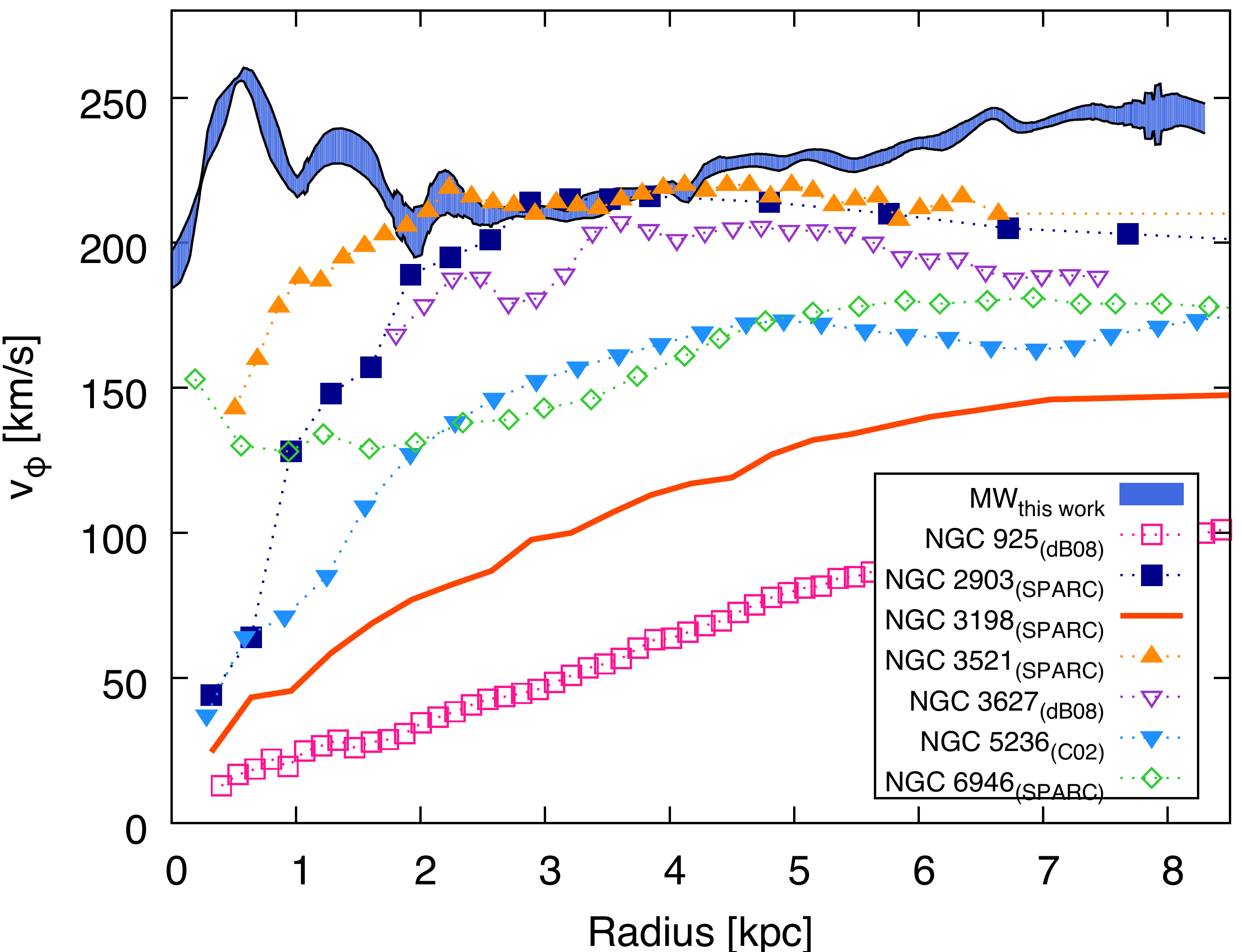}
\caption{Comparison between the rotation curve of the Milky Way (shaded region) and those of a sample of nearby barred spirals (symbols and lines) from the SPARC sample (NGC 2903, NGC 3198, NGC 3521, NGC 6946), from \citet[][dB08]{deBlok+08} (NGC 925, NGC 3627) and from \citet[][C02]{Crosthwaite+02} (NGC 5236). Symbols and colours are the same as in Fig.\,\ref{MWvsnearby}.}
\label{MWvsnearbyVrot}
\end{figure}

In Fig.\,\ref{MWvsnearbyVrot} we compare the \hi\ rotation curve of the Milky Way with those of the other galaxies in our sample.
The rotation curves of the external galaxies are taken from the Spitzer Photometry \& Accurate Rotation Curves (SPARC) database \citep{SPARC} (NGC 2903, NGC 3198, NGC 3521, NGC 6946), from \citet{deBlok+08} (NGC 925, NGC 3627) and from \citet{Crosthwaite+02} (NGC 5236).
NGC 3184 and NGC 3351 are part of the THINGS sample but have not been studied by de Blok et al., given their low inclination. 
Despite their variety in dynamical range, all rotation curves in our sample of external galaxies (except for NGC 6946) show a shallow rise in the inner parts followed by a flattening in the outer regions.
The Galactic rotation curve follows instead a different trend: its steep rise followed by wiggles and a rapid decline in the innermost $\sim2.5$ kpc strikes as a unique feature.
As discussed in Section \ref{effectofbar}, the Galactic rotation curve at $R\!\lesssim\!2.5\kpc$ is strongly contaminated by non-circular motions triggered by the bar and it is hardly representative for the circular velocity in that region.
It is hard to compare the impact of non-circular motions on the derived rotation curve of the Milky Way with that on the rotation curves of external galaxies, given the completely different technique adopted to extract the rotation velocities.
It well possible that the `true' rotation curve of our Galaxy, once cleaned from non-circular motions, may appear similar to some of those in our sample (for instance, like that of NGC 3521).

\section{Conclusions}\label{conclusions}
We have modelled the kinematics and the distribution of the atomic and CO-bright molecular hydrogen in the inner ($R\!<\!\rsun$) disc of the Milky Way. 
We have assumed that the Galactic disc can be decomposed into a series of concentric and co-planar rings, and that the gas in each ring can be fully described by four parameters: the rotation velocity, the velocity dispersion, the midplane density and the scale height.
The model is axisymmetric, but we have modelled the receding (QI) and the approaching (QIV) sides of the Galaxy separately.
Non-circular motions are not included in our model.

The parameters of each ring are fitted to the \hi\ data of the LAB survey \citep{Kalberla+05} and to the CO data of \citet{Dame+01}. 
Our fitting strategy consisted in simulating the various \hi\ and CO line-profiles in the regions of interest and comparing them with the observed profiles.
The fit proceeds iteratively from the outermost sight-lines ($l\!=\!90\de$ for QI or $l\!=\!270\de$ for QIV) to the innermost one ($l\!=\!0\de$), so that the parameters of each ring depend on those determined for the other rings at larger radii.
With our method, for a given ring a) all parameters can be fit simultaneously to the data without assuming a-priori a radial profile for the gas density or velocity dispersion, b) the terminal velocity in the line profile is not related a-priori to the gas at the tangent point radius.
To our knowledge, this is the first time that such a novel approach is used to model both the \hi\ and the H$_2$ in the Galaxy.

Our results can be summarized as follows:
\begin{itemize}
\item our disc model reproduces very well the $l\!-\!v$ diagrams for both the \hi\ and the CO emissions in the inner Galaxy up to a few degrees above the midplane. The agreement with the \hi\ is more evident as the atomic gas is distributed more smoothly within the disc.
\item the approaching and the receding sides of the inner Galaxy show very similar rotation, velocity dispersion and scale height profiles, suggesting a large degree of axisymmetry in the inner disc. Discrepancies between the two sides occur mainly at $R\!<\!2.5\kpc$ and are most likely induced by the Galactic bar.
\item The \hi\ and H$_2$ rotation curves overlaps remarkably well with each other. They rise in the centre and reach a flat part at $R\!>\!6.5\kpc$ (for $\vsun\!=\!240\kms$). Some bumps visible in the molecular gas rotation are likely caused by the patchy distribution of molecular clouds in the disc.
\item The \hi\ (H$_2$) velocity dispersion peaks at the Galactic centre with a value of $\sim30\kms$ ($\sim20\kms$), decreases rapidly in the innermost $1.5\kpc$ ($1\kpc$) and remains constant at larger radii, with a typical value of $8.9\kms$ ($4.4\kms$). In the solar neighborhood we derive a value of $\sim7.8\kms$ ($\sim3.7\kms$).
\item The \hi\ (H$_2$) scale-height is fairly flat for $R\!>\!3\kpc$, with a Gaussian HWHM of $202\pm28\pc$ ($64\pm12\pc$).
\item The \hi\ and H$_2$ surface density profiles show several common kpc-scale features: they both peak at Galactocentric radii of $4.5$ and $6.4\kpc$, and show a deep depression at $0.5\!<\!R\!<\!2.5\kpc$.
The peaks represent either genuine, axisymmetric ring-like overdensities within the disc, or local features due to the line-of-sights intercepting the Galactic spiral arms.
The depression is most likely caused by the Galactic bar.
The profiles are in good agreement with those determined by \citet{bm98}, but are in tension with other determinations, in particular with the \hi\ surface density derived by \citet{KalberlaDedes08}.
\item We estimate an \hi\ (H$_2$) surface density in the solar neighborhood of $4.5\pm0.7\msunsqp$ ($1.3\pm0.7\msunsqp$).
\item When compared with nearby barred spiral galaxies, the gas distribution in the Milky Way is not exceptional. In particular, it seems analogous to that of other SB galaxies like NGC 3198 and NGC 3351.
\end{itemize}

Additionally, we have built a simple model for the Galactic ISM and demonstrated that a spin temperature of $\sim150\K$ is optimal to describe the typical \hi\ opacity in the LAB dataset.
Neglecting the \hi\ opacity leads to underestimate the \hi\ mass within the inner Galaxy by $\sim30\%$. 

We have found that the \hi\ emission is better described by assuming two components, the first with high density ($\sim0.4$ cm$^{-3}$ on the midplane) and low velocity dispersion ($8\kms$), and the second with low density ($\sim0.04$ cm$^{-3}$ on the midplane) and high velocity dispersion ($15\!-\!20\kms$). 
We argue that the second component originates either from non-circular motions induced by non-axisymmetric features in the disc (like spiral arms), or from star formation feedback.
In the former scenario, this component would serve the sole purpose of absorbing insufficiencies in our model, and therefore would be artificial.

We have stressed that the gas flow in a barred galaxy like the Milky Way results in a asymmetric, non-stationary gas distribution and kinematics in the regions of the bar.
For this reason, we discourage the reader to trust our rotation velocity measurements in the innermost $\sim2.5\kpc$.

\begin{acknowledgements}
AM would like to thank Renzo Sancisi for critically reading this manuscript and for providing constructive comments, and Erwin de Blok for providing the rotation curves and the surface density profiles of THINGS galaxies. JMvdH acknowledges support from the European Research Council under the European Union's Seventh Framework Programme (FP/2007-2013)/ERC Grant Agreement no. 291531.
The authors thank an anonymous referee for insightful comments.
\end{acknowledgements}


\bibliographystyle{aa} 
\bibliography{aa} 

\appendix
\section{The $N_{\rm HI}(v)-T_{\rm B}$ relation in the Galaxy for a realistic ISM}\label{appendix_NHI_Tb}
In this Appendix we study the relation between the intrinsic \hi\ column density per unit velocity, $N_{\rm HI}(v)$, and the observed brightness temperature, $T_{\rm B}$, for a light-cone that intersects the Galactic disc.
The medium swept up by this light-cone is modelled as an ensemble of clouds with properties that are typical for the interstellar medium of the Milky Way.
The brightness temperature profile is derived by assuming spatial and velocity resolutions analogous to those of the LAB survey.

In Appendix \ref{A_multiphase_clumpy} we describe the radiative transfer equations for a generic ensemble of \hi\ clouds with different sizes, temperatures and densities.
Next, in Appendix \ref{multiphase_galaxy} and \ref{cloudkinematics} we set up a simple model for the Galactic ISM that uses a limited number of free parameters to fully describe the cloud internal properties, overall distribution and kinematics for any generic sight-line.
Given a set of free parameters, \hi\ line profiles can be derived as shown in Appendix \ref{syntheticprofiles}.
Finally, the $N_{\rm HI}(v)-T_{\rm B}$ relation as resulting from our modelling is discussed in Appendix \ref{results}.

\subsection{Radiative transfer in a multiphase clumpy medium}\label{A_multiphase_clumpy}
As we discussed in Section \ref{lineprofile}, eq.\,(\ref{radiative}) and (\ref{tau}) give the solution to the equation of radiative transfer and the optical depth for an homogenous and isothermal \hi\ layer that fills uniformly the telescope beam.

We now consider the case of two overlapping layers of \hi, one in the foreground (layer A) and one in the background (layer B), both filling the telescope beam.
Given that the emission from layer B is partially absorbed by layer A, the $T_{\rm B}$ is not simply given by the sum of the two components. 
Instead we have
\begin{equation}\label{Tb_overlap_2}
T_{\rm B}(v) = T_{\rm s,A}\,(1-e^{-\tau_{\rm A}(v)}) + T_{\rm s,B}\,(1-e^{-\tau_{\rm B}(v)})e^{-\tau_{\rm A}(v)}\,.
\end{equation}
It is trivial to generalise eq.\,(\ref{Tb_overlap_2}) to a number $N$ of overlapping layers.
Consider the layers ordered along the line of sight such that layer $i+1$ is located behind layer $i$ (where $1<i<N$), then
\begin{equation}\label{Tb_overlap}
T_{\rm B}(v) = \displaystyle\sum_{i=1}^{N} T_{{\rm s},i}\,(1-e^{-\tau_i(v)})e^{-\sum_{j=1}^{i-1}{\tau_j}}\,.
\end{equation}

Consider now the case of a spherical, homogeneous and isothermal cloud of \hi\ with radius $R_{\rm cl}$ and \hi\ mass $M_{\rm cl}$.
Assume that the sphere is unresolved, i.e., $R_{\rm cl}<R_{\rm beam}$, where $R_{\rm beam}$ is the equivalent radius of the telescope beam (in physical units).
Then brightness temperature will be given by the right-hand side of eq.\,(\ref{radiative}) diluted by a factor $\Omega=(R_{\rm cl}/R_{\rm beam})^2$.
In the case of a collection of N clouds that \emph{do not} shade each other along the line of sight, $T_{\rm B}$ will be
\begin{equation}\label{Tb_ensamble_a}
T_{\rm B}(v) = \displaystyle\sum_{i=1}^{N} {\Omega_i T_{{\rm s},i}\,(1-e^{-\tau_i(v)})}
\end{equation}
where $\tau_i(v)$, the optical depth of the $i-th$ cloud, is still given by eq.\,(\ref{tau}) being $N_{\rm HI}$ the cloud surface density (thus $\propto M_{\rm cl} R_{\rm cl}^{-2}$).

Finally, by using both eq.\,(\ref{Tb_overlap}) and eq.\,(\ref{Tb_ensamble_a}), it is possible to derive the $T_{\rm B}$ for an ensemble of $N$  clouds that partially shade each others:
\begin{equation}\label{Tb_final}
T_{\rm B}(v) = \displaystyle\sum_{i=1}^{N} \Omega_i T_{{\rm s},i}\,(1-e^{-\tau_i(v)})F_i(v)
\end{equation}
where $F_i(v)$ is a correction term due to foreground absorption from clouds in front of cloud $i$ (at velocity $v$), and is given by
\begin{equation}\label{foreground}
F_i(v) =  \displaystyle\sum_{k=1}^{m} \delta_k \exp{\left(-\displaystyle\sum_{j<i}\tau_j(v)\right)}
\end{equation}
where the first sum is extended to all $m$ `shaded' sectors in which cloud $i$ is partitioned by the foreground clouds (see example below), and $\delta_k$ is the area of sector $k$ divided by the area of cloud $i$ (so that $\sum_{k=1}^{m}\delta_k=1$).
Note that $F\le1$ by construction.

\begin{figure}[tbh]
\begin{center}
\includegraphics[width=0.35\textwidth]{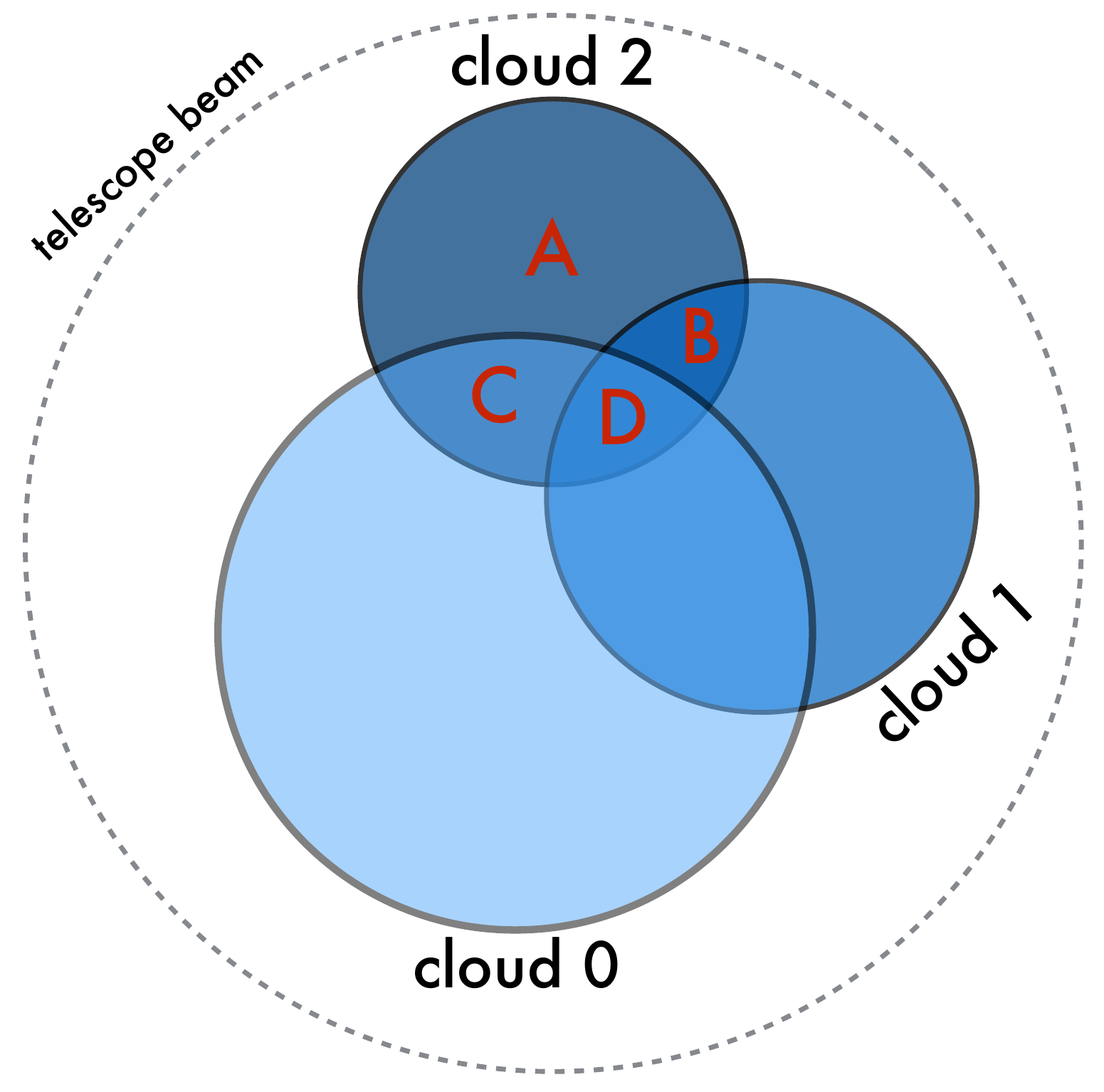}
\caption{Distribution of clouds in the example discussed in the text. Cloud 0 is in the foreground, cloud 2 is in the background.}
\label{foreground_fig}
\end{center}
\end{figure}

The foreground correction depends on how the clouds are distributed in space.
As an example, we compute $F$ for the background cloud in a system of three clouds that are partially shading each other and have the same velocity (Fig.\,\ref{foreground_fig}).
The cloud in the background (cloud 2) is divided into sectors A, B, C and D by the two foreground clouds. 
Sector A is unshaded, sectors B and C are shaded by clouds 1 and 0 respectively, while sector D is shaded by clouds 1 and 0 simultaneously.
Thus the foreground correction for cloud 2 will be given by
\begin{equation*}
F_2(v) = \delta_A + \delta_B e^{-\tau_1(v)} + \delta_C e^{-\tau_0(v)} + \delta_D e^{-\left(\tau_0(v)+\tau_{1}(v)\right)}
\end{equation*}
and $\delta_A + \delta_B + \delta_C + \delta_D = 1$\,.

Note that, for a generic distribution of clouds, computing the various $\delta_k$ terms in eq.\,(\ref{foreground}) is not trivial.
In the calculations presented here, we found more practical to sample randomly each cloud $i$ with a number $\eta$ of points, and study which foreground clouds shade the various regions of cloud $i$ point-by-point.
We adopt $\eta\!=\!250$, which we found to be the best compromise between accuracy and computational speed.

Eq.\,(\ref{Tb_final}) can be used to predict any \hi\ brightness temperature profile \emph{if} a model for the cloud distribution in mass, radius, temperature, location and velocity is provided. 
We discuss such a model below. 

\subsection{A simple model for the neutral ISM}\label{multiphase_galaxy}
The model that we adopt is a simplified version of that of \citet[][hereafter MO77]{McKeeOstriker77}.
We assume that the \hi\ is made by a collection of homogeneous and isothermal clouds, in pressure equilibrium with each other at the ISM pressure of $P_{\rm ISM}/k_{\rm B}\!=\!3000\K\cmmc$ \citep{Wolfire+03}.
These clouds follow distributions in temperature, size, masses and velocity that we discuss below.
To simplify the calculation, even though these clouds are distributed in a 3D space, we assume that they effectively behave as \emph{discs} of constant surface density, and have no physical depth. 

\begin{figure}[t]
\begin{center}
\includegraphics[width=0.49\textwidth]{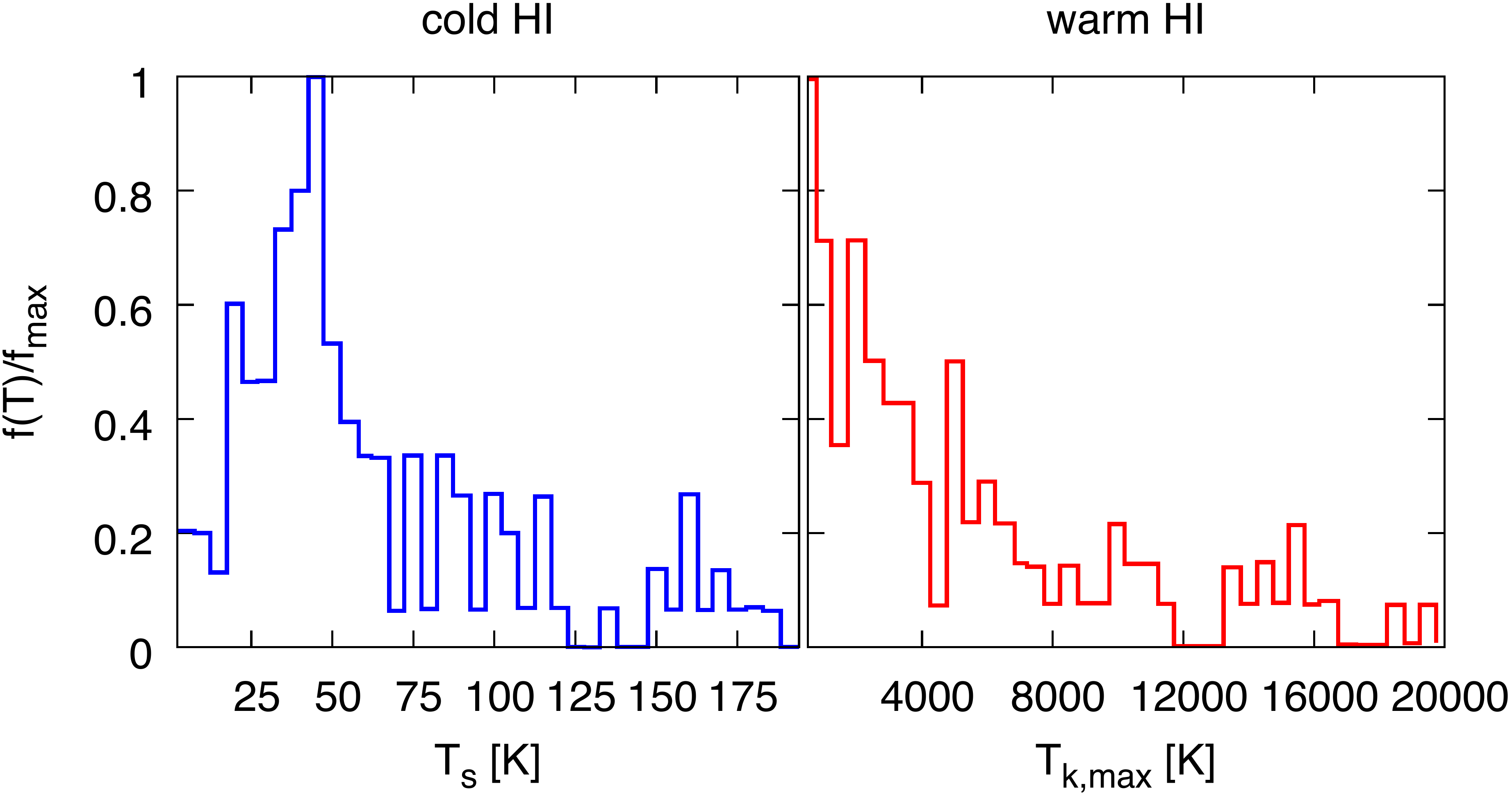}
\caption{Spin temperature distribution for the cold \hicap\ (\emph{left panel}) and maximum kinetic temperature distribution for the warm \hicap\ (\emph{right panel}) as derived by \citet{HT03b}.}
\label{Ts_distrib}
\end{center}
\end{figure}

Each cloud is made by two components: an outer envelope of warm material with optically thin properties, and a cold optically thick core.
In the MO77 model, the typical temperatures of these components are $8000\K$ for the warm envelope and $80\K$ for the cold core.
However, \citet[][hereafter HT03]{HT03b} studied in detail the interstellar \hi\ emission and absorption lines in the solar neighborhood using a Gaussian decomposition technique, finding a broad range of temperatures for both the warm and the cold neutral medium. 
The temperature distributions found by HT03 are shown in Fig.\,\ref{Ts_distrib}.
Note that spin temperatures can be measured only for the coldest (absorbing) \hi\ clouds, whereas only an upper limit to the \hi\ kinetic temperature $T_{\rm K}$ can be inferred for warmer clouds from the broadening of their emission profile.
Spin and kinetic temperatures are virtually identical for $T_{\rm S}\!\lesssim\!1000\K$, but $T_{\rm K}\!\ge\!T_{\rm S}$ at higher $T_{\rm S}$ \citep{Liszt01}.
However, for $T_{\rm S}\!>\!1000\K$ and for the column density range that we investigate here, the \hicap\ optical depth is sufficiently small that even overestimating $T_{\rm S}$ by thousands of $K$ produces little variation in the line profiles. 
Therefore, in the following we make no distinction between $T_{\rm S}$ and $T_{\rm K}$ and assume that clouds in our model follow the temperature distributions shown in Fig.\,\ref{Ts_distrib}.

In the MO77 model, two additional envelopes of warm and hot ionised gas surround each cloud, and the mass ratio between the different components is set by the detailed balance between the cloud physical properties and an external ionising UV and X-ray background. 
We do not include such level of detail in our model. 
Instead, we assume that the mass fraction of the warm neutral gas, $f_w=m_w/(m_w+m_c)$, is constant for each cloud.
HT03 found that the average warm-to-total \hi\ ratio in the solar neighborhood is $\sim0.6$, remarking that this is likely a lower limit.
The size of the warm clouds follows a power law distribution with a slope of $-3.6$, according to the observations of \citet{Miville+03}, who pointed out that this is in agreement with the Kolmogorov prediction for an incompressive turbulent fluid.
Consistently with the MO77 model, we assume that the minimum and maximum radii for the warm clouds are $R_{\rm min}\!=\!2.1\pc$ (below which clouds are easily disrupted by supernova shockwaves) and $R_{\rm max}\!=\!10.8\pc$ (above which clouds are gravitationally unstable) respectively.
Assuming a uniform size distribution in the same range has very little impact on our results.
Given the steepness of the size distribution, the value of $R_{\rm max}$ does not affect our results: we attempted to use $R_{\rm max}\!=\!100\pc$ and found no variations in the $N_{\rm HI}(v)-T_{\rm B}$ relation for our `fiducial' model (see below).
Our results are somewhat more sensitive to the minimum cloud size: for instance, using $R_{\rm min}\!=\!1\pc$ gives a slightly more `optically thin' $N_{\rm HI}(v)-T_{\rm B}$ relation.
It is difficult to define precise boundaries for the cloud sizes, as they depend on the local condition of the ISM.
Most importantly, we show below that the main driver of the $N_{\rm HI}(v)-T_{\rm B}$ relation is the warm gas fraction, while all the other variables seem to play a secondary role.
For these reasons, we decide to fix the cloud size boundaries to the values proposed by MO77.

Given that the temperature, pressure and size for a given warm cloud are set, then its mass $m_w$ will be $\propto P_{\rm ISM} T_{\rm S}^{-1}R_{\rm cl}^3$ and its optical depth is $\propto P_{\rm ISM} R_{\rm cl} T_{\rm S}^{-2}$.
The choice of $f_w$ sets the cold core mass $m_c$ and then its size and optical depth.
We verified that, with these choices for the temperature and size distributions, virtually all warm (cold) components have $\tau\ll1$ ($\tau>1$).

\subsection{Cloud distribution and kinematics} \label{cloudkinematics}
In order to derive a brightness temperature profile via eq.\,(\ref{Tb_final}), we must set up a virtual observation for our model of neutral interstellar medium described above.
For this purpose we consider a synthetic light-cone that, radiating from an observer placed within the Galactic disc at $R\!=\!\rsun$, intersects the ISM of the inner Galaxy at latitude $b\!=\!0$.
For simplicity, we consider a Galaxy model with a constant rotation velocity $v_{\phi}(R)=\vsun$ and a constant \hi\ density $n_{\rm HI}$. 

At a chosen longitude $l$, the maximum line-of-sight distance within the inner disc is $d_{\rm max}(l)=2\rsun\cos(l)$ (so $0\!<\!d\!<\!d_{\rm max}$) and, at this distance, the total volume encompassed by our synthetic light-cone will be $\frac{1}{3}\pi d^3_{\rm max}\tan^2(\alpha/2)$, where $\alpha$ is the FWHM angular resolution of our virtual observation.
We distribute our warm and cold clouds randomly within our light-cone until its mean density reaches the value of $n_{\rm HI}$.
The physical parameters of the clouds are extracted randomly from the distributions discussed in the previous Section (see also Appendix \ref{syntheticprofiles}).
A cloud at a given distance $d$ along the light-cone will have line-of-sight velocity given by eq.\,(\ref{vlos}).
For simplicity, we treat the cold core-warm envelope as two separate discs that share the same centre and velocity.

Then we need another step.
If all \hi\ in a cloud was at velocity $v_0$ given by eq.\,(\ref{vlos}), then the line profile would be a delta function.
In reality, thermal and turbulent motions broaden the gas velocity distribution around $v_0$. 
In our calculation, what matters is how the \hi\ optical depth of a given cloud is distributed in velocity, for which we assume
\begin{equation}\label{tauv} 
\tau(v) = \tau_0\,\mathcal{N}\left(v - v_0, (\sigma^2_{\rm th}+\sigma^2_{\rm turb})^\frac{1}{2}\right)
\end{equation}
where $\tau_0$ is the integrated optical depth of the cloud (proportional to the total surface density divided by $T_{\rm S}$), $\mathcal{N}(v,\sigma)$ is a normal distribution with standard deviation equal to $\sigma$, $\sigma_{\rm th}$ is the thermal velocity dispersion of the gas (equal to $\sqrt{k_{\rm B} T_{\rm S}/m_{\rm p}}$, where $k_{\rm B}$ is the Boltzmann constant and $m_{\rm p}$ is the proton mass) and $\sigma_{\rm turb}$ is an additional turbulent component.
We have verified that a different prescription for turbulence, based on cloud-to-cloud relative motions rather than on the internal motions of clouds, gives results similar to the current implementation.

\subsection{Building synthetic \hi\ profiles} \label{syntheticprofiles}
We briefly summarise the steps required to compute the \hi\ brightness temperature profile for a given Galactic longitude $l$, and therefore to study the $N_{\rm HI}(v)-T_{\rm B}$ relation.

First, we distribute \hi\ clouds within a light-cone as follows.
\begin{itemize}
\item Given the longitude and the beam FWHM of the virtual observation, we define the geometry of the light-cone within which clouds will be randomly distributed.
\item A warm neutral cloud is created at a random position within the light-cone. The cloud size and temperature are extracted randomly from the distributions discussed in Appendix \ref{multiphase_galaxy}. Given that $P_{\rm ISM}$ is fixed, the cloud mass, surface density and optical depth are set univocally. The location of the cloud sets its line-of-sight velocity (given a flat rotation curve). The cloud temperature, along with the chosen internal turbulence $\sigma_{\rm turb}$, sets the cloud velocity dispersion.
\item A cold neutral cloud is created at the same location, with a temperature extracted randomly from the distribution discussed in Appendix \ref{multiphase_galaxy}. The choice of $f_w$, the warm-to-total \hi\ ratio, sets univocally all the other cloud parameters.
\item The previous two steps are repeated until the \hi\ volume density within the light-cone reaches the desired value, $n_{\rm HI}$. The properties of each cloud are stored. The total number of clouds varies depending on the model parameters and on the longitude (for instance, the `fiducial' model shown in Appendix \ref{results} at $l\!=\!70\de$ uses $\sim10^4$ clouds).
\end{itemize} 
At this stage, an array $N_{\rm HI}(v_k)$ is built by using the surface densities, the line-of-sight velocities and the velocity dispersions of all clouds in the light-cone. 
We focus on the velocity range $-30\!<\!v_k\!<\!300\kms$ and adopt a velocity bin $(v_{k+1}-{v_k})$ of $1\kms$.

Next, we compute the brightness temperature profile as the sum of the contributions of each individual cloud, as shown by eq.\,(\ref{Tb_final}).
Each cloud $i$, however, contributes only to a limited portion of the profile, and specifically only around its line-of-sight velocity $v_i$.
Therefore, for each cloud we focus on the velocity range [$v_i-3\sigma_i,v_i+3\sigma_i$], being $\sigma_i$ the cloud velocity dispersion.
The following steps must be repeated for all clouds in the light-cone and, for a given cloud, for all velocity bins $v_k$ in the appropriate velocity range.
\begin{itemize}
\item We first identify a subsample of clouds that can potentially contribute to the foreground absorption.
These clouds are selected as those that a) are located between cloud $i$ and the observer, and b) have line-of-sight velocity that differs from $v_k$ by less than 3 times their velocity dispersion. 
\item Cloud $i$ is then randomly sampled by 250 points, which are used to compute $F_i(v)$ via eq.\,(\ref{foreground}) (see Appendix \ref{A_multiphase_clumpy} for further details). The terms $\tau_j(v_k)$ in eq.\,(\ref{foreground}) are given by eq.\,(\ref{tauv}). Note that, in computing the overlaps between clouds, projection effects must be taken into account.
\item The contribution of cloud $i$ to the $T_{\rm B}$ at velocity $v_k$ can be now computed. Note that, in this context, the quantity $\Omega_i$ in eq.\,(\ref{Tb_final}) represents the ratio between the \emph{angular} size of cloud $i$ and the beam FWHM.
\end{itemize}
Once all clouds have been processes, the resulting $T_{\rm B}$ profile is smoothed to a velocity resolution of $2\kms$.

As inputs, our model requires the warm-to-total \hi\ ratio $f_w$, the \hi\ volume density $n_{\rm HI}$, the turbulent component of the velocity dispersion $\sigma_{\rm turb}$ and the beam FWHM.
We fix the latter to $0.6\de$, as here we compare our synthetic \hi\ profiles with those of the (hanning-smoothed) LAB survey at full ($0.6\de$) angular resolution.
We note that, for this value of the beam, the largest light-cone intersecting the inner Galaxy - which is obtained at $l\!=\!0\de$ - has base radius (farthest away from the observer) of $87\pc$, about half of the \hi\ scale height at $R=\rsun$.
This implies that, in our model, we can neglect the vertical gradient in gas density.

\subsection{Results} \label{results}
\begin{figure*}[tbh]
\begin{center}
\includegraphics[width=0.9\textwidth]{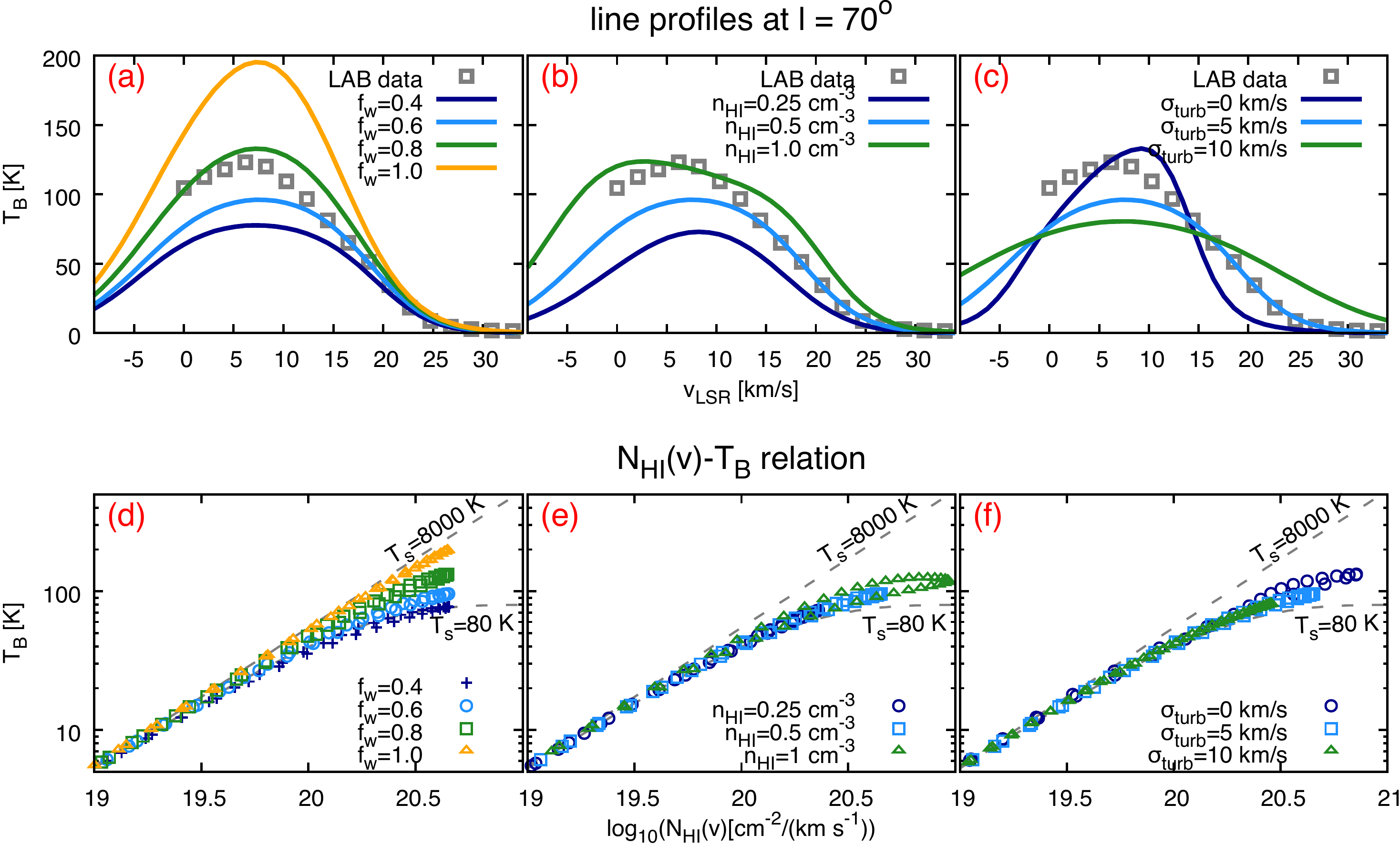}
\caption{The effect of varying the ISM model parameters on the \hicap\ line profiles (top panels) and on the $N_{\rm HI}(v)-T_{\rm B}$ relation (bottom panels) for the $l\!=\!70\de$ sight-line. In all cases, the light-cone aperture is $\alpha\!=\!0.6\de$. Panels $(a)$ and $(d)$ show the effect of varying the warm-to-total gas fraction $f_w$, panels $(b)$ and $(e)$ show the effect of varying the mean \hicap\ density in the cone $n_{\rm HI}$, while panels $(c)$ and $(f)$ show the effect of varying the intra-cloud turbulence $\sigma_{\rm turb}$. The grey squares in the top panels show the observed \hicap\ line profile from the LAB survey. In the bottom panels, the two dashed lines show the analytic solutions, given by eq.\,(\ref{radiative}), for two isothermal layers of \hicap\ at the spin temperature of $8000\K$ or $80\K$.}
\label{comparison}
\end{center}
\end{figure*}

\begin{figure}[tbh]
\begin{center}
\includegraphics[width=0.4\textwidth]{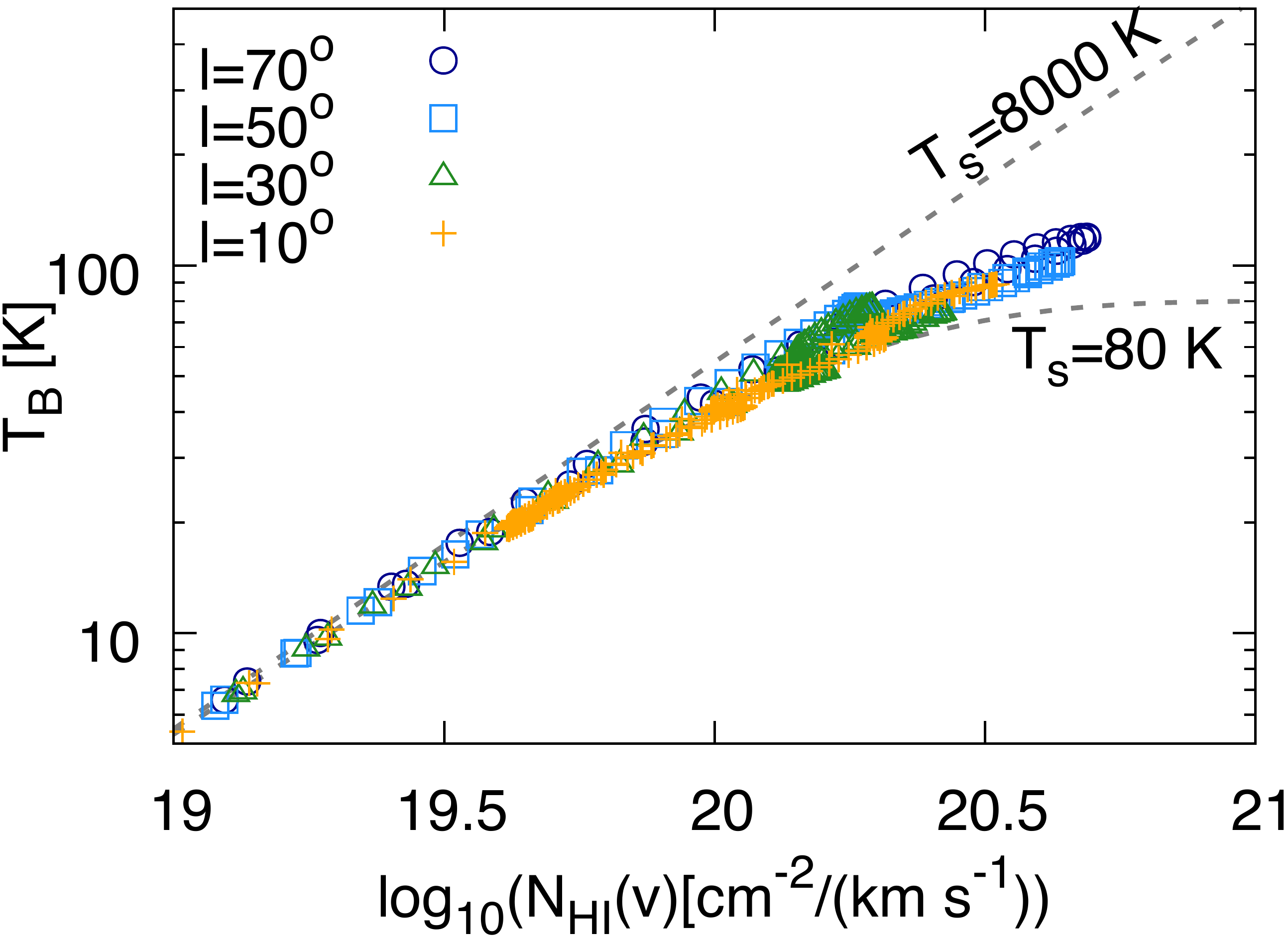}
\caption{$N_{\rm HI}(v)-T_{\rm B}$ relation for different sight-lines predicted by our fiducial ISM model. The light-cone aperture is $\alpha\!=\!0.2\de$.
The two dashed lines show the analytic solutions, given by eq.\,(\ref{radiative}), for two isothermal layers of \hicap\ at the spin temperature of $8000\K$ and $80\K$. The choice of the sightline produces little scatter in the relation.}
\label{NHIvsTB_l}
\end{center}
\end{figure}

\begin{figure}[tbh]
\begin{center}
\includegraphics[width=0.4\textwidth]{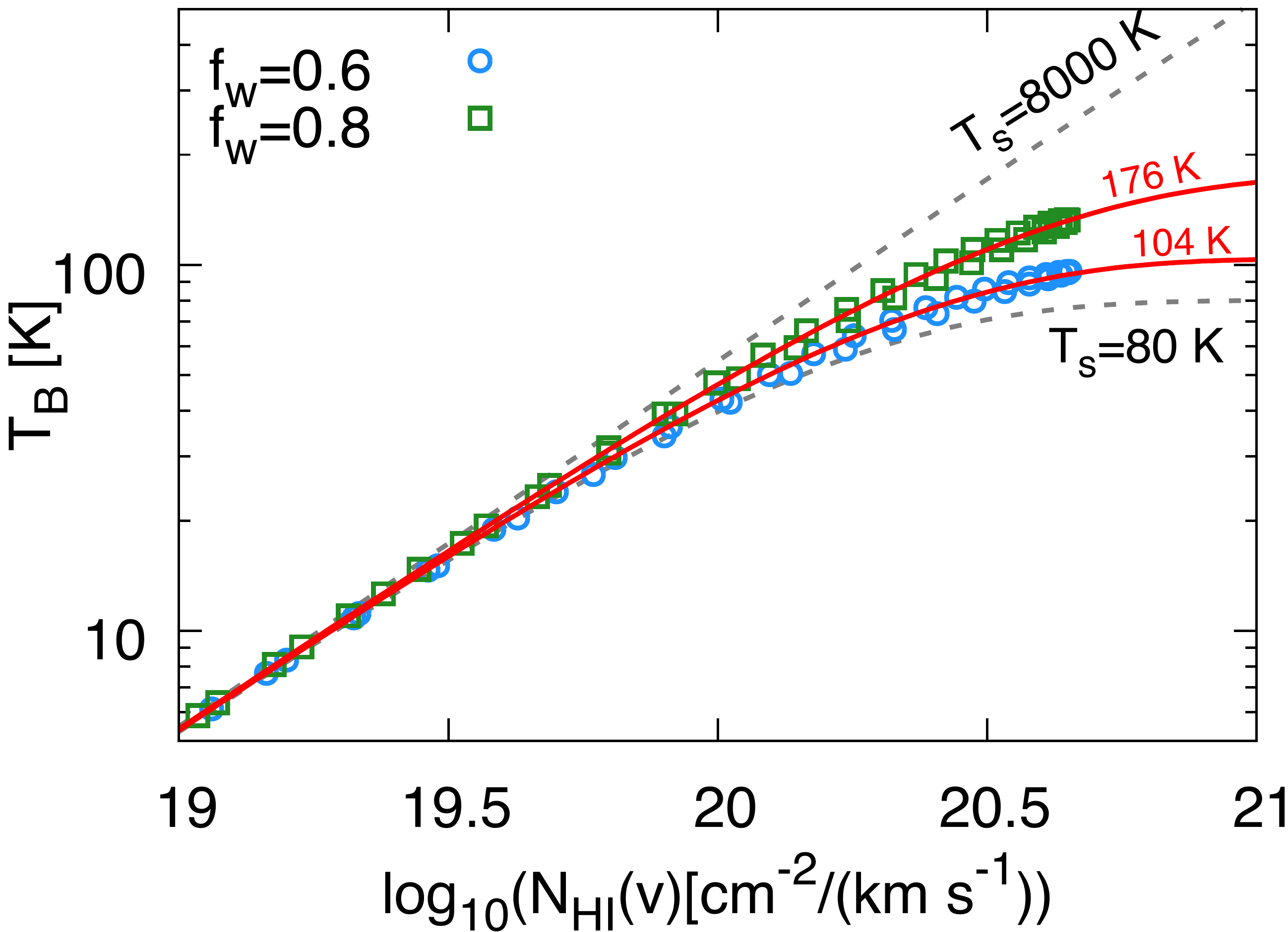}
\caption{$N_{\rm HI}(v)-T_{\rm B}$ relation for the two models ($f_w\!=\!0.6,0.8$) that better reproduce the LAB data at $l\!=\!70\de$. 
The other model parameters are $n_{\rm HI}\!=\!0.5\cmmc$, $\sigma_{\rm turb}\!=\!5\kms$. The red solid lines show the analytic predictions for two isothermal \hicap\ layers at $T_{\rm S}=104,176\K$ that provide the best-fit to these relations. 
As a reference, the predictions for two layers at $T_{\rm S}$ of $8000\K$ and $80\K$ are shown with dashed lines.}
\label{NHIvsTB_final}
\end{center}
\end{figure}

We now show how varying the parameters of our model affects the shape of the \hi\ line profiles and the relation between the intrinsic \hi\ column density per unit velocity and the brightness temperature.
We define a `fiducial' model with parameters $f_w=0.6$, $n_{\rm HI}=0.5\cmmc$, $\sigma_{\rm turb}=5\kms$ and study how the line profile and the $N_{\rm HI}(v)-T_{\rm B}$ relation change by varying one parameter at a time.
We focus on the latitude $l\!=\!70\de$, which gives a sufficiently large light-cone ($5.7\kpc$ in length) to ensure a robust cloud statistics: in the tests below, the number of clouds ranges from $5$ to $22$ thousands.

Fig.\,\ref{comparison} shows the effect of varying the model parameters ($f_w, n_{\rm HI}, \sigma_{\rm turb}$) on the \hi\ line profiles (top panels) and on the $N_{\rm HI}(v)-T_{\rm B}$ relation (bottom panel).
Remarkably, in all cases the latter falls in between the predictions for two isothermal layers of \hicap\ at the spin temperature of $8000\K$ and $80\K$, two values often quoted as representative for the temperatures of the warm and the cold phases of the atomic hydrogen.
As shown by panels (a) and (d), increasing $f_w$ boosts both the amplitude of the line profile and the brightness temperature at a given $N_{\rm HI}(v)$.
We remark that $0.6$ is likely to be a lower limit to the warm-to-total \hi\ ratio (HT03).
For $f_w\!=\!1$, the $N_{\rm HI}(v)-T_{\rm B}$ relation follows closely that of a single-phase warm ($\sim10^4\K$) homogeneous medium.
Although we do not aim to fit the observed line profile, we note that the models with $f_w\!=\!0.6$ or $0.8$ are those that better represent the LAB data.
The gas properties are likely to vary as a function of the distance from the Galactic centre, thus fluctuations in the value of $f_w$ across the disc are expected.
However, the model with $f_w\!=\!1$ not only fits the data poorly, but also largely exceeds the maximum brightness temperature measured in the LAB survey ($152\K$).  
This suggests that a cloud model for the Galactic ISM with no cold cores is inconsistent with the observations.

As shown by panels (b) and (e) of Fig.\,\ref{comparison}, increasing or decreasing $n_{\rm HI}$ produces large variations in the amplitude of the line profile but, interestingly, does not affect dramatically the $N_{\rm HI}(v)-T_{\rm B}$ relation. 
Increasing the cloud internal turbulence has the predictable effect of smearing the line profile, but has no impact on the $N_{\rm HI}(v)-T_{\rm B}$ relation, as shown by panels (c) and (f).
Note that the model with $\sigma_{\rm turb}\!=\!5\kms$ is the one that better reproduces the tail in the observed line profile.
This is consistent with our results based on fitting the LAB data (Section \ref{resultshi}), as the inferred \hi\ velocity dispersion never drops below $5\kms$ (even for the low-$\sigma$ component alone in our two-component model, see Fig.\,(\ref{result_2c_240})).

One may ask whether or not the $N_{\rm HI}(v)-T_{\rm B}$ relation depends significantly on the sight-line chosen.
Unfortunately, the computational time increases dramatically by decreasing $l$ (i.e., by increasing the light-cone length) and this makes unpractical the study of different sight-lines.
Reducing $\alpha$, the light-cone aperture, decreases the cone volume and allows us to study different sight-lines, but the resulting model can not be considered as representative for the LAB data.
In Fig.\,\ref{NHIvsTB_l} we compare the $N_{\rm HI}(v)-T_{\rm B}$ relations derived at the longitudes varying between $10\de$ and $70\de$ for a model that has our fiducial parameters, but with $\alpha\!=\!0.2\de$. 
Overall, the effect of varying the sight-line is not dramatic, as the scatter in the $N_{\rm HI}(v)-T_{\rm B}$ relation visible in Fig.\,\ref{NHIvsTB_l} is comparable to that produced by varying $n_{\rm HI}$ (see panel (e) in Fig.\,\ref{comparison}).

The above tests indicate that the warm-to-total \hi\ ratio ($f_w$) is the main driver for the $N_{\rm HI}(v)-T_{\rm B}$ relation, whereas the ISM density, the cloud turbulence and the adopted sight-line play only a secondary role.
From our analysis, it seems the values of $f_w$ between 0.6 and 0.8 are more consistent with the observations.
In Fig.\,\ref{NHIvsTB_final} we show that eq.\,(\ref{radiative}) provides an excellent fit to the $N_{\rm HI}(v)-T_{\rm B}$ relation ($l\!=\!70\de$, $\alpha\!=\!0.6\de$) with $f_w\!=\!0.6$ and $0.8$.
The resulting best-fit spin temperatures are $104\K$ and $176\K$ respectively, or $140\K$ as a mean.
Remarkably, this spin temperature range brackets the maximum brightness temperature measured in the LAB data ($152\K$).
Given that $T_{\rm B}\!\le\!T_{\rm S}$, in our calculation we decided to fix the \hi\ spin temperature to the value of $152\K$.

\section{Testing the method} \label{testing}
We test here the goodness of our modelling method (Section \ref{method}).
We perform a number of tests based on building mock \hi\ datacubes from axisymmetric, idealized gas distributions, and then running our fitting technique in an attempt to retrieve the original input parameters.

\begin{figure*}[tb]
\centering
\includegraphics[width=0.9\textwidth]{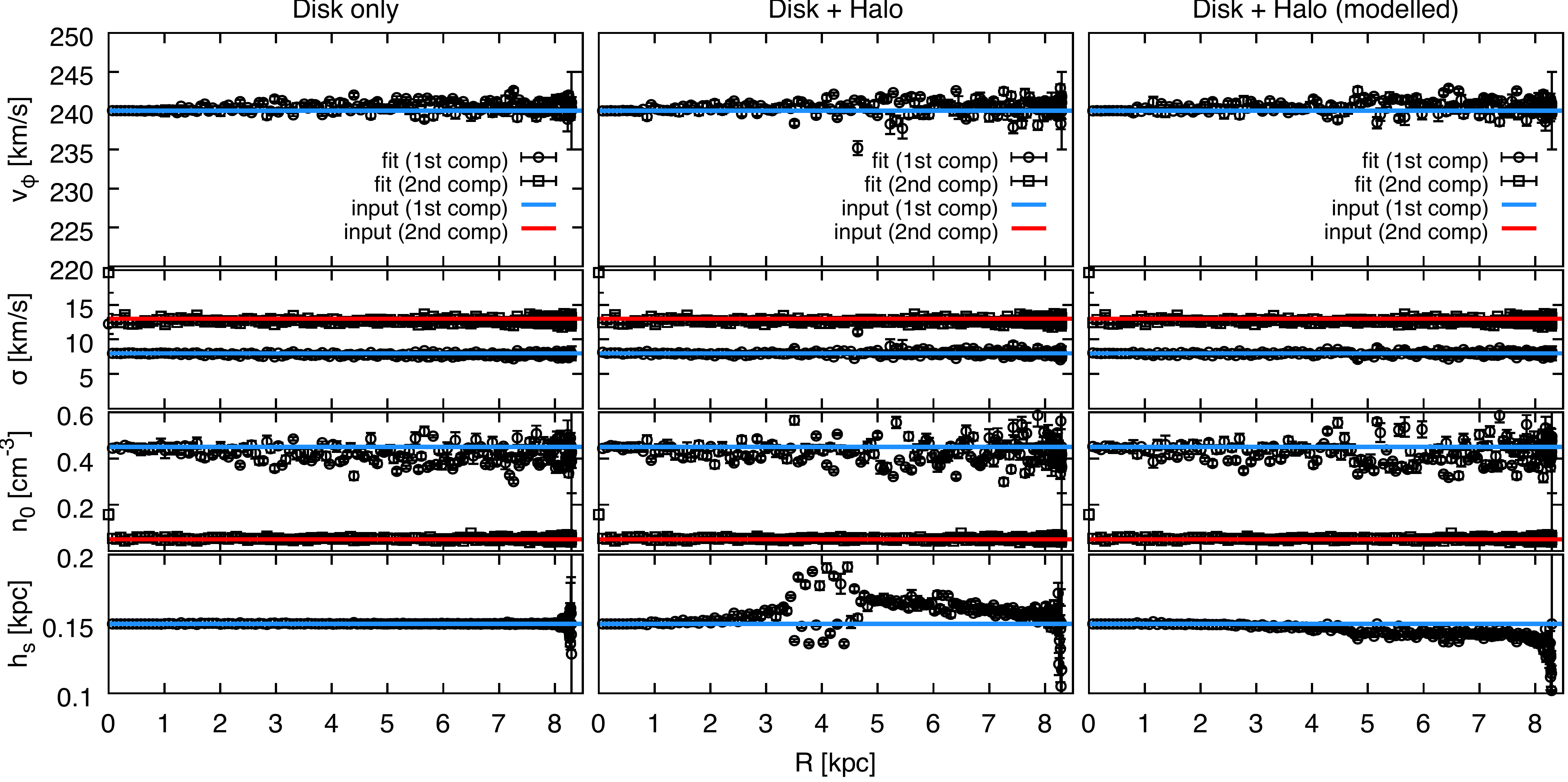}
\caption{\hicap\ properties recovered by applying our modelling technique on mock \hicap\ observations of idealised, axisymmetric systems. \emph{Left panel}: an idealised disc with flat profiles of $v_\phi$, $\sigma$, $n_0$ and $h_s$. The solid lines show the profiles used as input for the creation of the mock datacube. The system has two components, one with high-density and low $\sigma$ (blue line), and one with low-density and highe $\sigma$ (red line). The points with errorbars show the system parameters as recovered by our modelling technique. \emph{Middle panel}: the same system of the left panel, but with the additional presence of a slow-rotating and inflowing extra-planar \hicap\ (see text). Our modelling technique overpredicts the scale height of the disc. \emph{Right panel}: as in the middle panel, but now our modelling technique takes into account the presence of the extra-planar \hicap\ and recovers the input disc scale height.}
\label{mock_test}
\end{figure*}

As a first test, we consider the simple case of an \hi\ disc where all radial profiles for the various parameters are flat. 
We set $v_{\phi}\!=\!240\kms$, $h_S\!=\!0.15\kpc$ $n_{0,A}\!=\!0.45\cmmq$ cm$^{-3}$, $n_{0,B}\!=\!0.05$ cm$^{-3}$, $\sigma_{0,B}\!=\!8\kms$ and $\sigma_{0,B}\!=\!13\kms$, where A and B represents two components with different densities and velocity dispersions but with the same rotation velocity and scale height (see Section \ref{HI2comp}).
We build a mock \hi\ observation of this model and we run our fitting method on the derived mockcube in order to verify whether we are able to recover all the input parameters.
The leftmost panel of Fig.\,\ref{mock_test} shows that this is indeed the case, as the fit values (points with error bars) match the original input values (solid lines) very well. 
In particular, the scale height is recovered with almost no uncertainties, whereas $n_0$  shows some scatter for the densest component.

In our second test, we consider a mock system built by two components: the same \hi\ disc of the previous test, and an additional component representing a layer of slow rotating, inflowing extra-planar \hi\ (see Section \ref{extrapl}).
We model the latter with the following density profile \citep[from][]{Oosterloo+07}:
\begin{equation}\label{halodistrib}
n(R,z) = n_0\left(1+\frac{R}{R_g}\right)^\gamma \exp\left(-\frac{R}{R_g}\right)\exp\left(-\frac{z}{h_g(R)}\right)
\end{equation}
where
\begin{equation}\label{scaleheight}
h_g(R) = h_0 + \left(\frac{R}{h_R}\right)^\delta\ , \delta\ge0.
\end{equation}
We assume $R_g\!=\!1.08\kpc$, $h_0\!=\!175\pc$, $h_R\!=\!10.01\kpc$, $\gamma\!=\!6.30$, $\delta\!=\!1.89$ \citep[see][]{Marasco+13}, and $n_0=5\times10^{-5}$ cm$^{-3}$, so that the total \hi\ mass of this component is $\sim3\times10^8\msun$ \citep{MarascoFraternali11}.
As for the kinematics, we assume that this component rotates slower with increasing height from the midplane, with a vertical lag of $15\kmskpc$ with respect to the \hi\ in the disc, and that it is inflowing onto the disc with a speed of $30\kms$ and $20\kms$ in the radial and vertical direction respectively \citep{MarascoFraternali11}.
We build a mock observation of this model and we attempt to recover disc scale height \emph{without} making any correction for the presence of the additional component. 
Our results, shown in the middle panel of Fig.\,\ref{mock_test}, suggests that in the presence of an extra-planar \hi\ component the disc scale height at $R\!>\!3\kpc$ is slightly overestimated with respect to the input value. 
The clear peak visible around $R\!=\!4\kpc$ (or $l\!\sim\!30\de$) is due to the fact that, at this longitude, the line profiles at $|b|\!>\!0\de$ are largely contaminated by extra-planar \hi\ emission. 
This can produce a spurious increase in \hi\ surface density and total mass.

A possible way to correct for this effect is to include in our model the additional extra-planar component that we used to produce the mock observations in the first place, and proceed normally to derive the scale height of the \hi\ disc.
This procedure allows us to recover the input scale height profile much better, as shown in the rightmost panel of Fig.\,\ref{mock_test}.
We use this procedure on the LAB data in Section \ref{extrapl}.

\section{Re-scaling the model to a different $\vsun$} \label{Vsun220}
It is straightforward to re-scale our models to a different value of $\vsun$, the rotation velocity of the Solar circle, as it affects solely the rotation curve.
Once the parameters of the model are set, any choice of $\vsun$ would produce \emph{exactly the same} mock datacube if the model rotation curve $v_\phi(R)$ is re-scaled as follows:
\begin{equation}
v_\phi\,'(R) = v_\phi(R) + \frac{R}{\rsun}(\vsun' - \vsun)
\end{equation}
where $v_\phi\,'(R)$ is the new rotation curve and $\vsun' \equiv v_\phi\,'(\rsun)$.

\begin{figure*}[tbh]
\centering
\includegraphics[width=0.9\textwidth]{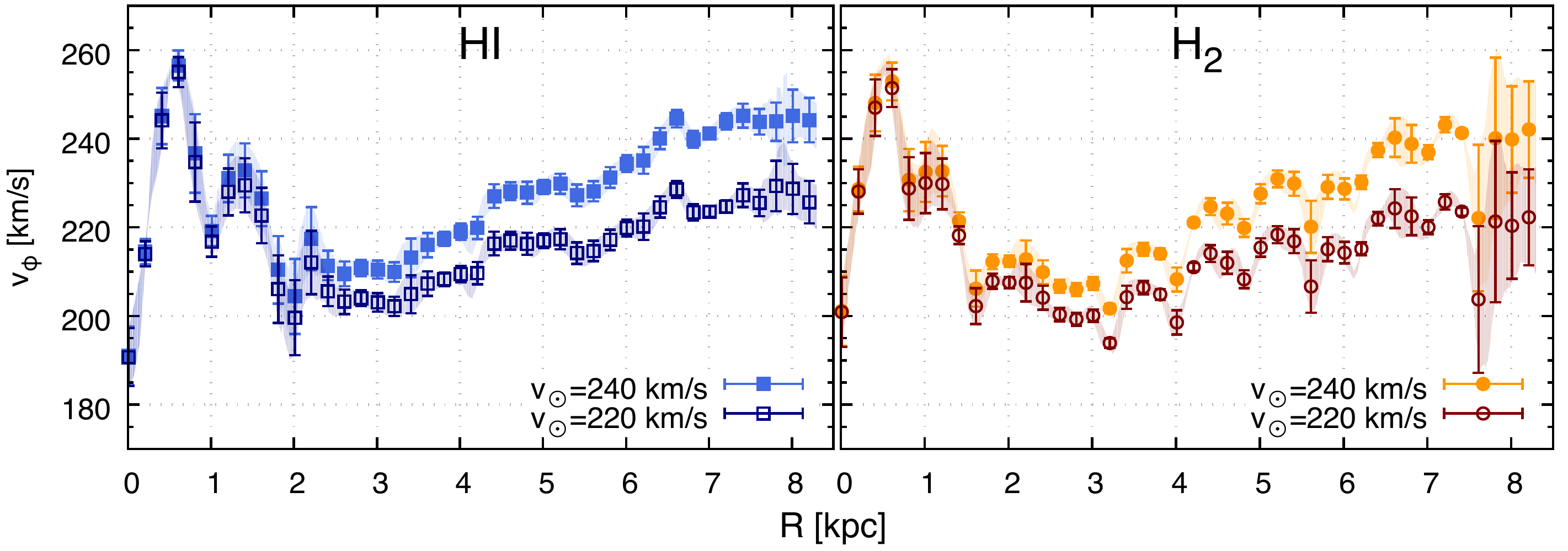}
\caption{Comparison between the Galactic rotation curves derived for $\vsun\!=\!240\kms$ (filled symbols with error bars) and  $\vsun\!=\!220\kms$ (empty symbols with error bars). The left (right) panel shows the atomic (molecular) hydrogen.}
\label{HI_vs_H2_220}
\end{figure*}

As an example, Fig.\,\ref{HI_vs_H2_220} illustrates how the \hi\ and H$_2$ Galactic rotation curve vary by using $\vsun\!=\!220\kms$ rather than $\vsun\!=\!240\kms$, the value used in the main text\footnote{Note that, after their computation, the rotation curves are smoothed to a resolution of $0.2\kpc$ and so the rotation velocity at $R\!=\!\rsun$ may slightly differ from $\vsun$.}.
Interestingly, with this lower value of $\vsun$ the rotation curve flattens at smaller radii ($\sim4\kpc$ instead of $6.5\kpc$).
We reiterate that, for gas density, velocity dispersion and scale height profiles like those in Fig.\,\ref{HI_vs_H2}, the model with this new rotation curve fits the data as well as the previous one.

\end{document}